\documentclass[preprint]{aastex}
\usepackage{graphicx}
\usepackage{epsfig}
\usepackage{amsmath}
\usepackage{natbib}
\usepackage[dvips]{color}
\usepackage{ifthen}
\newboolean{DeluxeTables}
\setboolean{DeluxeTables}{False} 

\newcommand{\ecomment}[1]{}
\newcommand{\bee}{\begin{equation}}
\newcommand{\eee}{\end{equation}}

\newcommand{\msun}{M$_{\odot}$}
\newcommand{\pc}{{\rm pc}}
\newcommand{\kpc}{{\rm kpc}}
\newcommand{\Myr}{{\rm Myr}}
\newcommand{\kms}{{\rm km~s}^{-1}}

\newcommand{\rstar}{R_{c}}
\newcommand{\sfe}{\varepsilon_{SF}}
\newcommand{\fbnd}{f_{b}}

\newcommand{\tgas}{t_{gas}}

\newcommand{\sigv}{\sigma_v}
\newcommand{\mseg}{F_{seg}}
\newcommand{\rsc}{R_{sc}}
\newcommand{\tcross}{\tau_c}
\newcommand{\lfuv}{L_{\rm FUV}}
\newcommand{\ffuv}{{\cal F}_{\rm FUV}}
\newcommand{\csec}{\langle\sigma\rangle}

\newcommand{\beStatTable}[3]{
 \newpage
\ifthenelse{\boolean{DeluxeTables}}{\begin{deluxetable*}{rcccccccccccc}}{\begin{deluxetable}{rcccccccccccc}}
 \tablewidth{0pt}
 \tablecaption{General Cluster Statistics, #1 \label{tab:#2}}
 \tablecolumns{9}
 \rotate
 \tablehead{
 \colhead{#3} & \colhead{$F_{b}$} & \multicolumn{2}{c}{$Q$} & \multicolumn{2}{c}{$R_{1/2}$ [pc]} & \multicolumn{2}{c}{$v$ [km/s]} & \multicolumn{2}{c}{$\sigma_V$ [km/s]} & \multicolumn{2}{c}{$\beta$} \\
 & (at 10 Myr) & (0-5 Myr) & (5-10 Myr) & (0-5 Myr) & (5-10 Myr) & (0-5 Myr) & (5-10 Myr)& (0-5 Myr) & (5-10 Myr)& (0-5 Myr) & (5-10 Myr)}
 \startdata
}

\newcommand{\eTable}{
 \enddata
\ifthenelse{\boolean{DeluxeTables}}{\end{deluxetable*}}{\end{deluxetable}}
}

\newcommand{\beRadTable}[4]{
\ifthenelse{\boolean{DeluxeTables}}{\begin{deluxetable*}{rcccccccccccc}}{\begin{deluxetable}{rcccccccccccc}}
 \tablewidth{0pt}
 \tablecaption{Radial Profile, #1 \label{tab:#2}}
 \tablecolumns{12}
 \tablehead{ 
 \colhead{#4} & \colhead{$p$} & \colhead{$r_0$} & \colhead{#3} & \colhead{$p$} & \colhead{$r_0$} & \colhead{#3} & \colhead{$p$} & \colhead{$r_0$} & \colhead{#3} & \colhead{$a$} & \colhead{\# Points}\\ & \multicolumn{3}{c}{(0-10 Myr)} & \multicolumn{3}{c}{(0-5 Myr)} & \multicolumn{3}{c}{(0-5 Myr)} & & &}
 \startdata
}

\newcommand{\beMassTable}[4]{
\ifthenelse{\boolean{DeluxeTables}}{\begin{deluxetable*}{rcccccccccccc}}{\begin{deluxetable}{rcccccccccccc}}
 \tablewidth{0pt}
 \tablecaption{Mass Profile, #1 \label{tab:#2}}
 \tablecolumns{12}
 \tablehead{ 
 \colhead{#4} & \colhead{$p$} & \colhead{$r_0$} & \colhead{#3} & \colhead{$p$} & \colhead{$r_0$} & \colhead{#3} & \colhead{$p$} & \colhead{$r_0$} & \colhead{#3} & \colhead{$a$} & \colhead{\# Points}
\\ & \multicolumn{3}{c}{(0-10 Myr)} & \multicolumn{3}{c}{(0-5 Myr)} & \multicolumn{3}{c}{(0-5 Myr)} & & &}
 \startdata
}

\newcommand{\beRMinTable}[4]{
\ifthenelse{\boolean{DeluxeTables}}{\begin{deluxetable*}{lccccccccccc}}{\begin{deluxetable}{lccccccccccc}}
 \tablewidth{0pt}
 \tablecaption{Close Approach Profiles, #1 \label{tab:#2}}
 \tablecolumns{12}
 \rotate
 \tablehead{
 \colhead{#4} & \colhead{$\Gamma_0$} & \colhead{$\gamma$} & \colhead{#3} & \colhead{$\Gamma_0$} & \colhead{$\gamma$} & \colhead{#3} & \colhead{$\Gamma_0$} & \colhead{$\gamma$} & \colhead{#3} & \colhead{$r_0$} & \colhead{\# Points} \\
 & \multicolumn{3}{c}{(0-10 Myr)}& \multicolumn{3}{c}{(0-5 Myr)}& \multicolumn{3}{c}{(5-10 Myr)} & &}
 \startdata
}

\newcommand{\beSStatTableQ}[4]{
\ifthenelse{\boolean{DeluxeTables}}{\begin{deluxetable*}{rcccc}}{\begin{deluxetable}{rcccc}}
 \tablewidth{0pt}
 \tablecaption{#4 as a Function of #2\label{tab:#3}}
 \tablecolumns{5}
 \tabletypesize{\footnotesize}
 \tablehead{
 & \multicolumn{2}{c}{($\rstar \sim N^{1/2}$)} & \multicolumn{2}{c}{($\rstar \sim N^{1/4}$)} \\
 & ($Q_i = 0.04$) & ($Q_i = 0.5$) & ($Q_i = 0.04$) & ($Q_i = 0.5$) \\
  \colhead{#2} & #1 & #1 & #1 & #1 }
 \startdata
}

\newcommand{\beSStatTableNThree}[4]{
\ifthenelse{\boolean{DeluxeTables}}{\begin{deluxetable*}{rccc}}{\begin{deluxetable}{rccc}}
 \tablewidth{0pt}
 \tablecaption{#4 as a Function of #2\label{tab:#3}}
 \tablecolumns{4}
 \tabletypesize{\footnotesize}
 \tablehead{
 & ($N=300$) & ($N=1000$) & ($N=2000$) \\
  \colhead{#2} & #1 & #1 & #1 }
 \startdata
}

\newcommand{\beSStatTableNTwo}[4]{
\ifthenelse{\boolean{DeluxeTables}}{\begin{deluxetable*}{rcc}}{\begin{deluxetable}{rcc}}
 \tablewidth{0pt}
 \tablecaption{#4 as a Function of #2\label{tab:#3}}
 \tablecolumns{3}
 \tabletypesize{\footnotesize}
 \tablehead{
 & ($N=300$) & ($N=1000$) \\
  \colhead{#2} & #1 & #1  }
 \startdata
}

\newcommand{\beRminTableQ}[2]{
\ifthenelse{\boolean{DeluxeTables}}{\begin{deluxetable*}{rcccccccc}}{\begin{deluxetable}{rcccccccc}}
 \tablewidth{0pt}
 \tablecaption{Interaction Rate Parameters as a Function of #1\label{tab:#2}}
 \tablecolumns{9}
 \tabletypesize{\footnotesize}
 \tablehead{
 & \multicolumn{4}{c}{($\rstar \sim N^{1/2}$)} & \multicolumn{4}{c}{($\rstar \sim N^{1/4}$)} \\
 & \multicolumn{2}{c}{($Q_i = 0.04$)} & \multicolumn{2}{c}{($Q_i = 0.5$)} & \multicolumn{2}{c}{($Q_i = 0.04$)} & \multicolumn{2}{c}{($Q_i = 0.5$)} \\
  \colhead{#1} & $\Gamma_0$ & $\gamma$ & $\Gamma_0$ & $\gamma$ & $\Gamma_0$ & $\gamma$ & $\Gamma_0$ & $\gamma$ }
 \startdata
}

\newcommand{\beRminNThree}[2]{
\ifthenelse{\boolean{DeluxeTables}}{\begin{deluxetable*}{rcccccc}}{\begin{deluxetable}{rcccccc}}
 \tablewidth{0pt}
 \tablecaption{Interaction Rate Parameters as a Function of #1\label{tab:#2}}
 \tablecolumns{7}
 \tabletypesize{\footnotesize}
 \tablehead{
 & \multicolumn{2}{c}{($N=300$)} & \multicolumn{2}{c}{($N=1000$)} & \multicolumn{2}{c}{($N=2000$)}\\
  \colhead{#1} & $\Gamma_0$ & $\gamma$ & $\Gamma_0$ & $\gamma$ & $\Gamma_0$ & $\gamma$ }
 \startdata
}

\newcommand{\beRminNTwo}[2]{
\ifthenelse{\boolean{DeluxeTables}}{\begin{deluxetable*}{rcccc}}{\begin{deluxetable}{rcccc}}
 \tablewidth{0pt}
 \tablecaption{Interaction Rate Parameters as a Function of #1\label{tab:#2}}
 \tablecolumns{5}
 \tabletypesize{\footnotesize}
 \tablehead{
 & \multicolumn{2}{c}{($N=300$)} & \multicolumn{2}{c}{($N=1000$)} \\
  \colhead{#1} & $\Gamma_0$ & $\gamma$ & $\Gamma_0$ & $\gamma$ }
 \startdata
}

\newcommand{\beRTableQ}[2]{
\ifthenelse{\boolean{DeluxeTables}}{\begin{deluxetable*}{rcccccccccccc}}{\begin{deluxetable}{rcccccccccccc}}
 \tablewidth{0pt}
 \tablecaption{Radial Profile Parameters as a Function of #1\label{tab:#2}}
 \tablecolumns{13}
 \tabletypesize{\footnotesize}
 \tablehead{
 & \multicolumn{6}{c}{($\rstar \sim N^{1/2}$)} & \multicolumn{6}{c}{($\rstar \sim N^{1/4}$)} \\
 & \multicolumn{3}{c}{($Q_i = 0.04$)} & \multicolumn{3}{c}{($Q_i = 0.5$)} & \multicolumn{3}{c}{($Q_i = 0.04$)} & \multicolumn{3}{c}{($Q_i = 0.5$)} \\
  \colhead{#1} & $p$ & $r_0$ & $a$ & $p$ & $r_0$ & $a$ & $p$ & $r_0$ & $a$ & $p$ & $r_0$ & $a$}
 \startdata
}

\newcommand{\beRNThree}[2]{
\ifthenelse{\boolean{DeluxeTables}}{\begin{deluxetable*}{rccccccccc}}{\begin{deluxetable}{rccccccccc}}
 \tablewidth{0pt}
 \tablecaption{Radial Profile Parameters as a Function of #1\label{tab:#2}}
 \tablecolumns{10}
 \tabletypesize{\footnotesize}
 \tablehead{
 & \multicolumn{3}{c}{($N=300$)} & \multicolumn{3}{c}{($N=1000$)} & \multicolumn{3}{c}{($N=2000$)}\\
  \colhead{#1} & $p$ & $r_0$ & $a$ & $p$ & $r_0$ & $a$ & $p$ & $r_0$ & $a$ }
 \startdata
}

\newcommand{\beRNTwo}[2]{
\ifthenelse{\boolean{DeluxeTables}}{\begin{deluxetable*}{rcccccc}}{\begin{deluxetable}{rcccccc}}
 \tablewidth{0pt}
 \tablecaption{Radial Profile Parameters as a Function of #1\label{tab:#2}}
 \tablecolumns{7}
 \tabletypesize{\footnotesize}
 \tablehead{
 & \multicolumn{3}{c}{($N=300$)} & \multicolumn{3}{c}{($N=1000$)}\\
  \colhead{#1} & $p$ & $r_0$ & $a$ & $p$ & $r_0$ & $a$ }
 \startdata
}

\begin{document}
\title{Dynamical Evolution of Young Embedded Clusters: \\
A Parameter Space Survey}

\author{Eva-Marie Proszkow}
\affil{Michigan Center for Theoretical Physics, Physics Department,
University of Michigan, Ann Arbor, MI 48109; {\rm emproszkow@gmail.com} }
\author{Fred C. Adams}
\affil{Michigan Center for Theoretical Physics, Physics Department,
University of Michigan, Ann Arbor, MI 48109; {\rm fca@umich.edu } \\
Astronomy Department, University of Michigan, Ann Arbor, MI 48109}

\shorttitle{Dynamical Evolution of Young Clusters}
\shortauthors{Proszkow \& Adams}

\begin{abstract}

This paper investigates the dynamical evolution of embedded stellar
clusters from the protocluster stage, through the embedded
star-forming phase, and out to ages of 10 Myr --- after the gas has
been removed from the cluster.  The relevant dynamical properties of
young stellar clusters are explored over a wide range of possible star
formation environments using $N$-body simulations.  Many realizations
of equivalent initial conditions are used to produce robust
statistical descriptions of cluster evolution including the cluster
bound fraction, radial probability distributions, as well as the
distributions of close encounter distances and velocities. These
cluster properties are presented as a function of parameters
describing the initial configuration of the cluster, including the
initial cluster membership $N$, initial stellar velocities, cluster
radii, star formation efficiency, embedding gas dispersal time, and
the degree of primordial mass segregation.  The results of this
parameter space survey, which includes $\sim25,000$ simulations,
provide a statistical description of cluster evolution as a function
of the initial conditions.  We also present a compilation of the FUV
radiation fields provided by these same cluster environments.  The
output distributions from this study can be combined with other
calculations, such as disk photoevaporation models and planetary
scattering cross sections, to ascertain the effects of the cluster
environment on the processes involved in planet formation.

\end{abstract}


\section{Introduction} \label{sec:Intro}

The formation of stars and planets constitutes a fundamental problem
in astrophysics. Although a working theory of star formation has been
constructed over the past two decades \citep[e.g.,][]{Shu1987ARAA},
much of the theoretical development applies specifically to the
formation of isolated stars. In contrast, recent observational work
underscores the fact that most star formation takes place in embedded
stellar groups and clusters \citep[e.g.,][]{Lada2003ARAA,
  Porras2003AJ, Megeath2004ApJS, Allen2007prpl}.  Given that most
stars form in clusters, we are faced with the following overarching
question: If stars form in clusters, how does the background cluster
environment affect star formation and the accompanying process of
planet formation? The fundamental goal of this paper is to help
address the second portion of this question.

In rough terms, the cluster environment can influence star and planet
formation through two channels: through direct dynamical effects
(e.g., scattering events between cluster members) and through the
background radiation fields produced by massive stars in the
cluster. For the latter issue, the massive stars tend to reside in the
cluster center, so that radiation exposure depends on the radial
locations of the constituent solar systems, and these locations are
determined dynamically. As a result, both scattering interactions and
radiation exposure depend on the dynamical evolution of the cluster.
To investigate these issues, this paper presents an extensive
parameter space study of the dynamics of young embedded clusters
spanning a wide range of initial conditions and other properties.

Star-forming clusters can be viewed in two ways. One can consider the
cluster itself as an astrophysical object, and study its properties as
a function of time. For example, we can track the fraction of stars
retained, the half-mass radius, the virial parameter, and other
variables as the cluster is born, lives, and dies.  On the other hand,
we can focus on the effects of the cluster environment on its
constituent solar systems.  Although this paper presents results
relevant to both points of view, we concentrate on the latter.

This paper will focus on clusters with membership sizes $N$ in the
range $N$ = 100 -- 3000. The current observational surveys in the
solar neighborhood indicate that the majority of star formation takes
place within clusters within this size range \citep{Lada2003ARAA,
  Allen2007prpl}. Large clusters ($N \gtrsim 10^4$) are relatively
rare and their dynamics are well-studied (e.g., Portegies Zwart et
al. 1999; see also Heggie \& Hut 2003).  Small systems ($N \lesssim
100$) do not have a large impact on star/planet formation
\citep{Adams2001ApJ}, except for few-body effects that have already
been considered \citep{Sterzik1998AA}. This paper thus works in an
intermediate regime of parameter space, between the extremes where
clusters are highly disruptive \citep[e.g.,][]{Bate2003MNRAS} and more
isolated cases where individual collapse events take place unimpeded.

This parameter space survey will perform dynamical calculations
spanning the first 10 Myr of cluster evolution. This time scale is
comparable to the typical lifetime for gaseous disks
\citep{Hernandez2007ApJ}, the time required to form gas giant planets
\citep{Lissauer2007prpl}, and the lifetime of massive stars. In
addition, the gaseous component of embedded young clusters is observed
to disperse in only 3 -- 5 Myr \citep[e.g.,][]{Allen2007prpl}, so that
the clusters expand appreciably by the time they reach an age of 10
Myr (Bastian et al. 2008), and this expansion leads to reduced
interaction rates \citep[hereafter APFM]{Adams2006ApJ}.  As a result,
embedded clusters exert their greatest effects on forming solar
systems during their first 10 Myr.  Notice also that protostellar
collapse takes place over a much shorter time scale, only about 0.1
Myr, so that the cluster environment has relatively little time to
affect the star formation process {\it per se}. The most important
effects of the background environment act on circumstellar disks and
planet formation, i.e., forming solar systems.

We note that a great deal of dynamical work on young clusters has been
done previously \citep[e.g., ][and many others]{Lada1984ApJ,
  Rasio1995ApJ, Kroupa1995, Boily2003MNRAS665, Boily2003MNRAS673}.  In
addition to expanding the parameter space under consideration, this
work differs from previous dynamical studies of young clusters in
several ways.  The clusters considered here, with $N$ = 100 -- 3000,
are highly chaotic so that different, but equivalent, realizations of
the system can produce different dynamical results (e.g., different
numbers of close approaches). In addition, the sampling of the stellar
initial mass function is not complete for these intermediate-sized
clusters. To address these issues, we must adopt a statistical
approach.  In order to characterize the dynamics of these systems, we
run multiple equivalent realizations of the simulations to build up
robust distributions of the output measures, e.g., the distributions
of closest approaches and distributions of radial positions (which
determine radiation exposure).  Previous work (APFM,
\citealt{Proszkow2009ApJ, Proszkow2008PhDT}; see also
\citealt{Malmberg2007MNRAS}) indicates that for these moderate-sized
clusters and intermediate time scales one needs $\sim100$ realizations
for each set of initial conditions to obtain robust statistics. As a 
result, this survey of parameter space includes the results from 
$\sim25,000$ numerical simulations.  

The initial conditions for the $N$-body simulations of this paper also
differ from most previous studies. Past studies often assume that the
initial phase space variables of the stars are close to virial
equilibrium. In many regions, however, pre-stellar clumps are observed
to move subsonically before the clumps form stars (Peretto et al.
2006; Walsh et al. 2004; Andr{\'e} 2002); these data imply that newly
formed stars begin their dynamical evolution with subvirial speeds.
Recent results from the {\it Spitzer Space Telescope} show that the
class I objects (protostars) are segregated from the class II objects
(star/disk systems) in embedded clusters, but their positions are
highly correlated with each other and with the gas ridges (T. Megeath,
private communication); this result suggests that the protostars are
not moving in a fully dynamical manner. As shown previously (APFM),
and reinforced in this study, subvirial starting states have a
significant impact on the resulting cluster properties.

This paper is organized as follows. In Section \ref{sec:NumSims}, we
outline the parameter space that the $N$-body simulations explore,
including membership size $N$, stellar profiles, initial velocities,
radial sizes, star formation efficiency, gas removal timescale, and
the degree of primordial mass segregation. The results from these
simulations are then presented in Section \ref{sec:Results}, including
both the parameters that describe cluster properties (e.g., fraction
of bound stars, mass distributions, and number density distributions)
and parameters that affect solar system disruption (e.g., rate of
close encounters). The effects of these cluster environments on the
solar systems forming within them is addressed in Section
\ref{sec:Effects}, including a discussion of the radiation fields
produced by these cluster systems.  We conclude in Section
\ref{sec:Conclusion} with a summary of our results and a discussion of
their implications.

\section{Numerical Simulations of Young Embedded Clusters}\label{sec:NumSims}

Throughout this work, a modified version of the NBODY2 code developed
by \cite{Aarseth2001NewA} is employed to numerically calculate the
dynamics in young stellar clusters from the embedded stage out to ages
of $10~\Myr$. The modifications made to this code allow for the
cluster's initial conditions to be more like those observed in young
stellar clusters.  More specifically, these modifications allow us to
specify the form and time evolution of the embedding gas, provide
differing degrees of initial mass segregation, and define the geometry
and velocity structure of the stellar distribution.  Additional
modifications are implemented to produce the output parameters of
interest in this study, including closest approach distributions,
velocity distributions, and mass profiles.

In addition to focusing on clusters with initial conditions similar to
those found in nearby young clusters, this work is distinguished by
its statistical character. The $N$-body problem is by nature chaotic,
so that clusters with the similar initial configurations can produce
dissimilar results. In order to produce robust statistical
descriptions of cluster evolution, we must carry out $100$
realizations of the cluster simulations for each initial condition
configuration considered in this study. Specifically, for a given set
of initial cluster conditions (i.e., cluster membership, radius,
velocity distribution, etc.), $100$ simulations are completed using a
different random number seed to sample the relevant distributions
(e.g., stellar positions, IMF).  The resulting output parameters are
then averaged over the set of realizations to provide a statistical
description of how a similar cluster is likely to evolve. In this
context, a ``similar cluster'' is one produced by an independent
realization of the initial conditions, which are sampled from the same
distribution of values.

We note that the version of the $N$-body code used for these
simulations (NBODY2) does not include binaries in the initial
conditions (see Aarseth 1999, 2001); we also use softening parameter
$\epsilon$ = 0.001 (which under-resolves binaries that may form during
the course of cluster evolution).  With this value of $\epsilon$, the 
code resolves stellar encounters down to $\sim100$ AU, the smallest 
approach distance considered here. Since this integration package is 
relatively fast, it can produce many realizations for each set of
initial conditions, as required here to obtain good statistics. In
dense and/or long-lived clusters, binaries can affect the energetics
of the system by absorbing and storing energy in their orbits. In the
systems of interest here, however, the densities are low and the
evolutionary times are short, so that binaries have relatively little
impact on cluster evolution. In other words, interactions are
sufficiently rare and sufficiently distant, so that binarity has a
relatively small effect on overall energy budget (for further
discussion, see Kroupa 1995, Kroupa et al. 2003).  In our previous
work (APFM), we checked this approximation for consistency by running
a parallel set of simulations including binaries (using the NBODY6
integration package; Aarseth 1999) and found that the results were
essentially unchanged. This same test case implies that our choice of
softening parameter does not greatly affect the global results. As an
additional check on binary effects, we can use the distributions of
closest approaches resulting from our ensemble of simulations; these
results also indicate that that binary interactions are not
energetically important.

In this section we outline the standard initial conditions used in the
simulated clusters.  Specifically, we discuss the qualities most
commonly observed in nearby young embedded clusters and identify these
qualities as the center of our parameter space.  These initial
conditions thus define the typical cluster, and the parameter space
survey is conducted by varying one or more of the initial conditions
at a time.  In the following discussion, in Section \ref{sec:Input},
we enumerate the variables of interest. The particular range of
parameter space investigated in this survey is then outlined in
Section \ref{sec:Pspace}. The results of the cluster simulations are
presented in Section \ref{sec:Results}, where we also discuss the
implications for planet formation within these clusters.

\subsection{Specification of Input Variables}\label{sec:Input}

\emph{Cluster Membership, $N$.}  As outlined above, in this study we
consider intermediate-sized clusters with stellar memberships ranging
from $N = 100$ to $3000$ \citep{Lada2003ARAA, Allen2007prpl}. This
range corresponds to that observed in the solar neighborhood, where we
have the most direct window into the star formation process.

\emph{Initial Mass Function.}  The shape of the stellar initial mass
function (IMF) observed in young stellar clusters is almost universal
for clusters with more than $N \sim 100$ members \citep{Lada2003ARAA}.
In our simulations, stellar masses are sampled from the log-normal
analytic fit to the standard IMF of \cite{Miller1979ApJS} presented by
\cite{Adams1996ApJ}. With this choice, the average stellar mass in a
cluster is $0.5$~\msun (the average stellar mass is somewhat higher
than the median stellar mass which is roughly $\sim0.3$~\msun),
consistent with observations of young stellar clusters
\citep{Muench2002ApJ, Luhman2003ApJ}.

\emph{Cluster Radius, $\rstar$.}  Stars are initially distributed
within the cluster radius $\rstar$ according to the density profiles
described below. The cluster radius $\rstar$ is taken to be a function
of the cluster membership $N$, the scaling radius $\rsc$, and the
power-law index $\alpha$, so that
\bee
\rstar = \rsc \left(\frac{N}{300}\right)^{\alpha}.
\label{eq:RofNparam}
\eee
This membership-radius relationship is observed in young clusters in
the solar neighborhood \citep{Lada2003ARAA, Porras2003AJ}, and typical
values of the parameters are $\rsc$ = $1~\pc$ and $\alpha$ = $1/2$. 
Thus, a cluster with $N = 300$ stars typically has a radius of $1~\pc$.

\emph{Initial Stellar Profile.}  Many young embedded clusters display
degrees of central concentration (see Lada \& Lada 2003, and
references therein).  The simulated clusters in this study are
correspondingly centrally condensed and have initial stellar density
distributions of the power-law form $\rho_{\ast} \sim r^{-1}$.
Density profiles of this form are consistent with the observed density
profiles of the embedding gas in cluster-forming cores (Jijina et al.
1999; see also below). For simplicity, we take the initial profiles to
be spherically symmetric.  However, we note that irregular initial
distributions (Goodwin \& Whitworth 2004) can have interesting
effects, e.g., accelerated mass segregation.

\emph{Mass Segregation.}  Young stellar clusters exhibit varying 
degrees of mass segregation, even though the clusters themselves are
not old enough to have undergone dynamical mass segregation (their age
is generally less than a relaxation time, e.g., Bonnell \& Davies
1998).  However, recent work shows that mass segregation can occur 
more rapidly when the clusters have subvirial initial conditions and
clumpy substructure (McMillan et al. 2007, Allison et al. 2009,
Moeckel \& Bonnell 2009).  For most of this survey, the simulated
clusters contain minimal mass segregation implemented by a
straightforward algorithm: At the start of the simulation, the most
massive star in the cluster is relocated to the center of the cluster,
while the positions of the rest of stars are not correlated with mass.

\emph{Initial Stellar Velocities.}  Kinematic observations of the
youngest stellar objects (Class I sources) and of starless dense cores
indicate that these objects are moving at speeds that are a fraction
of the virial speed in many young clusters \citep{Walsh2004ApJ,
  Andre2002ApSS, Peretto2006AA,Kirk2006ApJ}.  In the cluster
simulations, initial stellar velocities are sampled from a spatially
isotropic distribution and then scaled by the initial virial ratio of
the cluster $Q_i$. The virial ratio is defined as $Q \equiv |K/W|$,
where $K$ is the total kinetic energy and $W$ is the total potential
energy of the cluster.  A cluster that is in virial equilibrium has a
virial ratio $Q = 0.5$.  Most simulations considered in this study are
initialized with a virial ratio $Q_i = 0.04$ which results in stellar
velocities that are approximately one-third of the virial velocity of
the cluster and is consistent with the kinematic observations of young
embedded clusters.

\emph{Star Formation History.} The stars in the simulated clusters
have a spread in formation times of $\Delta t = 1~\Myr$.  The
formation time of each star is sampled from a uniform distribution
over the range from $0$ to $1~\Myr$, independent of position within
the cluster and independent of stellar mass.  We assume that the
forming stars are tied to their formation site (they do not move
dynamically) until the collapse phase is complete, i.e., until the
star is formed.  The stars are included in the simulations as static
point masses until their formation time.  After they form, the stars
are allowed to move through the gravitational potential of the cluster
with an initial velocity sampled from the distribution described
above.

\emph{Embedding Gas Profile.}  Extremely young stellar clusters (with
ages less than $\sim3$ Myr) are almost always associated with a
molecular cloud core \citep{Leisawitz1989ApJS}. These cores are often 
centrally concentrated \citep{Larson1985MNRAS, Myers1993ApJ,
  Jijina1999ApJS}. In the simulated clusters this embedding gas is
represented as a static gravitational potential with a Hernquist
profile \citep{Hernquist1990ApJ}, with potential, density, and mass
profiles of the form
\bee 
\Psi = \frac{2 \pi G \rho_0 r_s^2}{1 + \xi }, \quad 
\rho = \frac{\rho_0}{\xi (1 + \xi)^3}, \quad {\rm and} 
\quad M = \frac{M_\infty \xi^2}{(1+\xi)^2},
\label{eq:hqprofileparam} 
\eee 
where $\xi \equiv r/r_s$. Here the parameter $r_s$ is a scale length,
which is chosen to be equal to the cluster radius, i.e., $r_s = \rstar$.  
In the inner limit, this profile has the form of $\rho \sim r^{-1}$,
and outside of the cluster the density profile matches onto a
force-free background.  We are thus neglecting external forces on the
cluster (e.g., galactic tides).

\emph{Star Formation Efficiency.}  Estimates of the star formation
efficiency $\sfe$ in young star-forming regions vary from $\sim10\%$
to $50\%$ \citep{Lada2003ARAA}.  In our simulated clusters, a standard
star formation efficiency of $\sfe = 33\%$ is assumed.  This value 
corresponds to a total stellar mass $M_{T\ast}$ in the cluster that is
one half of the mass of the embedding gas $M_{gas}$ (from equation
[\ref{eq:hqprofileparam}], $M_{gas} = M_\infty/4$ is the effective gas
mass within the cluster radius $\rstar$).

\emph{Gas Removal History.}  Although young stellar clusters are
associated with embedding molecular gas, the gas is quickly dispersed
from the cluster by a collection of processes, including stellar winds
from young stars, ionizing radiation from massive stars, and other
processes.  Clusters with ages greater than $\sim5~\Myr$ are rarely
associated with molecular gas.  In the cluster simulations, the depth
of the potential well associated with the embedding gas is reset to
zero instantaneously at a time denoted as $\tgas$.  The gas removal
mechanism is thus assumed to rapidly disperse gas from the vicinity of
the cluster at this epoch.

\emph{Stellar Evolution.} This parameter space survey considers only
the first 10 Myr of cluster evolution and does not include stellar
evolution effects. We expect these corrections to be relatively small
for the cases considered here: Note that a star with initial mass
$M_\ast = 20 M_\odot$ is expected to burn hydrogen for 8.13 Myr and
then burn helium for another 1.17 Myr (Woosley et al.  2002). The
total time before exploding as a supernova is thus about 9.3 Myr. As a
result, only those stars with masses $M_\ast > 20 M_\odot$ will finish
their evolution during the 10 Myr window considered here. And only
about 1 star in 800 (or 1000, depending on the IMF) has a starting
mass this large. Stellar evolution thus has a small effect on the mass
budget of the clusters.  Although supernovae would have a large impact
on the gas content of clusters, the gas is removed earlier, as
indicated by observations (and presumably driven by other mechanisms).

\subsection{Parameter Space Overview}\label{sec:Pspace}

As described in Section \ref{sec:Input}, a large number of initial
parameters must be specified in order to characterize a cluster at the
start of a simulation. As a result, the parameter space available for
studying the evolution of embedded stellar clusters is extremely
large. In this current work, we target our parameter space survey on
embedded cluster environments similar to those observed in our solar
neighborhood \citep[Gutermuth et al. 2009, in
  preparation]{Lada2003ARAA, Allen2007prpl, Megeath2004ApJS}, with an
extrapolation to somewhat larger clusters.  In this section we
identify the range of parameter space for which our survey is
conducted.  As described below, we perform many different series of
simulations, where each series explores the effects of one (or more)
specific parameters(s).  It is important to note that this range,
while motivated by observations of nearby clusters, does not
necessarily encompass all of the possible initial conditions spanned
by these these cluster environments.  The range of parameter space
surveyed and the initial conditions assumed in the simulated clusters
are summarized in Table \ref{tab:Initial}.

\ifthenelse{\boolean{DeluxeTables}}{\begin{deluxetable*}{lcccc}}{\begin{deluxetable}{lcccc}}
	\tablewidth{0pt}
  \tabletypesize{\small}
  \tablecaption{Initial Conditions for Parameter Space Survey\label{tab:Initial}}
  \tablehead{
  \colhead{Experiment Series} & \colhead{Parameter} & \colhead{Parameter Range} 
& \colhead{Variations} & \colhead{\# Sims} } 
  \startdata
Cluster Membership  & $N$ & $100 - 3000$ & $Q = 0.04, 0.5$ & 6,200\\
  & & & $\rstar = (N/300)^{1/2}~\pc$ & \vspace{0.05 in} \\
Cluster Membership  & $N$ & $100 - 3000$ & $Q = 0.04, 0.5$ & 6,200\\
  & & & $\rstar = (N/300)^{1/4}~\pc$ & \vspace{0.05 in} \\ 
Virial Ratio & $Q_i$ & $0.025 - 0.5$  & $N = 300, 1000, 2000$ & 6,000\\
  & & & $\rstar = (N/300)^{1/2}~\pc$ & \vspace{0.05 in} \\ 
Radius Scaling Factor & $\rsc$ & $0.33 - 3.0~\pc$ & $N = 300, 1000, 2000$ & 2,700\\
  $\qquad$  & & & $Q = 0.04$ & \vspace{0.05 in} \\ 
Star Formation Efficiency & $\sfe$ & $0.1 - 0.5$ & $N = 300, 1000$ & 1,600\\
  $\qquad$ & & & $Q = 0.04$ & \vspace{0.05 in} \\ 
Mass Segregation & $\mseg$ & $1/N - 0.99$ & $N = 300, 1000, 2000$ & 2,100\\
  & & & $Q = 0.04$ & \\
  \enddata
\ifthenelse{\boolean{DeluxeTables}}{\end{deluxetable*}}{\end{deluxetable}}

\subsubsection{Cluster Membership}\label{sec:ClusterN}

We perform a series of simulations to study the effect that stellar
membership has on the dynamics of young embedded clusters.  We
consider spherical clusters embedded in centrally concentrated gas
potentials with a star formation efficiency $\sfe = 0.33$.  The
stellar membership in the simulated clusters ranges from $N = 100$ to
$3000$.  Clusters of this size roughly span the range of young
clusters observed in the solar neighborhood \citep{Lada2003ARAA,
  Porras2003AJ}.  Motivated by observations of young stellar objects
with subvirial velocities, this study considers embedded clusters with
both subvirial and virial initial velocity distributions.  The
subvirial and virial clusters have $Q_i = 0.04$ and $0.5$
respectively.  The subvirial clusters thus have initial stellar
velocities that are approximately $1/3$ of the virial velocity.

It is possible that the index $\alpha$ appearing in the
membership-radius relation (equation [\ref{eq:RofNparam}]) takes on
different values for different cluster samples.  For example, the
value $\alpha = 1/2$ is a reasonable fit to the observed data within
approximately $2~\kpc$ of the Sun \citep{Lada2003ARAA, Porras2003AJ},
where this sample contains intermediate-sized clusters with $N
\lesssim 2000$.  In environments with star formation rates much higher
than that of the solar neighborhood, a significant amount of activity
occurs in clusters more massive than those found in our solar
neighborhood (e.g., Chandar et al. 1999, Pfalzner 2009).  These
extremely massive young clusters, some which are thought to be
progenitors of globular clusters, contain as many as $N \sim 10^6$
stars and have sizes on the order of $\rstar \sim 10~\pc$
\citep{Mengel2008AA}. If we extend the cluster membership-radius
relation out to stellar memberships as high as $N \sim 10^6$, the
choice of $\alpha = 1/2$ would overestimate the cluster radius by a
factor of $\sim5$.  A power-law index of $\alpha =1/4$ more closely
reproduces the observed data points over the full range of $N$. In
this study, we investigate the evolution of intermediate-sized
clusters using both $\alpha = 1/2$ and $\alpha = 1/4$ power-law
indices in the cluster membership-radius relation.  In both cases, we
chose $\rsc = 1.0~\pc$ so that the power-law passes through the point
where $N = 300$ and $\rstar = 1.0~\pc$.

These two choices of the index $\alpha$ result in clusters whose
average number density varies differently as a function of cluster
membership.  Specifically, substituting the membership-radius relation
into the equation for average number density $n_0$ gives the relation
\bee
n_0 \sim \frac{N}{\rstar^3} \sim \frac{N}{N^{3\alpha}} = N^{1-3\alpha}.
\eee
For the choice $\alpha =1/2$ the average stellar density decreases as
a function of $N$, whereas for $\alpha = 1/4$ the stellar density is
an increasing function of $N$.  In the results summarized in Section
\ref{sec:Results}, many of the trends observed as a function of
cluster membership $N$ are more fundamentally trends in average
stellar density as a function of $N$.

\begin{figure}
\epsscale{1.0}
\plotone{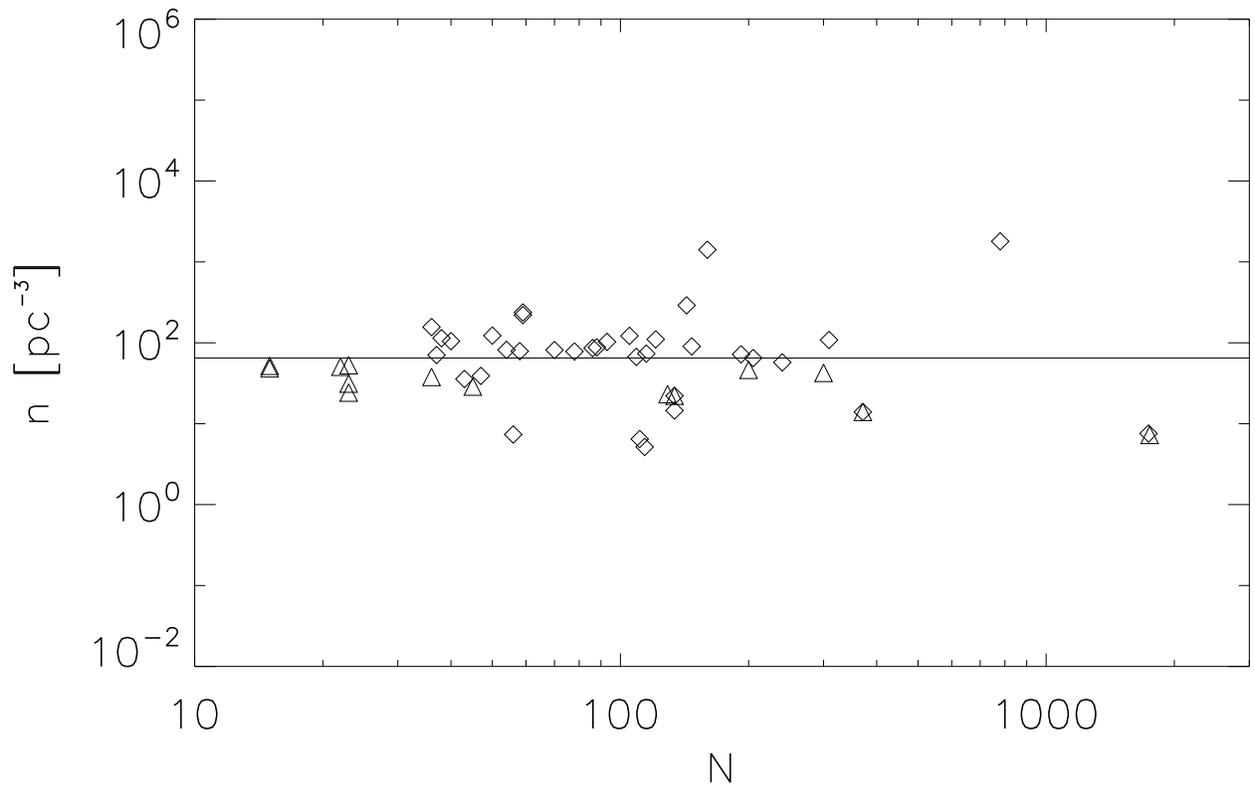} 
\caption{Average number density (in stars per cubic parsec) of
  clusters in the solar neighborhood plotted as a function of cluster
  membership $N$. Diamonds indicate data from the catalogs presented
  in \cite{Lada2003ARAA} and triangles indicate data from the catalog
  of \cite{Carpenter2000AJ}. The solid horizontal line indicates the
  median number density for this sample, $n_M$ = 65~pc$^{-3}$.}
\label{fig:ClusterDensities} 
\end{figure}

Note that an intermediate value of the index, $\alpha = 1/3$, implies
a constant stellar density.  This benchmark density value is $n_0 \sim
100$ stars $\pc^{-3}$. Figure \ref{fig:ClusterDensities} displays the
average number densities found in clusters in the solar neighborhood.
The data are taken from the cluster catalogs of \cite{Lada2003ARAA}
(diamonds) and \cite{Carpenter2000AJ} (triangles).  Number densities
are calculated assuming spherical symmetry in the stellar clusters.
Nearby young clusters may have higher densities in the cluster cores
\citep{Hillenbrand1998ApJ, Gutermuth2005ApJ, Teixeira2006ApJL}, but
their average stellar densities are relatively constant. The 
horizontal line in the figure shows a constant density reference 
with the median value of the data set, $n_M$ = 65 \pc$^{-3}$. 

\subsubsection{Initial Virial Parameter}\label{sec:VirialQ}

As discussed in Section \ref{sec:Input}, recent observations of young
embedded clusters indicate that stars are formed with initial
velocities lower than the virial speed of the cluster.  During the
early evolution of a subvirial cluster, the average stellar velocities
increase as individual stars fall through the global potential well of
the cluster.  Stars with initially subvirial velocities thus trade
potential energy for kinetic energy during the early phases of cluster
evolution.  On somewhat longer timescales, interactions share energy
among stellar orbits and the cluster approaches virial equilibrium.
Here we present a series of numerical experiments designed to
investigate the effect that the initial virial ratio $Q_i$ has on the
evolution of clusters.  While the simulations completed as a part of
the cluster membership parameter study make a rough comparison between
subvirial and virial clusters, this set of simulations samples a much
wider range of virial ratios $0.025 \leq Q_i \leq 0.5$ with much
higher resolution.  Simulations are completed for clusters with
initial membership $N = 300, 1000$, and $2000$ that have a
membership-radius relation characterized by $\alpha = 1/2$, similar to
that observed in the solar neighborhood.

One question that this study attempts to address is: How small must
the initial virial parameter $Q_i$ be in order for cluster evolution
to differ significantly from that of a cluster in virial equilibrium?
Our results indicate that even moderately subvirial clusters display
characteristics significantly different from virial clusters (see 
Section \ref{sec:Results}).  For instance, the bound fraction in a 
cluster with initial virial parameter $Q_i = 0.35$ is almost $50\%$
larger than in a virial cluster with $Q_i = 0.5$. A value of 
$Q_i = 0.35$ corresponds to an average stellar velocity that is 
approximately $84\%$ of the virial velocity of the cluster. As a 
result, initial stellar velocities can be an appreciable fraction 
of the virial value and still lead to significant differences. 

\subsubsection{Cluster Scaling Radius}\label{sec:ClusterR}

The cluster membership-radius relation presented in equation
(\ref{eq:RofNparam}) depends on both the power-law index $\alpha$ and
the fiducial scaling parameter $\rsc$ that sets the radius of a
cluster with $N = 300$ members. Although this scaling relationship is
robust, the radius of observed clusters still contains significant
scatter as a function of stellar membership $N$ (see Figure 1 of
APFM).  Some of this scatter results from the observational difficulty
of determining the outer radius of a cluster as the surface density
approaches that of the background sky. In addition, it is difficult to
determine cluster radii for non-spherical clusters and clusters with
small memberships \citep{Gutermuth2005ApJ, Allen2007prpl}.

A series of simulations are completed to investigate how cluster
evolution varies with different values of scaling parameter $\rsc$.
Specifically, cluster simulations are completed for scaling radii in
the range $1/3~\pc \leq \rsc \leq 3~\pc$, where we use power-law index
$\alpha = 1/2$ in the membership-radius relation. The clusters are
assumed to have stellar memberships $N = 300, 1000$, and $2000$ and
subvirial initial velocities with $Q_i = 0.04$.  Changing the scaling
radius $\rsc$ effectively changes the average stellar density in a
cluster, and many of the trends observed in the cluster evolution as a
function of scaling radius $\rsc$ are linked to this change in density. 
 
\subsubsection{Star Formation Efficiency}\label{sec:SFE} 

The star formation efficiency (SFE) of a region is defined as $\sfe
\equiv M_{T\ast}/(M_{T\ast}+M_{gas})$, where $M_{T\ast}$ and $M_{gas}$
are the total stellar and gaseous mass contained in the region,
respectively.  The stellar mass $M_{T\ast}$ thus corresponds to the
value after the interval $\Delta t$, i.e., after star formation has
been completed.  Estimates of the SFE for young embedded clusters in
the solar neighborhood range between $0.1$ and $0.3$
\citep{Lada2003ARAA, Allen2007prpl}.  These efficiencies are
significantly higher than the SFEs of entire giant molecular clouds,
which have typical SFEs less than $\sim$0.05 \citep{Duerr1982ApJ,
  Evans1991IAUS}.  As a part of this parameter space study, we
complete a suite of cluster simulations in which the SFE is varied
over the range $0.1 \leq \sfe \leq 0.5$.  The clusters are assumed to
be in an initially subvirial state, and have stellar memberships of 
$N$ = 300 and 1000. The value of the star formation efficiency
parameter $\sfe$ is attained by varying the mass of the gas in the
simulated cluster (for fixed $N$).  The SFE of a cluster is a major
factor in determining the fraction of the stars in a cluster that
remain bound after the embedding gas is dispersed (due to outflows
from young stars or ionizing radiation from the most massive star; see
Section \ref{sec:Results}).

\subsubsection{Gas Removal Timescale}\label{sec:GasTime}

Although the youngest star-forming clusters are deeply embedded in
their natal molecular clouds, clusters with ages greater than $\sim10$
Myr are associated with relatively little molecular gas
\citep{Leisawitz1989ApJS}. In this series of simulations, we consider
the evolution of young clusters with gas removal times $\tgas$ in the
range from $1~\Myr$ to $7~\Myr$.  The gas is assumed to be removed
instantaneously at $\tgas$.  The study considers subvirial clusters
with stellar memberships $N = 300, 1000$, and $2000$.  A significant
fraction of the stars in a given cluster become gravitationally
unbound at the time of gas dispersal and the cluster begins to expand
radially outward.  As the cluster expands, the average density
decreases and close interactions between stellar members become less
frequent.  As a result, in addition to affecting how much of the
cluster remains gravitationally bound, the gas removal time $\tgas$
places important constraints on the close encounter rates in young
clusters.

\subsubsection{Mass Segregation}\label{sec:MassSeg}

Observations indicate that massive stars are preferentially found near
the center of both evolved open clusters and young embedded clusters.
Mass segregation in the evolved clusters can be explained by dynamical
theory: high mass stars lose energy to low mass stars through two-body
interactions and subsequently sink toward the cluster's center.  This
process takes place on timescales comparable to a cluster's dynamical
relaxation time, which is given by 
\bee
t_{relax} \approx \frac{N}{8 ~{\rm ln} ~N}\tcross \approx 
\frac{N \rstar}{8 ~{\rm ln} ~N \left<v\right>}, 
\eee 
where $\tcross$ is the cluster crossing time and $\left<v\right>$ is
the average stellar velocity \citep{Binney1987Book}. Open clusters
have typical ages 20 -- 500 Myr and thus are old enough for dynamical
mass segregation to have occurred. However, observations of mass
segregation in young embedded clusters are more difficult to explain
\citep{Bonnell1998MNRAS}.  A (logarithmically) average embedded
cluster in the solar neighborhood has $N \sim 300$, $\rstar \sim
1~\pc$, $\left<v\right> \sim 1~\kms$, and a corresponding relaxation
time of roughly $t_{relax} = 6.5~\Myr$.  As a result, dynamical
evolution is unlikely to be responsible for the mass segregation
observed in young clusters such as the Trapezium, NGC 2071, or NGC
2074 \citep{Hillenbrand1998ApJ, Lada1991ApJ, Bonnell1998MNRAS}. 
This finding suggests that the mass segregation is due to a primordial
tendency to form massive stars near the center of clusters.  We note 
that primordial mass segregation is naturally produced in embedded 
clusters through some proposed massive star formation scenarios. For
example, competitive accretion preferentially forms massive stars in 
the deepest part of the cluster potential well, near the center of the
cluster \citep{Bonnell2001MNRAS, Beuther2007prpl}.

One experiment in the parameter space survey explores the evolution of
clusters with varying amounts of primordial mass segregation.  We
define the primordial mass segregation parameter $\mseg$ as the
fraction of the cluster membership which has been ordered by mass at
the center of the cluster, $\mseg = N_{seg}/N$. More specifically, at
the start of the simulation, the stellar masses are sampled from a
standard IMF and stellar positions are sampled from a $\rho_\ast \sim
r^{-1}$ density profile (regardless of mass). After this initial
sampling, the most massive star in the cluster is moved from its
initial (randomly assigned) position to the center of the cluster.
This resulting state represents a cluster with minimal mass
segregation.  For values of $\mseg > 1/N$, additional mass segregation
is implemented by rearranging the stellar positions so that the
$N_{seg}$ most massive stars are located at the inner $N_{seg}$ radial
positions.  The mass segregation parameter is varied over the range
$1/N \leq \mseg \leq 0.99$ in subvirial clusters with $N$ = 300, 1000,
and 2000 members. 

\section{Summary of Results}\label{sec:Results}

This parameter space survey includes a large number ($\sim$25,000) of
$N$-body simulations. In this section, we package the results of these
numerical calculations by extracting relevant output variables. The
fraction of stars that remain bound to the cluster is discussed in
Section \ref{sec:boundfrac}. Next we determine the interaction rates
between cluster members (Section \ref{sec:Gammas}) and the
distributions of interaction speeds (Section \ref{sec:Velocity}).
Finally, we construct mass profiles and number density profiles in
Section \ref{sec:MassDists}.

\subsection{Bound Fraction}\label{sec:boundfrac}

Observational studies that compare the formation rates of embedded
clusters and open clusters have shown that the embedded cluster
formation rate is significantly higher than that of open clusters
\citep{Elmegreen1985ApJ, Battinelli1991MNRAS, Piskunov2006AA}.  This
discrepancy in the formation rates leads to the conclusion that while
most star formation occurs within clusters, only a fraction (about
$10\%$) of the stellar population is born within ``robust'' clusters
that are destined to become open clusters (which then live for
$\sim100~\Myr$ or longer). This result suggests that very few embedded
clusters remain gravitationally bound after their embedding molecular
gas is removed. The process by which gas removal leads to the
unbinding of a cluster has been denoted as ``infant mortality'' for
embedded clusters, and has been addressed via both analytical studies
\citep{Hills1980ApJ, Elmegreen1983MNRAS, Verschueren1989AA,
  Adams2000ApJ} and numerical simulations \citep{Lada1984ApJ,
  Geyer2001MNRAS, Boily2003MNRAS665, Boily2003MNRAS673}. Evidence that
this process is occurring in extragalactic young massive clusters has
been presented by \cite{Bastian2006MNRAS}.

An important output parameter explored in our simulations is the
fraction $\fbnd$ of stars that remain gravitationally bound as a
function of time.  The bound fraction is defined as $\fbnd \equiv
N_{bound}/N$ where $N$ is the initial stellar membership, and
$N_{bound}$ is the number of stars that have total energy (kinetic
plus potential) less than zero.  Throughout the embedded phase of
cluster evolution, the bound fraction remains equal to unity (note
that the timescale of the embedded phase is shorter than the
relaxation time).  The embedding gas potential is removed from the
simulated clusters instantaneously at time $\tgas$. This event
significantly reduces the depth of the potential well in which the
cluster members reside.  Rapid gas removal is an appropriate
approximation for gas expulsion due to high mass star formation, which
removes the embedding gas over timescales as short as $\sim10^4$ years
\citep{ Whitworth1979MNRAS}.

When the gravitational potential of the gas is removed from the
system, the high-velocity stars become gravitationally unbound while
the low-velocity stars remain bound to the cluster's gravitational
potential. As a result, the bound fraction $\fbnd$ decreases
significantly (by as much as $50\%$) over a short period of time.  
The fraction $\fbnd$ then continues to decrease (more slowly) until
the end of the simulations. Note that our temporal cutoff is chosen 
to be $10~\Myr$, but the clusters will continue to evolve and $\fbnd$ 
will continue to decrease on longer timescales.

Figures \ref{fig:Bound_t_01} and \ref{fig:Bound_t_02} display
$\fbnd$ as a function of time for the range of parameter space
surveyed in this study.  Each panel illustrates the temporal evolution
of $\fbnd$ for a specific cluster parameter, where the individual
curves correspond to different values of that cluster parameter.  For
example, in Figure \ref{fig:Bound_t_01} panel (a), the evolution
of $\fbnd$ is plotted for clusters with different star formation
efficiencies $\sfe$.  The top curve corresponds to a cluster with
$\sfe = 0.75$ and the bottom curve corresponds to a cluster with $\sfe
= 0.05$.  For each of the curves in Figures \ref{fig:Bound_t_01}
and \ref{fig:Bound_t_02}, the other initial conditions (i.e.,
cluster membership $N$, initial cluster radius $\rstar$, degree of 
mass segregation $\mseg$, etc.) are held constant. All of the panels
show a rapid decrease in $\fbnd$ at $t = 5~\Myr$ (except for Figure
\ref{fig:Bound_t_02} panel (a), which represents simulations in
which the gas removal timescale $\tgas$ is varied). As expected, the
downward jump in $\fbnd$ corresponds to the time at which the gas is
removed from the cluster.

\begin{figure}
\epsscale{1.0} 
\plotone{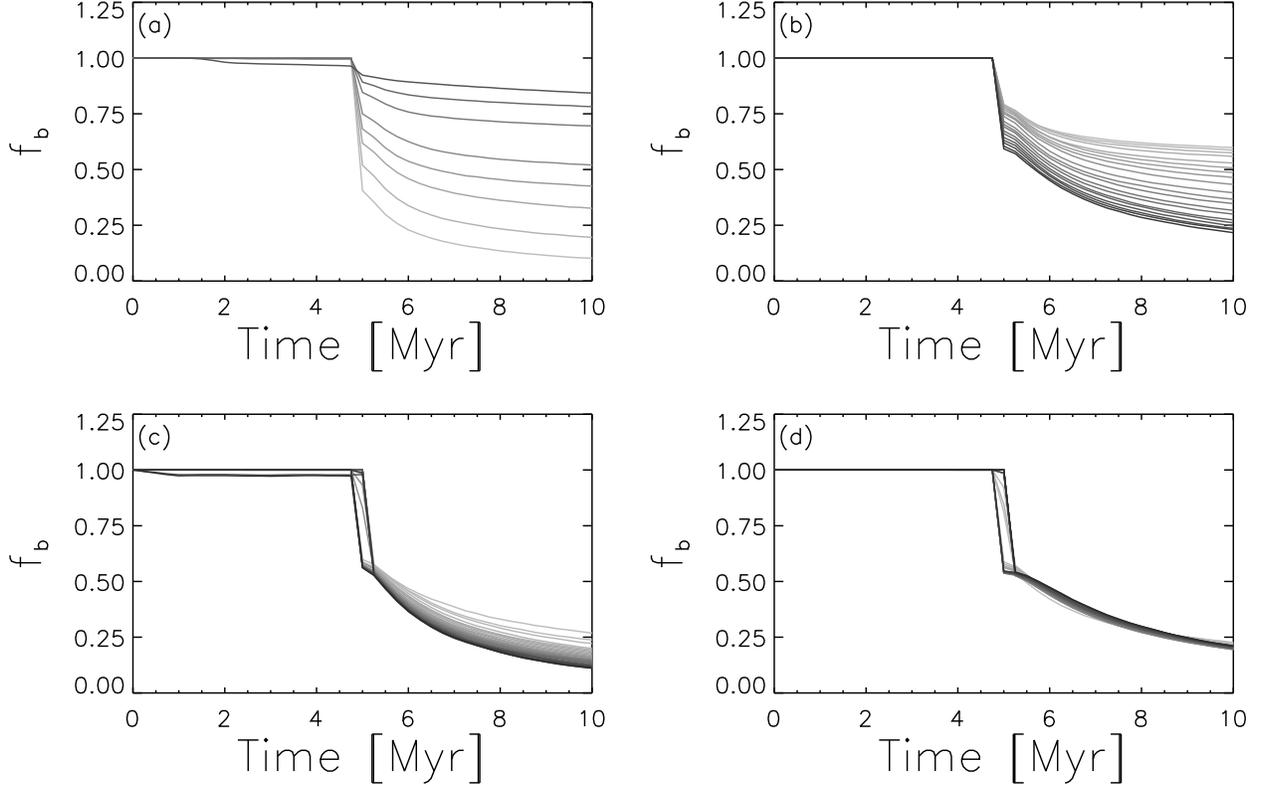} 
\caption{Fraction of stars $\fbnd$ that remain bound to the cluster's
  gravitational potential as a function of time, for all of 
  clusters included in the parameter space survey.  Each panel
  corresponds to a different cluster parameter that is varied: (a)
  Star formation efficiency where $\sfe$ ranges from $\sfe = 0.1$
  (lower curve) to $\sfe = 0.75$ (upper curve) (b) Initial virial
  ratio where $Q_i$ ranges from $Q_i = 0.025$ (upper curve) to $Q_i =
  0.5$ (lower curve), (c) Stellar Membership $N$ for scaling
  relationship $\rstar \sim N^{1/2}$ where $N$ ranges from $N = 100$
  (upper curve) to $N = 3000$ (lower curve), and (d) Stellar
  Membership $N$ for scaling relationship $\rstar \sim N^{1/4}$ where
  $N$ ranges from $N = 100$ to $N = 3000$. The individual curves
  correspond to clusters with different initial values of the cluster
  parameter of interest.  In all simulations the gas expulsion takes
  place at $\tgas = 5~\Myr$. Immediately after gas removal, a
  significant fraction of stars become unbound from the cluster.  As
  the cluster's evolution continues, the mass loss rate drops
  significantly and $\fbnd$ approaches a constant value.}
\label{fig:Bound_t_01} 
\end{figure}

\begin{figure}
\epsscale{1.0}
\plotone{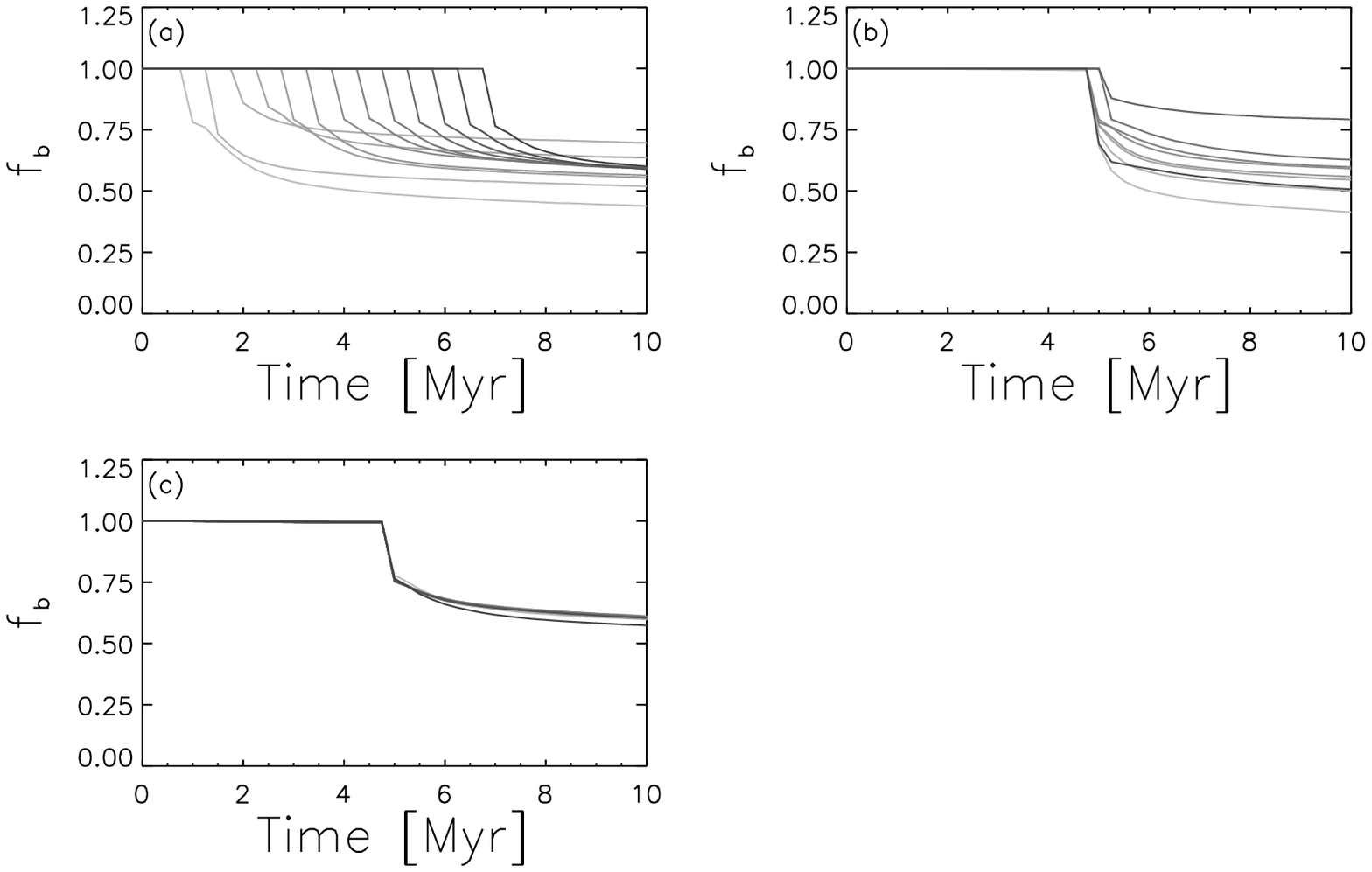} 
\caption{Continuation of Figure \ref{fig:Bound_t_01}.  Each panel
  corresponds to a different cluster parameter that is varied: (a) Gas
  removal time where $\tgas$ ranges from $\tgas = 1$~Myr (lightest
  curve) to $\tgas = 7$~Myr (darkest curve), (b) Cluster scaling
  radius where $\rsc$ ranges from $\rsc = 0.33~\pc$ (lightest curve)
  to $\rsc = 3~\pc$ (darkest curve) and (c) Degree of primordial mass
  segregation where $\mseg$ ranges from $\mseg = 0.01$ (lower curve)
  to $\mseg = 0.99$ (upper curve). The individual curves correspond to
  clusters with different initial values of the cluster parameter of
  interest.  In all simulations (except those in panel (a)), the gas
  expulsion takes place at $\tgas = 5~\Myr$. Immediately after gas
  removal, a significant fraction of stars become unbound from the
  cluster.  As the cluster's evolution continues, the mass loss rate
  drops significantly and $\fbnd$ approaches a constant value.}
\label{fig:Bound_t_02} 
\end{figure}

The value of the bound fraction at $t = 10~\Myr$ provides one measure
of how tightly bound a cluster remains after the embedding gas is
removed.  Figures \ref{fig:Bound_01} and \ref{fig:Bound_02}
display the value of the bound fraction at $t = 10~\Myr$ as a function
of the initial cluster parameter values for the range of parameters
considered in this survey.  Previous theoretical and numerical work
has identified a cluster's star formation efficiency $\sfe$ as the
most important parameter in determining whether or not a cluster will
remain gravitationally bound \citep{Hills1980ApJ, Elmegreen1983MNRAS,
  Lada1984ApJ}. In clusters with high SFEs, a large proportion of the
total cluster mass remains behind (in the form of stars) after the
embedding gas is removed. Clusters with high SFEs remain more tightly
bound after gas dispersal than clusters with low SFEs. In our cluster
parameter survey, we also find that the bound fraction $\fbnd$ at $t =
10~\Myr$ depends sensitively on the star formation efficiency $\sfe$
of the cluster.  Figure \ref{fig:Bound_01} panel (a) displays the
cluster bound fraction $\fbnd$ as a function of star formation
efficiency, $\sfe$.  The data is well fit by a power-law in $\sfe$:
\bee 
\fbnd = 2.23 \left(\sfe\right)^{1.2}
\qquad{\rm where} \qquad \sfe \leq 0.5 \, .  
\eee
This fit is shown as the solid curve (line) in the figure. 

\begin{figure}
\epsscale{1.0}
\plotone{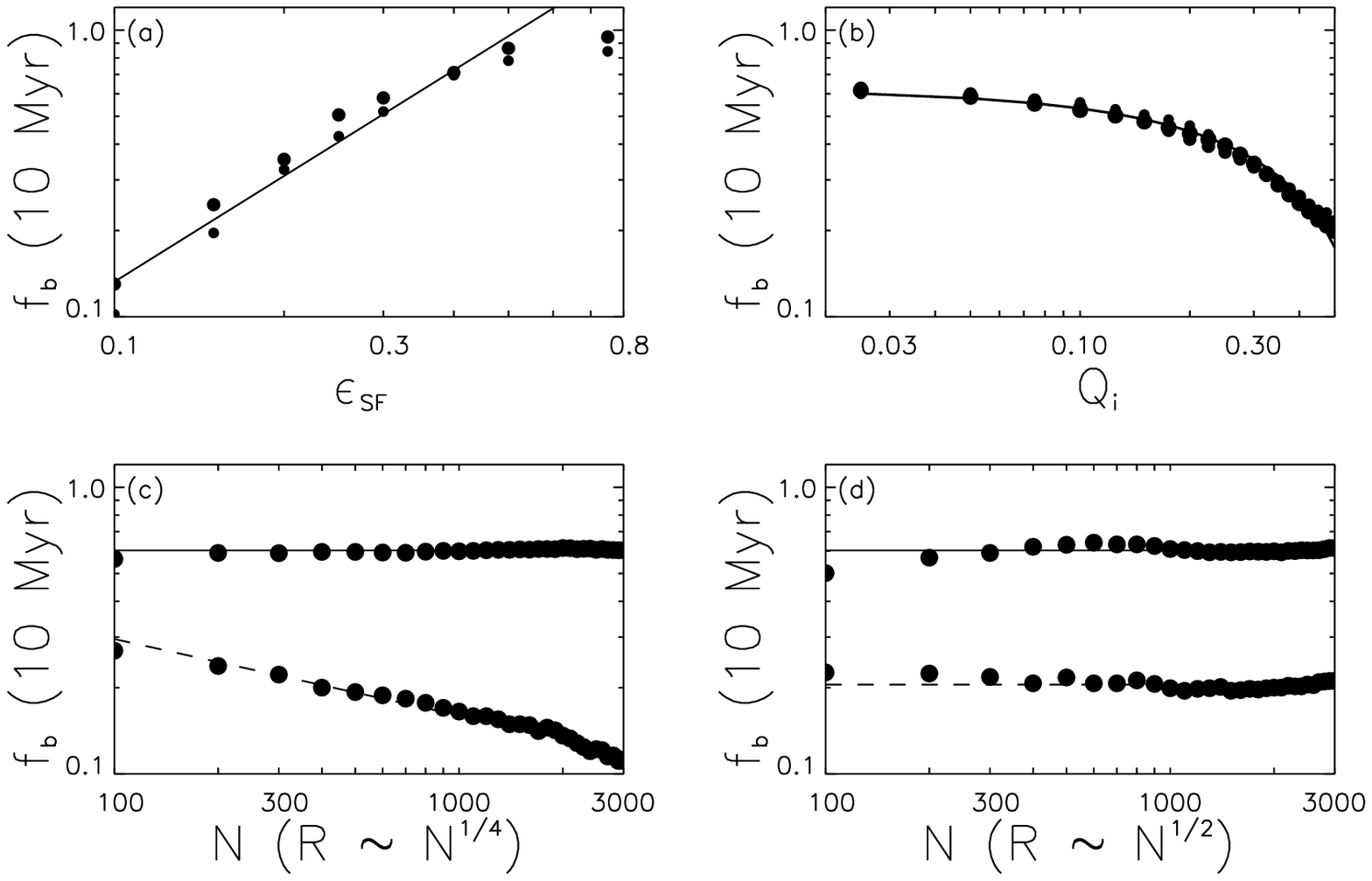} 
\caption{Cluster bound fraction $\fbnd$ at $t = 10~\Myr$ plotted as a
  function of cluster initial conditions, for all clusters included in
  the parameter space survey.  The best fit functions described in the
  text are displayed in panels (a) -- (d). The cluster parameter varied
  in each panel is as follows: (a) Star formation efficiency $\sfe$,
  (b) Initial virial ratio $Q_i$, (c) Stellar Membership $N$ for
  scaling relationship $\rstar \sim N^{1/2}$, (d) Stellar Membership
  $N$ for scaling relationship $\rstar \sim N^{1/4}$. Different curves
  and circle sizes indicate different initial conditions within each
  series of simulations. In panels (a) and (b) $N = 300, 1000$, and
  $2000$ are indicated by the small, medium, and large circles,
  respectively, though there are no major differences between the
  clusters as a function of size.  In panels (c) and (d), subvirial
  clusters are indicated by the solid curve while virial clusters are
  indicated by the dashed curve.}
\label{fig:Bound_01} 
\end{figure}

\begin{figure}
\epsscale{1.0}
\plotone{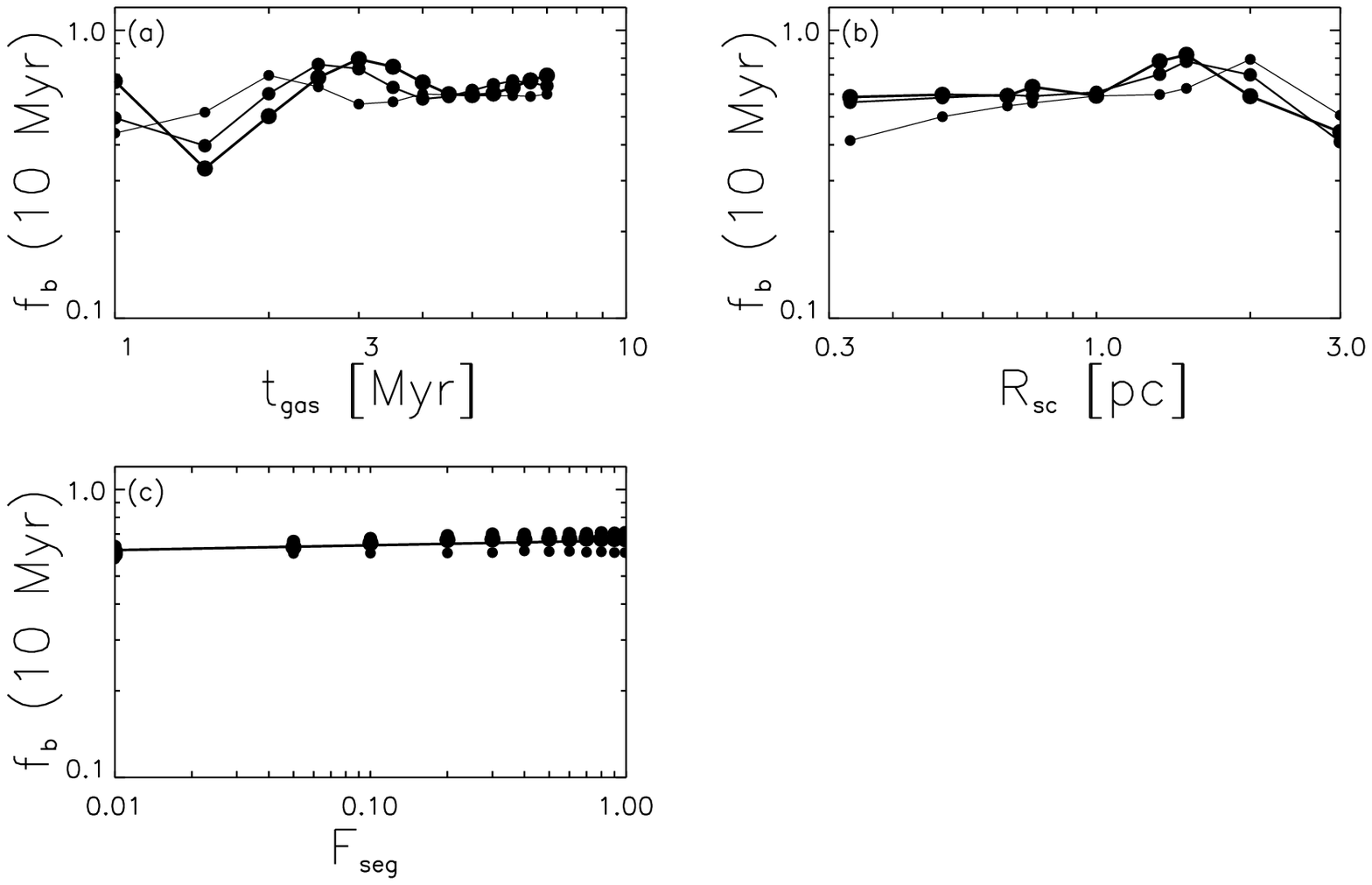} 
\caption{Continuation of Figure \ref{fig:Bound_01}. The cluster
  parameter varied in each panel is as follows: (a) Gas removal time
  $\tgas$, (b) Cluster scaling radius $\rsc$, and (c) Degree of 
  primordial mass segregation $\mseg$. Different curves and circle
  sizes indicate different initial conditions within each series of
  simulations. In panels (a), (b), and (c) $N = 300, 1000$, and $2000$
  are indicated by the small, medium, and large circles connected by
  thin, medium, and thick curves, respectively. In panel (c) there are
  no major differences between the clusters as a function of size and
  the best fit function described in the text is displayed.}
\label{fig:Bound_02} 
\end{figure}

In the suite of simulations used to investigate the effects of star
formation efficiency, the clusters are initially subvirial.  After gas
removal, subvirial clusters are more tightly bound than virial
clusters (APFM). For even relatively high star formation efficiencies
($\sfe = 0.3$) and small initial virial parameters, conditions which
produce the most tightly bound systems, the clusters are significantly
disrupted by gas removal and promptly lose $\sim40\%$ of their stars.
Star formation efficiencies larger than $\sim30\%$ are rarely observed
\citep[and references therein]{Lada2003ARAA}, and are difficult to
attain theoretically \citep{Matzner2000ApJ}. Our results are thus in
agreement with previous studies indicating that a significant fraction
of the stellar population is lost from a cluster during the gas
removal phase (e.g., Lada et al. 1984, Adams 2000, Boily \& Kroupa
2003ab).

As mentioned above, clusters with subvirial initial velocities are
more tightly bound than clusters with virial initial conditions.
Figure \ref{fig:Bound_01} panel (b) demonstrates this trend by
plotting the bound fraction as a function of the initial virial ratio.
The bound fraction decreases steadily with the initial virial ratio
$Q_i$ over the range considered, $0.01 \leq Q_i \leq 0.5$.  Gas
removal has a weaker effect on spherical clusters with subvirial
initial conditions because as a subvirial cluster collapses, more of
the stars spend more of their time inside of the embedding gas (which
is assumed to be static, i.e., not in a state of global collapse).
When the embedding gas is removed from the cluster, many of the
cluster members are interior to the gas and are less affected by the
change in potential.  The results from this suite of simulations again
indicate that a significant fraction of cluster members are lost due
to the change in the gravitational potential that occurs during the
dispersal of the natal gas. In the most tightly bound subvirial
clusters with $Q_i = 0.025$, approximately $40\%$ of stars become
unbound due to dispersal of the gas. 
  
In clusters with subvirial initial conditions, the bound fraction
remains constant as a function of cluster membership $N$, for both 
the $\rstar \sim N^{1/2}$ and the $\rstar \sim N^{1/4}$ cluster
membership-radius scaling relations.  This finding also holds true for
virial clusters that have cluster membership-size relations similar to
those observed in the solar neighborhood ($\rstar \sim N^{1/2}$).  The
bound fraction at $t=10~\Myr$ is plotted as a function of cluster size
$N$ in panels (c) and (d) of Figure \ref{fig:Bound_01}.  The upper
curves in these panels indicate the bound fraction at $10~\Myr$ in the
more tightly bound subvirial clusters, whereas the lower curves
correspond to the virial clusters.

In virial clusters with a lower power-law index $\alpha$ in the
cluster membership-size relation ($\rstar \sim N^{1/4}$), the bound
fraction decreases as a function of the cluster membership $N$.  The
bound fraction decreases roughly as $\fbnd \sim N^{-1/4}$ (see Figure
\ref{fig:Bound_01}, panel (c), lower curve).  This decrease in
$\fbnd(N)$ is due to a combination of effects arising from the
relationship between cluster radius and cluster membership defined by
equation (\ref{eq:RofNparam}). In clusters with $\rstar \sim N^{1/4}$, 
the mean velocity and velocity dispersion scale approximately as
$\left<v\right> \sim \sigv \sim \sqrt{GmN/\rstar} \sim N^{3/8}$.  The
velocity distributions in the clusters are nearly Gaussian during the
embedded phase (rather than perfectly Maxwellian as would be expected
in a collisionless isothermal sphere of stars).  As a result, the 
increased velocity dispersion in clusters with larger $N$ results in
more stars with velocities high enough to escape from the cluster
$v_{esc} \sim \sqrt{2GmN/\rstar}$.  In addition, the interaction rate
between cluster members increases with the stellar density, which
increases as $n \sim N^{1/4}$ in these clusters.  In virial clusters
with $\rstar \sim N^{1/2}$ (Figure \ref{fig:Bound_01}, panel (d),
lower curve), the bound fraction is roughly constant.  This trend
occurs because although the average velocity and velocity dispersions
increase as a function of cluster membership, the dependence on $N$ is
not as strong: $\left<v\right> \sim \sigv \sim N^{1/4}$.  In addition,
the stellar density actually decreases with $N$, $n \sim N^{-1/2}$; 
the competing effects of increased velocities and lower interaction 
rates are comparable and act to cancel each other out.

The bound fraction does not appear to be simply related to either the
gas removal timescale $\tgas$ or the scaling radius $\rsc$, i.e., the
bound fraction $\fbnd$ is not a monotonic function (see Figure
\ref{fig:Bound_02}, panels (a) and (b)). This result occurs
because in subvirial clusters, such as the ones considered in these
parameter space surveys, changing either the scaling radius or the gas
removal time affects the relationship between the gas removal time and
the initial collapse and relaxation time. The resulting $\fbnd$ is
sensitive to the particular dynamical state of the system at the time
of gas removal. For example, if the cluster is re-expanding (after its
initial collapse) when the gas is removed, many stars will have
trajectories that are directed radially outward and are thus more
likely to become gravitationally unbound.

Next, we consider the effects of primordial mass segregation. In
general, the effects of mass segregation saturate when more than
approximately $20\%$ of the stars are segregated by mass.  Mass
segregation only slightly affects the bound fraction, and clusters
with minimal mass segregation (where the largest star is located at
the cluster center) have slightly lower bound fractions than
clusters with $\mseg = 0.2$, as shown in panel (c) of Figure
\ref{fig:Bound_02}.

In summary, the results of this part of the study indicate that the
star formation efficiency $\sfe$ is the parameter that most
significantly affects the fraction of a cluster that remains bound
after the embedding gas is removed from the system. In addition, the
initial virial state of the cluster, as well as the specific dynamical
state at the time of gas dispersal, are important parameters in
determining how many members remain bound to the cluster.  We find
that in sufficiently subvirial clusters, $Q_i \lesssim 0.2$, the bound
fraction is not a sensitive function of the initial stellar density
(as indicated by the suite of simulations varying $N$ and the cluster
membership-radius relations), but rather is dominated by the fact that
the initial global collapse produces a cluster whose members reside
interior to the bulk of the embedding gas and thus are not strongly
affected by the gas removal.

Two caveats should be included in this discussion.  First, current
observations of young emerging clusters cannot determine whether a
cluster member is gravitationally bound or unbound from its host
cluster.  Over the first $\sim10 - 20~\Myr$, bound and unbound
clusters are visibly similar, and the results of simulations such as
those presented here are not easily compared directly to observations.
Second, this parameter space study focuses on the early evolution of
embedded clusters.  Additional dynamical evolution of the clusters on
timescales greater than $\sim10~\Myr$ will lead to even lower bound
fractions $\fbnd$ at later times.  As a result, the bound fractions
presented in this work should be considered as upper limits on the
expected bound fractions for clusters with older ages.

Finally we note that Tables 2 -- 7 provide compilations of the bound
fractions $\fbnd$ evaluated at time $t$ = 10 Myr for the simulations
presented here. Each table lists the bound fractions $\fbnd$ as a
function of a given input variable, including the stellar membership
$N$ (Table 2), initial virial parameter $Q_i$ (Table 3), cluster
scaling radius $\rsc$ (Table 4), star formation efficiency $\sfe$
(Table 5), gas removal time $\tgas$ (Table 6), and the mass
segregation parameter $\mseg$ (Table 7).

\subsection{Stellar Interaction Rates}\label{sec:Gammas}

A significant consequence of living in high density environments, such
as those found in young embedded clusters, is that close encounters
with other cluster members may be relatively frequent.  If these
interactions are sufficiently close, they can have important
ramifications for planet formation in circumstellar disks and for
solar system survival. During early stages of solar system formation,
encounters can disrupt protoplanetary disks and limit the mass
reservoir for planet formation \citep{Ostriker1994ApJ, Heller1993ApJ,
  Heller1995ApJ, Kobayashi2001Icar}.  At later times, close encounters
can disrupt planetary systems themselves, by significantly altering
the eccentricities of planets and, in sufficiently close encounters,
ejecting planets from the solar system entirely (Adams \& Laughlin
2001, APFM).

Throughout the cluster simulations, close encounters with distance of
closest approach less than $b = 10^4$~AU are recorded. Note that the
distance of closest approach is somewhat smaller than the impact
parameter of the encounter due to gravitational focusing.  A
cumulative distribution of the close encounters is then constructed,
and the interaction rate $\Gamma$ is calculated by averaging the
encounter distributions over the time span of interest (here we use
the embedded phase $t = 0 - \tgas$, the exposed phase $t = \tgas -
10~\Myr$, or the entire $10~\Myr$ interval).  Specifically, the
interaction rate $\Gamma(b)$ is defined as the number of close
encounters with distance of closest approach $r \leq b$ per star per
million years.  We find that the interaction rates have the form of
power-laws for encounters with closest approach distances less than
$\sim3000$~AU.  In other words, the interaction rate has the form
\bee
\Gamma = \Gamma_0\left(\frac{b}{1000\rm{AU}}\right)^{\gamma} \, ,
\label{eq:IntRate}
\eee
where $b$ is the distance of closest approach. The fiducial
interaction rate $\Gamma_0$ and the power-law index $\gamma$ are fit
to the cumulative closest approach distribution for each set of
cluster simulations.  The fiducial interaction rate $\Gamma_0$
corresponds to the number of encounters with impact parameter $b$ less
than $1000$~AU per star per million years.  The fiducial interaction
rate $\Gamma_0$ is displayed as a function of cluster initial
conditions in Figures \ref{fig:Gamma0_01} and \ref{fig:Gamma0_02}.

\begin{figure}
\epsscale{1.0}
\plotone{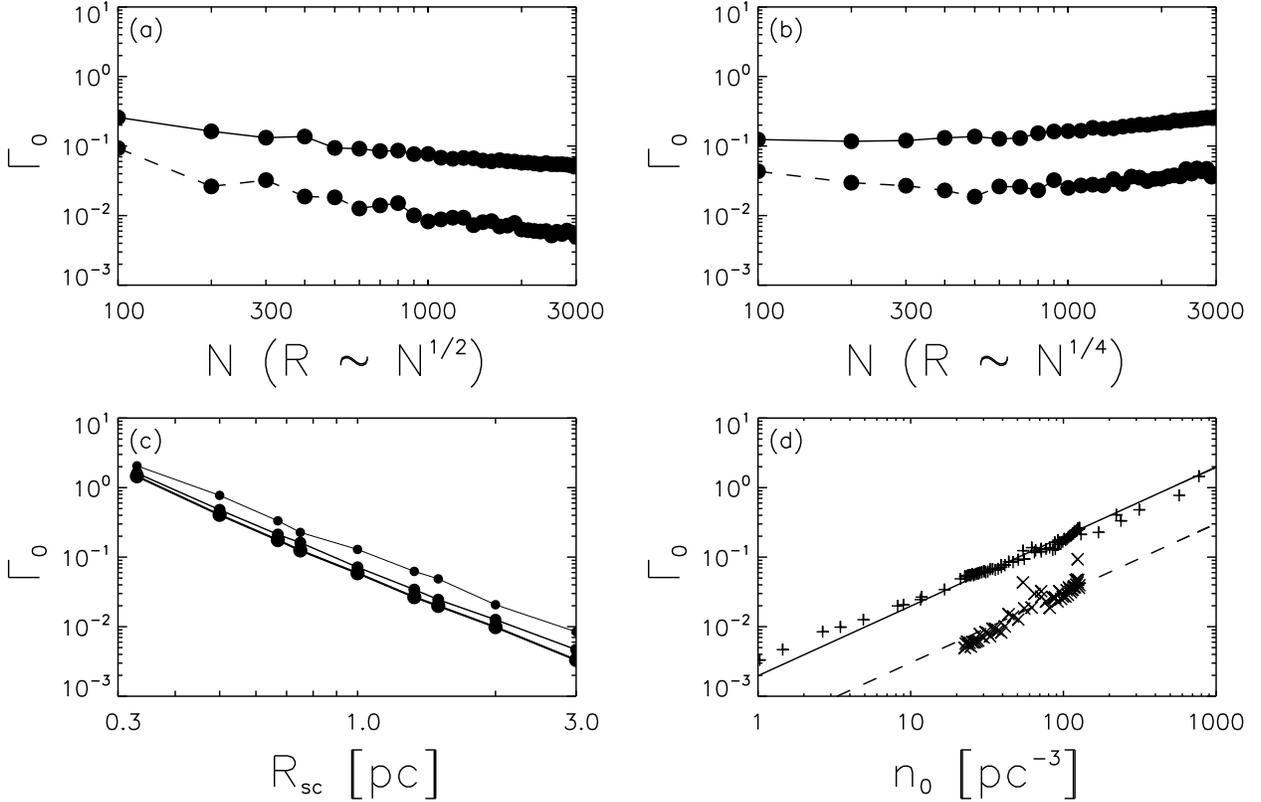} 
\caption{Fiducial interaction rate $\Gamma_0$ (in units of the number
  of interactions per star per Myr) is plotted as a function of
  initial cluster parameter, for all clusters included in the
  parameter space survey. The cluster parameter varied in each panel
  is as follows: (a) Stellar Membership $N$ for scaling relationship
  $\rstar \sim N^{1/2}$, (b) Stellar Membership $N$ for scaling
  relationship $\rstar \sim N^{1/4}$, and (c) Cluster scaling radius
  $\rsc$.  Panel (d) displays the combined data from panels (a) - (c)
  as a function of initial stellar number density $n_0$ and displays
  the trend $\Gamma_0 \sim n_0$ discussed in the text. In panels (a),
  (b), and (d) subvirial clusters are indicated by the solid curve
  while virial clusters are indicated by the dashed curve. In panel
  (c), clusters with $N = 300, 1000$, and $2000$ are indicated by the
  small, medium, and large circles 
  (\emph{thin, medium, and thick curves}), respectively.}
\label{fig:Gamma0_01} 
\end{figure}

\begin{figure}
\epsscale{1.0}
\plotone{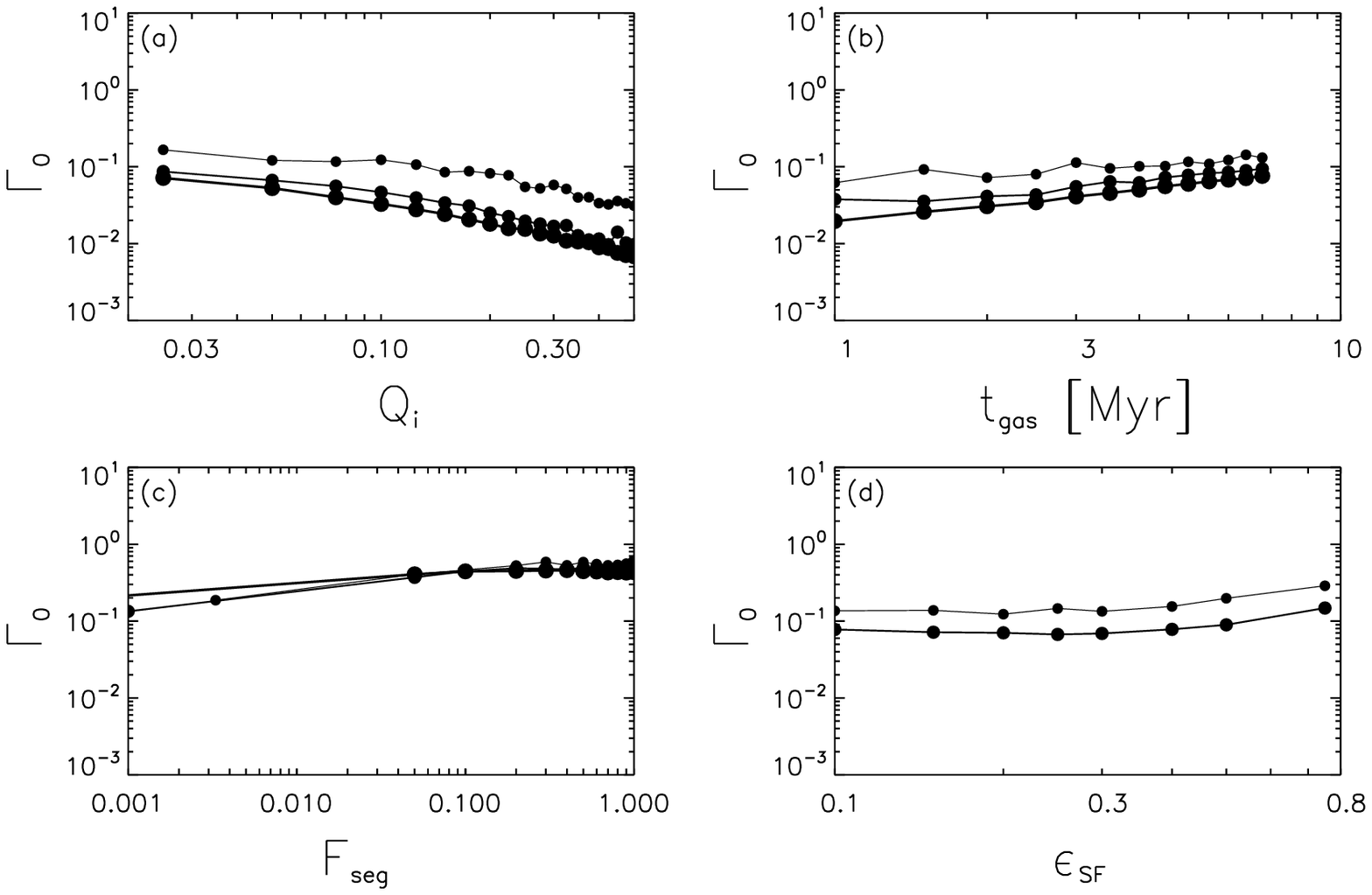} 
\caption{Continuation of Figure \ref{fig:Gamma0_01}. The cluster
  parameter varied in each panel is as follows: (a) Initial virial
  parameter $Q_i$, (b) Gas removal time $\tgas$, (c) Degree of 
  primordial mass segregation $\mseg$, and (d) Star formation
  efficiency $\sfe$.  Clusters with $N = 300, 1000$, and $2000$ 
  are indicated by the small, medium, and large circles 
  (\emph{thin, medium, and thick curves}), respectively.}
\label{fig:Gamma0_02} 
\end{figure}

\begin{figure}
\epsscale{1.0}
\plotone{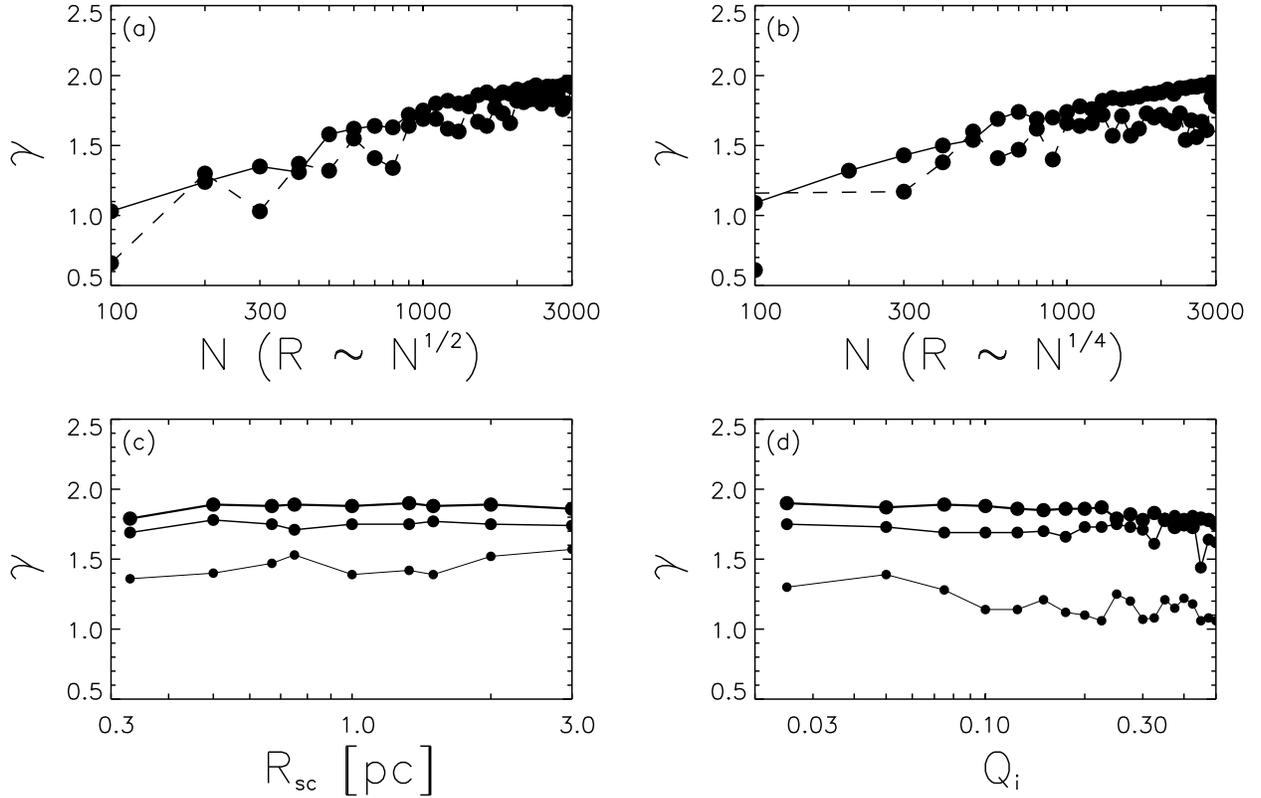}
\caption{The power-law index $\gamma$ for the interaction rates is
  plotted as a function of initial cluster parameter, for all clusters
  included in the parameter space survey.  The cluster parameter
  varied in each panel is as follows: (a) Stellar Membership $N$ for
  scaling relationship $\rstar \sim N^{1/2}$, (b) Stellar Membership
  $N$ for scaling relationship $\rstar \sim N^{1/4}$, (c) Cluster
  scaling radius $\rsc$, and (d) Initial virial ratio $Q_i$. In panels
  (a) and (b), subvirial clusters are indicated by the solid curve
  while virial clusters are indicated by the dashed curve. In panels
  (c), and (d) clusters with $N = 300, 1000$, and $2000$ are indicated
  by the small, medium, and large circles 
  (\emph{thin, medium, and thick curves}), respectively.}
\label{fig:gamma_01} 
\end{figure}

\begin{figure}
\epsscale{1.0}
\plotone{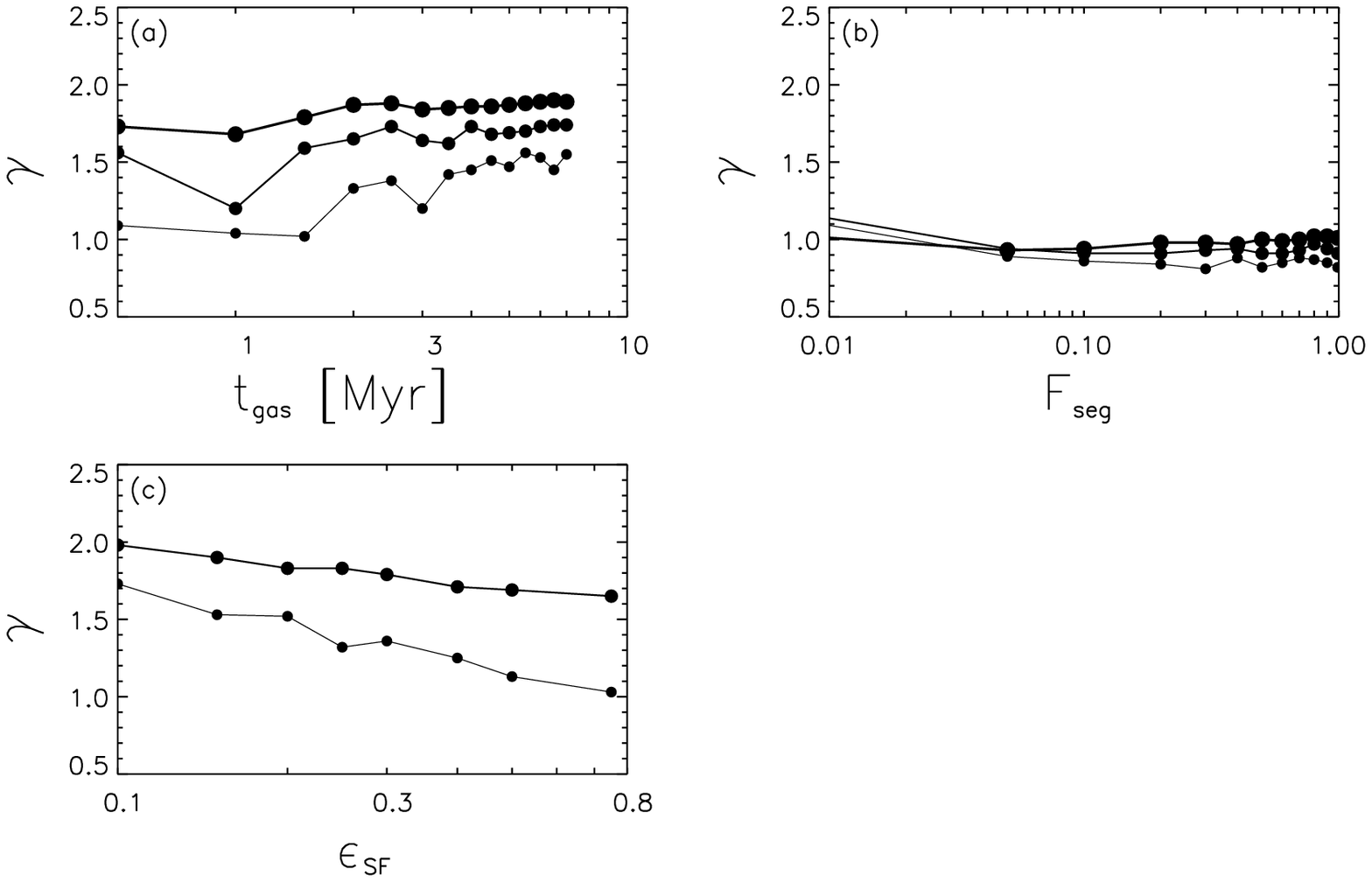} 
\caption{Continuation of Figure \ref{fig:gamma_01}.  The cluster
  parameter varied in each panel is as follows: (a) Gas removal time
  $\tgas$, (b) Degree of primordial mass segregation $\mseg$, and (c)
  Star formation efficiency $\sfe$. Clusters with $N = 300, 1000$, and
  $2000$ are indicated by the small, medium, and large circles
  (\emph{thin, medium, and thick curves}), respectively.}
\label{fig:gamma_02} 
\end{figure}

For the parameter space considered in this study, the interaction rate
$\Gamma$ depends most sensitively on a single parameter, the stellar
number density $n$.  The trends observed as a function of stellar
membership $N$ and cluster scaling radius $\rsc$ (see Figure
\ref{fig:Gamma0_01} panels (a) -- (c)) are more fundamentally trends
indicating how $\Gamma_0$ varies as a function of the average stellar
density $n$ (keep in mind that the stellar density varies with $N$ 
and $\rsc$). 

We can understand this behavior in simple terms as follows: Consider a
cluster with $N$ stars and radius $\rstar$. For simplicity, we ignore
the difference between impact parameters and distances of closest
approach $b$.  A star passing through the cluster will experience, 
on average, a number $\delta n$ of close encounters with impact 
parameters within the range $b$ to $b+db$ \citep{Binney1987Book},
where $\delta n$ is given by
\bee 
\delta n = \frac{2N}{\rstar^2}b \delta b \, .
\eee 
The crossing time in a cluster is given by $\tcross \approx \rstar/v$,
where $v$ is the average stellar velocity.  As a result, the star will
experience close encounters in the given annulus at the rate
\bee 
\delta \Gamma \approx
\frac{2N}{\rstar^2}\left(\frac{v}{\rstar}\right)b ~\delta b 
= 2 n v b ~\delta b \, .  
\eee
The interaction rate for all encounters with impact parameter less
than $b$ is then given by $\Gamma \sim n v b^2$. Since the mean
dynamical speeds are slowly varying over the regime of cluster
parameter space considered in this paper, the interaction rate 
$\Gamma \propto n$ as claimed.

\ecomment{\bee
\frac{1}{t_{coll}} = 4 \sqrt{\pi} n \sigma r^2_{coll} 
+ \rm{gravitational ~focusing ~term. ~(BT ~8-122)}
\eee
where we've used the relationship that $<v> =
\frac{4\sigma}{\sqrt{\pi}}$ for a Maxwellian distribution.}

In Figure \ref{fig:Gamma0_01} panel (d), the $\Gamma_0$ values are
plotted as a function of initial stellar density $n_0 = N/\rstar^3$
for all of the simulations with varying cluster membership $N$ and
scaling radius $\rstar$ (using the data from Figure \ref{fig:Gamma0_01} 
panels (a) -- (c)). The plus symbols indicate the interaction rate
$\Gamma_0$ for clusters with initially subvirial velocities, whereas
the interaction rates for clusters with initially virial velocities
are indicated by the x's. Straight lines (indicating a power-law with
index = $1$) are included in the panel; note that the numerically
determined data are roughly consistent with the simple scaling relation
$\Gamma_0 \sim n_0$.

Figures \ref{fig:gamma_01} and \ref{fig:gamma_02} present the values
of the index $\gamma$ that provide the best fit to the close encounter
distributions as a function of the initial cluster parameters.  The
value of $\gamma$ does not vary strongly as a function of the initial
conditions, but rather remains in the range $\sim1-2$.

Using the simple argument constructed above, the total rate $\Gamma_b$
of close encounters with impact parameter less than $b$ in a cluster
of membership size $N$ is given by
\bee
\Gamma_b \approx  0.122 \, \, N^{-1/2} 
\left(\frac{b}{1000 ~{\rm AU}}\right)^2 {\rm Myr}^{-1}  
\qquad {\rm for} \qquad \rstar  = 1.0 ~{\rm pc} 
\left(\frac{N}{300}\right)^{1/2} \, ,
\eee
and 
\bee
\Gamma_b \approx  0.0095 \, \, N^{1/4} 
\left(\frac{b}{1000 ~{\rm AU}}\right)^2 {\rm Myr}^{-1}  
\qquad {\rm for} \qquad \rstar  = 1.0 ~{\rm pc} 
\left(\frac{N}{300}\right)^{1/4} \, .
\eee 
These estimates are similar to the interaction rates found for the
virial clusters, although the fitted value of the index $\gamma$ is
slightly lower than $2$ (due to gravitational focusing) in the
numerically determined distributions (see Figures \ref{fig:gamma_01}
and \ref{fig:gamma_02}), and the fiducial interaction rate is somewhat
higher.

The subvirial clusters have interaction rates that are about 8 times
larger than the rates for virial clusters of the same starting density
(as defined by $n_0 \propto N/\rstar^3$). This trend is due to a
combination of the smaller effective cluster radius that a subvirial
cluster attains after initially collapsing and its higher bound fraction 
after gas dispersal.  During the embedded phase, subvirial clusters
($Q_i=0.04$) behave as if they have nearly zero-temperature starting
states and thus collapse to roughly $\sqrt{2}$ of their initial size.
This decrease in radius corresponds to an increase in density by a
factor of $2\sqrt{2}$.  In addition, subvirial clusters retain more of
their members after the gas is removed from the cluster --- we find
that the bound fraction is $\sim3$ times higher for the subvirial
starting states. Since the close encounter profiles are averaged over
the initial number of stars in the cluster, the interaction rates in
subvirial clusters will be roughly 3 times higher than in virial
clusters over the exposed phase of cluster evolution (due to increased
stellar retention).  Note that combining these two factors increases
the interaction rates by $\sim6\sqrt{2} \sim 8$. The results of the
parameter survey varying $Q_i$ also indicate that subvirial clusters
have higher interaction rates.  In Figure \ref{fig:Gamma0_02} panel
(a), the interaction rate clearly decreases as a function of initial
virial parameter $Q_i$.

We also find that the interaction rates are somewhat higher in
clusters that have more of their massive stars residing near the
cluster center (see Figure \ref{fig:Gamma0_02}, panel (c)).  This
finding is consistent with the modeling results of the observed
cluster NGC 1333 presented in APFM.  The fiducial interaction rate
$\Gamma_0$ in the simulated NGC 1333 cluster was approximately $5$
times higher than the $\Gamma_0$ calculated in equivalent subvirial
clusters with minimal mass segregation.  We suggested that the
increased interaction rate was due to the primordial mass segregation
observed in NGC 1333 (see Figure 15 of APFM).  For comparison,
subvirial clusters with $N$ = 300 stars and $\mseg = 0.05$ have
fiducial interaction rates that are about $5$ times larger than
those found in subvirial clusters with $N = 300$ members and minimal
mass segregation (see Figure \ref{fig:Gamma0_02}, panel (c), top
curve).

The average interaction rate also increases as a function of gas
removal time $\tgas$ as shown in Figure \ref{fig:Gamma0_02} panel (b).
This interaction rate is averaged over the $10~\Myr$ simulation time.
However, the majority of close encounters occur during the embedded
phase, and hence the average interaction rate increases as the length
of the embedded phase increases.  For clusters with embedded phases
lasting more than $\sim2~\Myr$, the rate of close encounters during
the embedded phase is roughly constant.  When the embedded phase lasts
less than $\sim2~\Myr$, the clusters have lower encounter rates due to
lower densities during the first $\sim1~\Myr$ while the subvirial
cluster is still contracting.  For completeness, we note that the
interaction rate does not display strong trends with varying
star-formation efficiency $\sfe$ (see Figure \ref{fig:Gamma0_02},
panel (d)).

For all of the simulations discussed in this section, Tables 8 -- 13
provide listings of the parameters ($\Gamma_0$, $\gamma$) that specify
the interaction rates through equation (\ref{eq:IntRate}).  Each table
provides the values of $\Gamma_0$ and $\gamma$ as a function of a
given input variable, including the stellar membership $N$ (Table 8),
initial virial parameter $Q_i$ (Table 9), cluster scaling radius
$\rsc$ (Table 10), star formation efficiency $\sfe$ (Table 11), gas
removal time $\tgas$ (Table 12), and the mass segregation parameter
$\mseg$ (Table 13). These interaction rates, as determined by the
values of $(\Gamma_0, \gamma)$, are one of the primary products of
this investigation.  They can be used to calculate interaction rates
as a function of closest approach distance for a wide variety of
cluster environments (see Section \ref{sec:Encounters} below for one
such application).

\subsection{Interaction Velocities}\label{sec:Velocity}

In addition to constructing the distribution of closest approach
distances associated with close encounters in the simulated clusters,
we also determine the distribution of encounter velocities.  The
distribution of encounter velocities provides additional information
regarding the effect that close encounters may have on the constituent
solar systems.  For example, the interaction cross sections depend on
the encounter speeds.

\begin{figure}
\epsscale{0.60}
\plotone{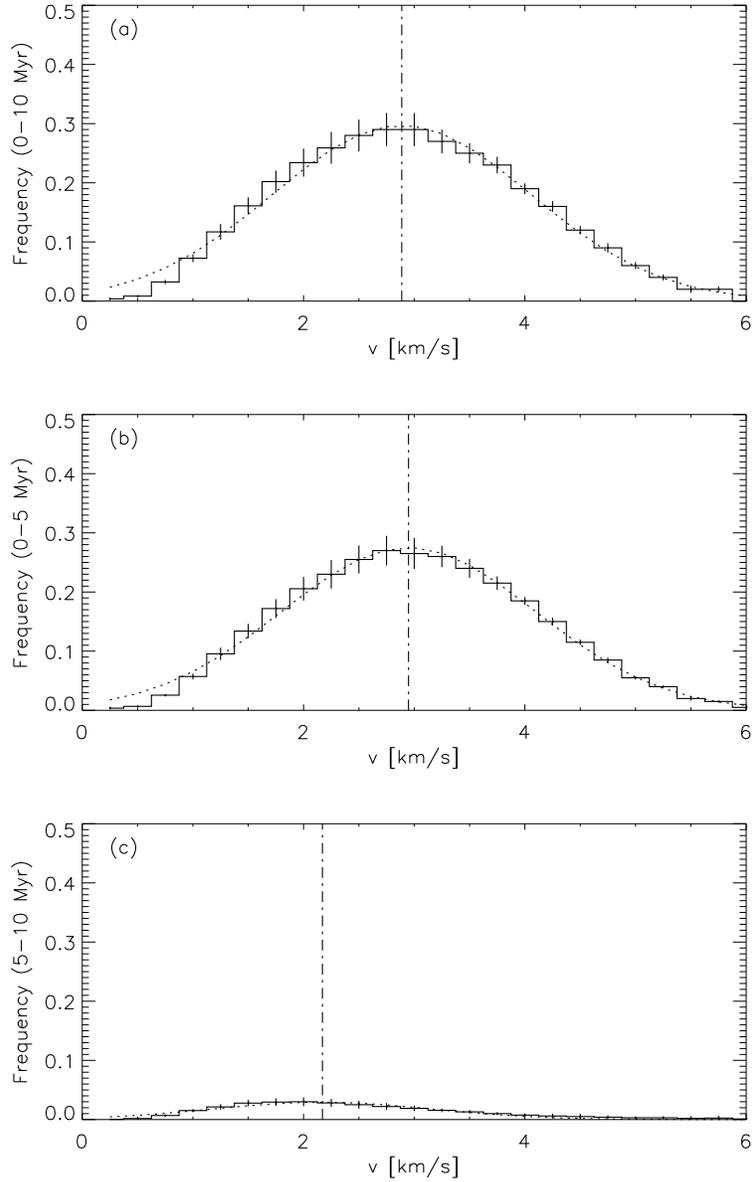}
\caption{The distribution of interaction velocities in a subvirial
  cluster with $N = 300$, $\rstar \sim N^{1/2}$, and $Q_i = 0.04$.
  The distribution averaged over the interval $t = 0-10~\Myr$ is
  presented in panel (a).  Panel (b) displays the time averaged
  distribution of interaction velocities during the embedded phase 
  ($t = 0-5~\Myr$) and panel (c) presents the distribution averaged 
  over the remainder of the cluster evolution ($t = 5-10~\Myr$).  The
  histogram binning size is $0.25~\kms$.  Error bars correspond to the
  dispersion (variance) within each velocity bin. The best fit 
  Gaussian curve is indicated by the dashed curve, and the mean of the
  Gaussian is indicated by the vertical dash-dot line.}
\label{fig:VelDist} 
\end{figure}

We define the encounter velocity $v_{enc}$ as the magnitude of the
relative velocities of the stars at the moment of closest approach.
We then create distributions of the frequency of encounter velocities
throughout the simulations. Figure \ref{fig:VelDist} presents the
resulting distribution of encounter velocities for a cluster with
radius $\rstar =1~\pc$, $N = 300$ stars, and subvirial initial speeds
($Q_i = 0.04$).  A binning size of $0.25~\kms$ has been used to
construct the histogram, and error bars are included on the
distribution to indicate the dispersion within each velocity bin.

We find that the encounter velocity distribution can be approximated
reasonably well by a Gaussian curve where the mean and the width (as
measured by the variance $\sigma^2$) of the Gaussian are varied to fit
the encounter velocity distribution for each particular set of initial
conditions.  The Gaussian fits to the velocity distributions in Figure
\ref{fig:VelDist} are indicated by the dashed curves.  We note that
the Gaussian fit slightly overestimates the number of low velocity
encounters with $v_{enc} \lesssim 0.5~\kms$ (and formally even
predicts a few encounters with negative velocities).  However, the
general shape and width of the distribution are well represented by
these gaussian forms.

In Figures \ref{fig:VelsNorm_01} and \ref{fig:VelsNorm_02}, the
mean encounter velocity $\left<v_{enc}\right>$ is plotted as a
function of the initial cluster parameter. The encounter velocity has
been normalized by the mean velocity within the cluster's half-mass
radius (the regime where most of the interactions occur within the
cluster).  The error bars indicate the normalized width (FWHM) of the
Gaussian that best fits the velocity distribution.  This figure
demonstrates that the normalized encounter velocity distributions do
not vary strongly as a function of the initial conditions, but rather
are a robust function of mean cluster velocity.

The encounter velocities are about twice the average velocity in the
interactive region of the cluster. This result is roughly consistent
with an analytic estimate of the relative velocities of cluster
members whose velocities are sampled from a Maxwellian distribution,
so that $\sqrt{v_{rel}^2} \sim \sqrt{2\left<v^2\right>}$
\citep{Binney1987Book}. The encounter velocities are somewhat larger
than those predicted by this estimate (about twice the mean velocity)
due in part to gravitational focusing.  Note that $b \sim 10^3$ AU is
a typical encounter distance, and that the orbit speed $v_{\rm orb}
\sim (GM_\ast/b)^{1/2} \sim 1$ km s$^{-1}$, so that the gravity of the
stars taking part in the encounter does matter.  The numerically
calculated distribution of interaction velocities includes only a
subsample of the relative velocities, because only encounters with
impact parameter $b \lesssim 10^4$~AU are included in the interaction
velocity distribution, and this subsample is likely to have somewhat
larger relative velocities.

\begin{figure}
\epsscale{1.0}
\plotone{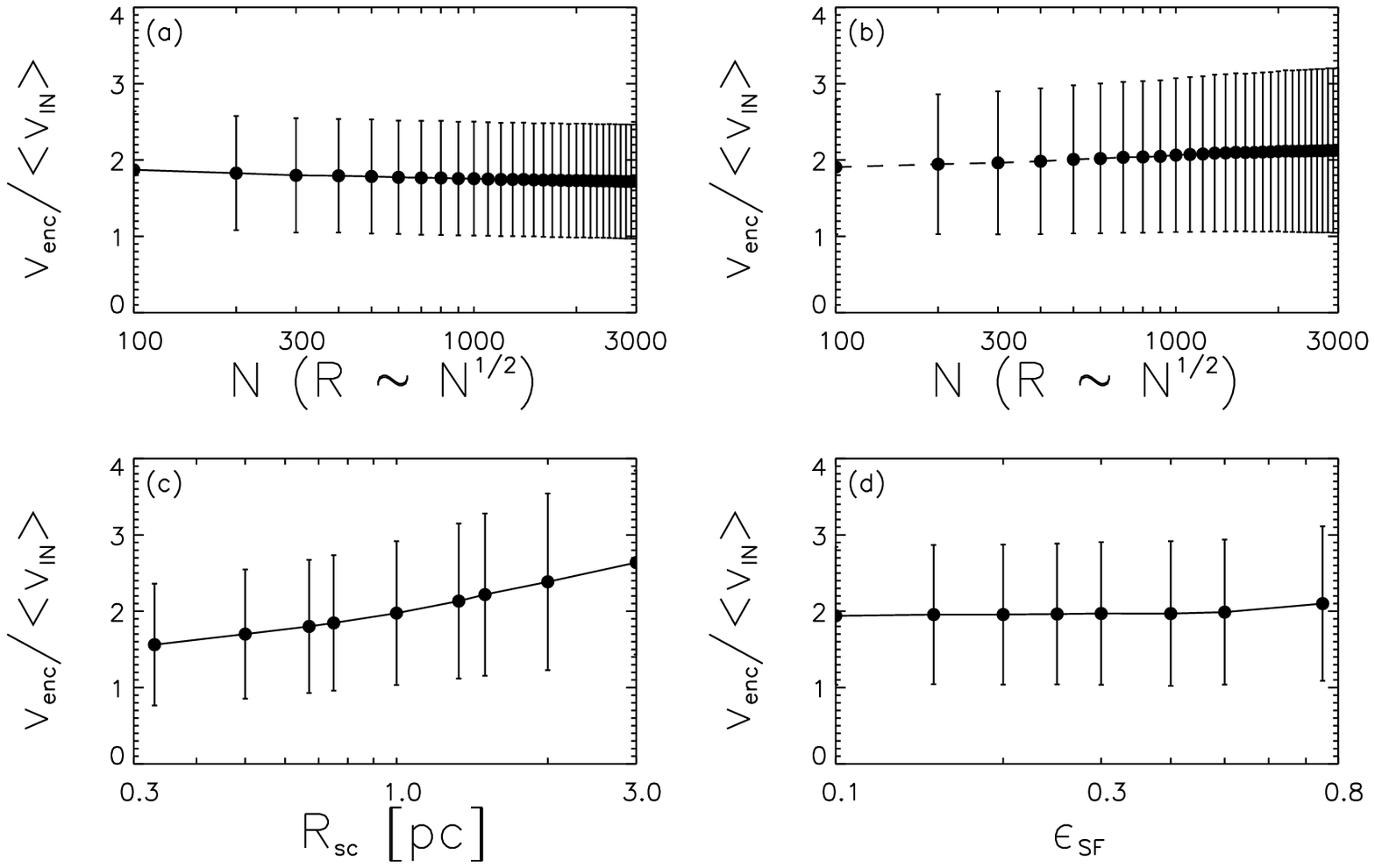} 
\caption{The parameters specifying the distribution of interaction
  velocities, specifically the mean $\mu$ and the FWHM of the
  distribution as a function of initial cluster parameter, for all
  clusters included in the parameter space survey. The mean
  interaction velocity $\mu$ (in $\kms$) is indicated by the data
  points and the FWHM is depicted by the error bars.  Velocities are
  scaled by the cluster mean velocity.  The distribution parameters
  $\mu$ and FWHM are presented as a function of (a) Stellar membership
  $N$ for virial clusters $Q_i = 0.5$ and $\rstar \sim N^{1/2}$, (b)
  Stellar membership $N$ for subvirial clusters $Q_i = 0.04$ and
  $\rstar \sim N^{1/2}$, (c) Initial cluster scaling radius $\rsc$,
  and (d) Star formation efficiency $\sfe$.}
\label{fig:VelsNorm_01} 
\end{figure}

\begin{figure}
\epsscale{1.0}
\plotone{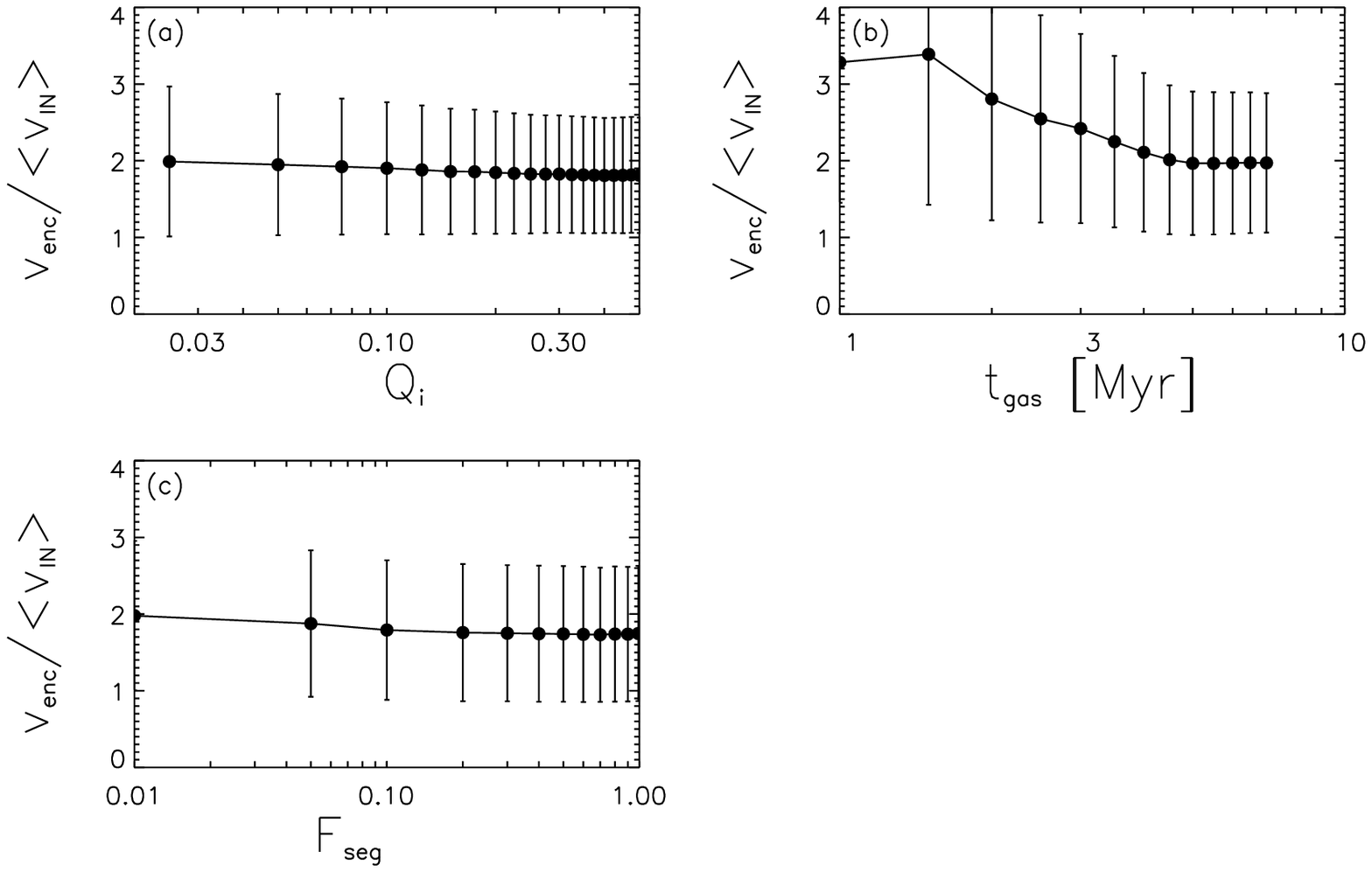} 
\caption{Continuation of Figure \ref{fig:VelsNorm_01}. The
  distribution parameters $\mu$ and FWHM are presented as a function
  of (a) Initial virial parameter $Q_i$, (b) Time of gas removal
  $\tgas$, and (c) Degree of primordial mass segregation $\mseg$.}
\label{fig:VelsNorm_02} 
\end{figure}

\subsection{Mass and Number Distributions}\label{sec:MassDists}

As a cluster evolves, interactions between stars, and between the
stars and the background gas potential, produce a distribution of
stellar positions and velocities.  The distribution of stars within a
cluster at a given time $t$ can be characterized by the cumulative
mass distribution $M(r,t)/M_{T\ast}$ or the cumulative number
distribution $N(r,t)/N$, where $M_{T\ast}$ and $N$ are the total
masses and numbers of the stars that are gravitationally bound to the
cluster at time $t$, respectively.  In the simulated clusters, each of
these distributions is calculated at intervals of $0.25$ Myr.  The
profiles are then averaged over the cluster lifetime and over 100
realizations of the cluster used to produce a statistical description
of the mass and number profiles.  We find that both of these
distributions may be fit by simple functions of the form
\bee
\frac{M(r)}{M_{T\ast}} = \left(\frac{\xi^a}{1+\xi^a}\right)^{p},
\label{eq:MofR}
\eee
\bee
\frac{N(r)}{N} = \left(\frac{\xi^a}{1+\xi^a}\right)^{p},
\label{eq:NofR}
\eee
where $\xi = r/r_0$ and the scale radius $r_0$ and index $p$ are free
parameters that are fit to the distributions observed in the simulated
clusters.  The parameter $a$ may also be varied to fit the data.  We
find that the choice $a =2$ gives the best fit for the subvirial
clusters and $a = 3$ gives the best fit for the initially virial
clusters (this finding is consistent with the results found in APFM). 
In the series of simulations where the initial virial parameter $Q_i$
is varied, the choice $a = 2$ works best for simulations for which
$Q_i < 0.25$, and $a = 3$ works best for those with $Q_i \geq 0.25$. 
The parameters $r_0$ and $p$ that provide the best fit for the radial
distributions (equation [\ref{eq:NofR}]) and the mass profiles
(equation [\ref{eq:MofR}]) are similar, although not identical, because 
stars of different masses have somewhat different radial profiles. 
 
These radial profiles of clusters provide insight into the general
evolution of a cluster, and perhaps more importantly, into the
expected radiation fields that young solar systems in the cluster will
experience.  Circumstellar disks and forming solar systems residing in
a cluster will be subjected to the FUV and EUV radiation fields
produced by the cluster population, and these fields are dominated by
the large stars in the cluster.  If these radiation fields are strong
enough, they are capable of photoevaporating circumstellar disks and
thereby preventing (or at least limiting) giant planet formation. The
massive young stars that produce the majority of the UV radiation are
often located near the center of the cluster
\citep{Hillenbrand1998ApJ, Bonnell1998MNRAS, Lada2003ARAA}. To first
approximation, the UV radiation is a cluster can be considered as a
point source at the center.  An understanding of the EUV and FUV
fields associated with young clusters combined with the average radial
distributions of stars in young clusters thus provides a framework with
which to predict how effectively cluster radiation can restrict planet
formation \citep{Johnstone1998ApJ, Fatuzzo2008ApJ}.

\begin{figure}
\epsscale{1.0}
\plotone{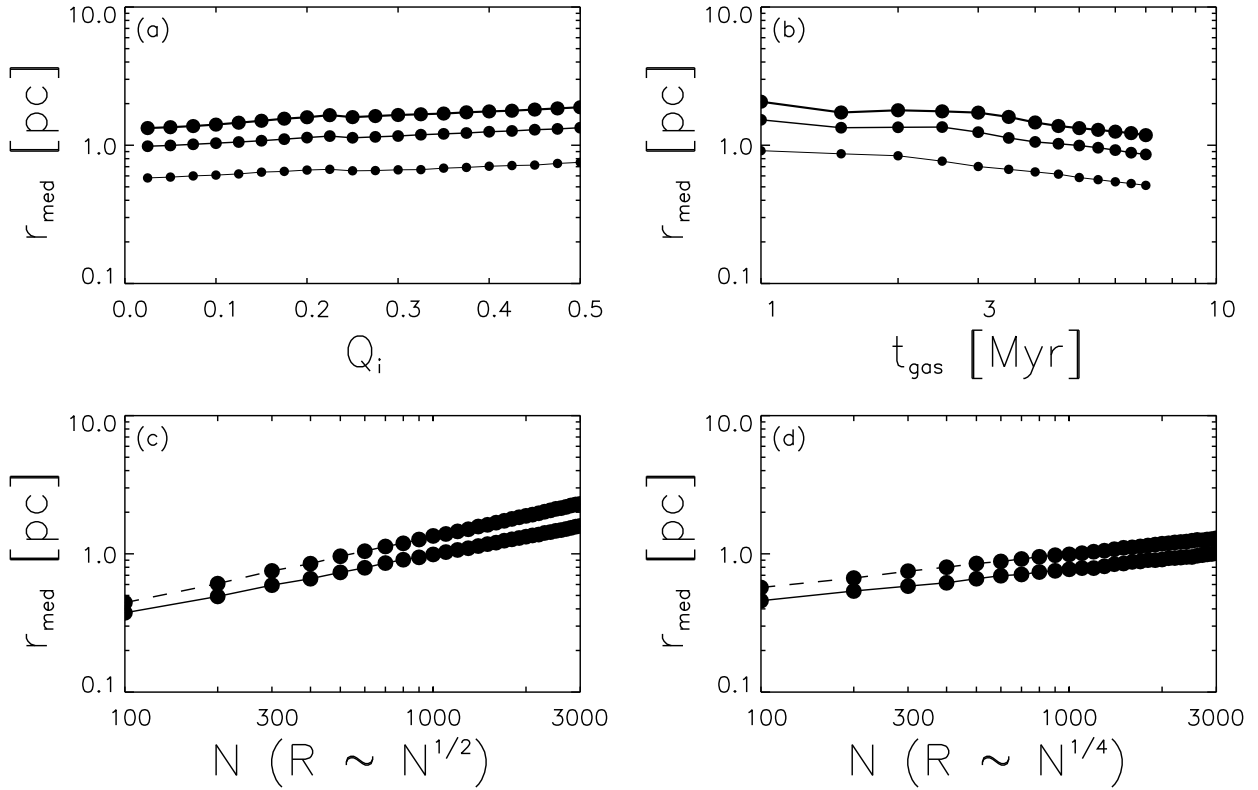} 
\caption{The median radius calculated from the radial profiles
  $N(r)/N$ as a function of initial cluster parameter, for all
  clusters included in the parameter space survey. The cluster
  parameter varied in each panel is as follows: (a) Initial virial
  ratio $Q_i$, (b) Gas removal time $\tgas$, (c) Stellar Membership
  $N$ for scaling relationship $\rstar \sim N^{1/2}$, and (d) Stellar
  Membership $N$ for scaling relationship $\rstar \sim N^{1/4}$. In
  panels (a) and (b) clusters with $N = 300, 1000$, and $2000$ are
  indicated by the small, medium, and large circles 
  (\emph{thin, medium, and thick curves}), respectively. In panels (c)
  and (d), subvirial clusters are indicated by the solid curve while
  virial clusters are indicated by the dashed curve.}
\label{fig:MedianRadius_01} 
\end{figure}

\begin{figure}
\epsscale{1.0}
\plotone{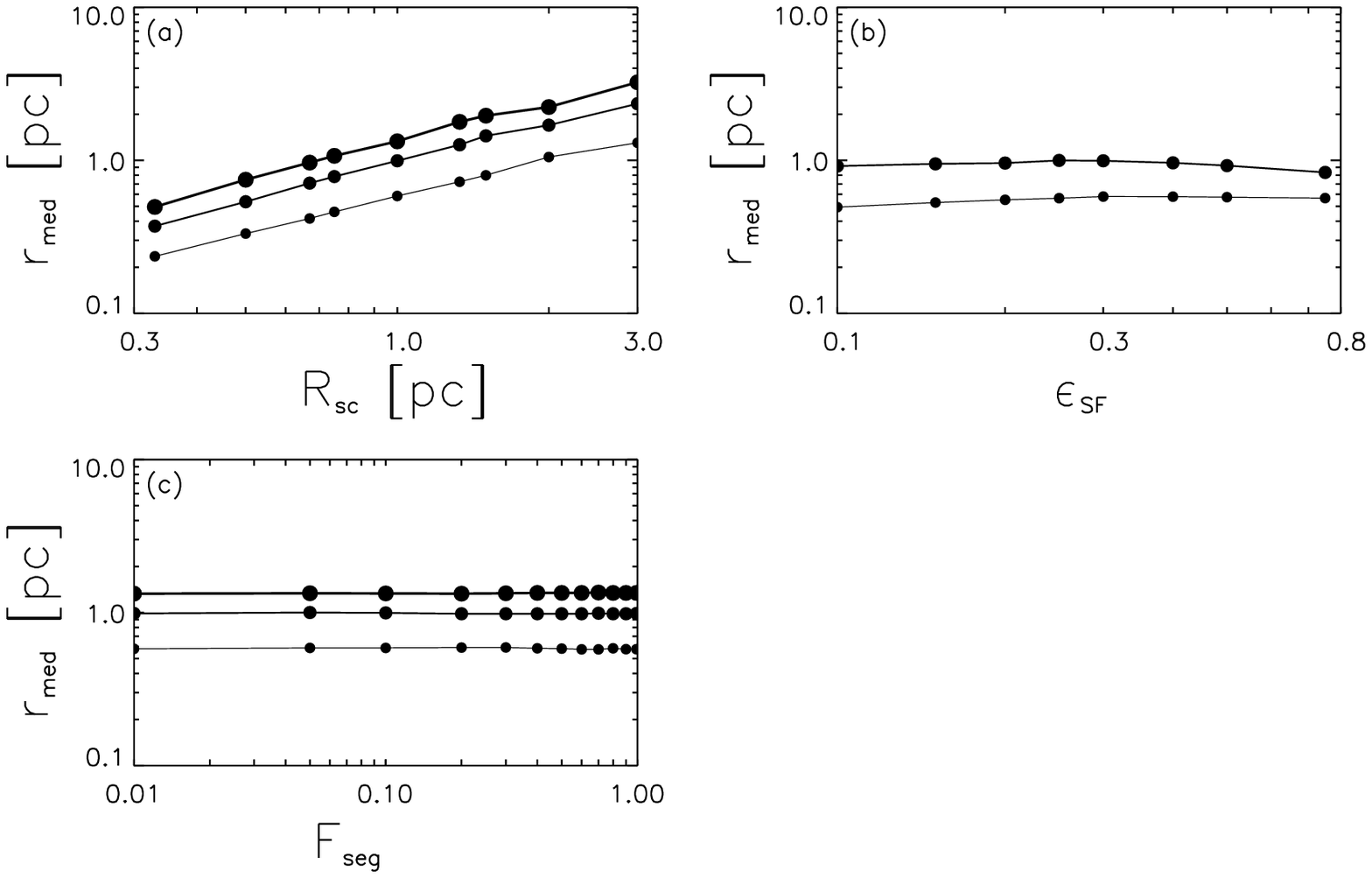} 
\caption{Continuation of Figure \ref{fig:MedianRadius_01}. The cluster
  parameter varied in each panel is as follows: (a) Cluster scaling
  radius $\rsc$, (b) Star formation efficiency $\sfe$, and (c) Degree
  of primordial mass segregation $\mseg$.  Clusters with $N$ = 300,
  1000, and 2000 are indicated by the small, medium, and large
  circles (\emph{thin, medium, and thick curves}), respectively.}
\label{fig:MedianRadius_02} 
\end{figure}

\begin{figure}
\epsscale{1.0}
\plotone{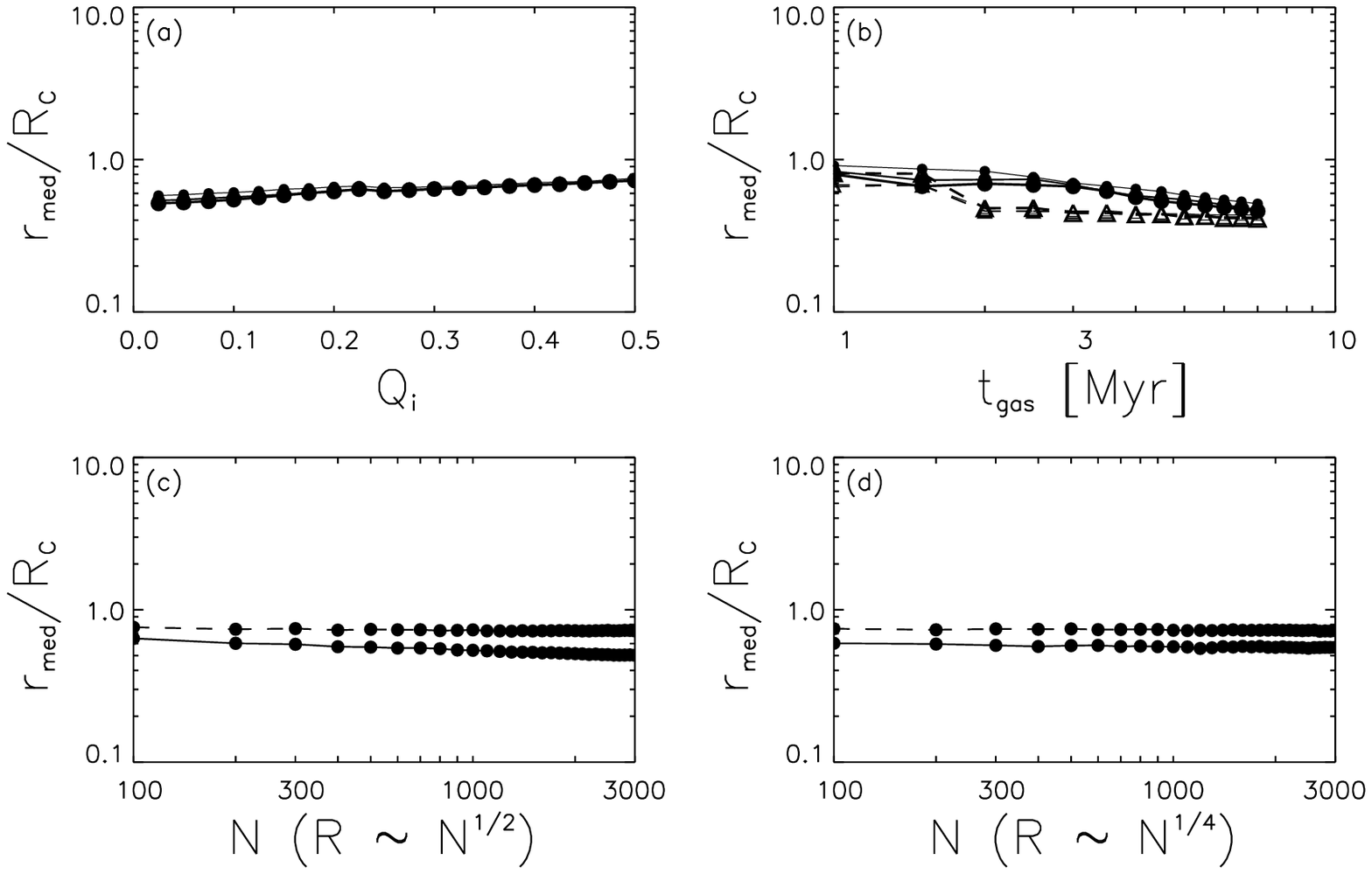} 
\caption{The median radius calculated from the radial profiles
  $N(r)/N$ normalized by the initial cluster radius $\rstar$ as a
  function of initial cluster parameter, for all clusters included in
  the parameter space survey. The cluster parameter varied in each
  panel is as follows: (a) Initial virial ratio $Q_i$, (b) Gas removal
  time $\tgas$, (c) Stellar Membership $N$ for scaling relationship
  $\rstar \sim N^{1/2}$, and (d) Stellar Membership $N$ for scaling
  relationship $\rstar \sim N^{1/4}$. In panels (a) and (b) clusters
  with $N = 300, 1000$, and $2000$ are indicated by the small, medium,
  and large circles (\emph{thin, medium, and thick curves}); in panel
  (b), the dashed curves (marked by open triangles) show the result
  for the normalized median radius averaged over the embedded phase
  only. In panels (c) and (d), subvirial clusters are indicated by the
  solid curves while virial clusters are indicated by the dashed
  curves.}
\label{fig:MedianRadiusNorm} 
\end{figure}

\begin{figure}
\epsscale{1.0}
\plotone{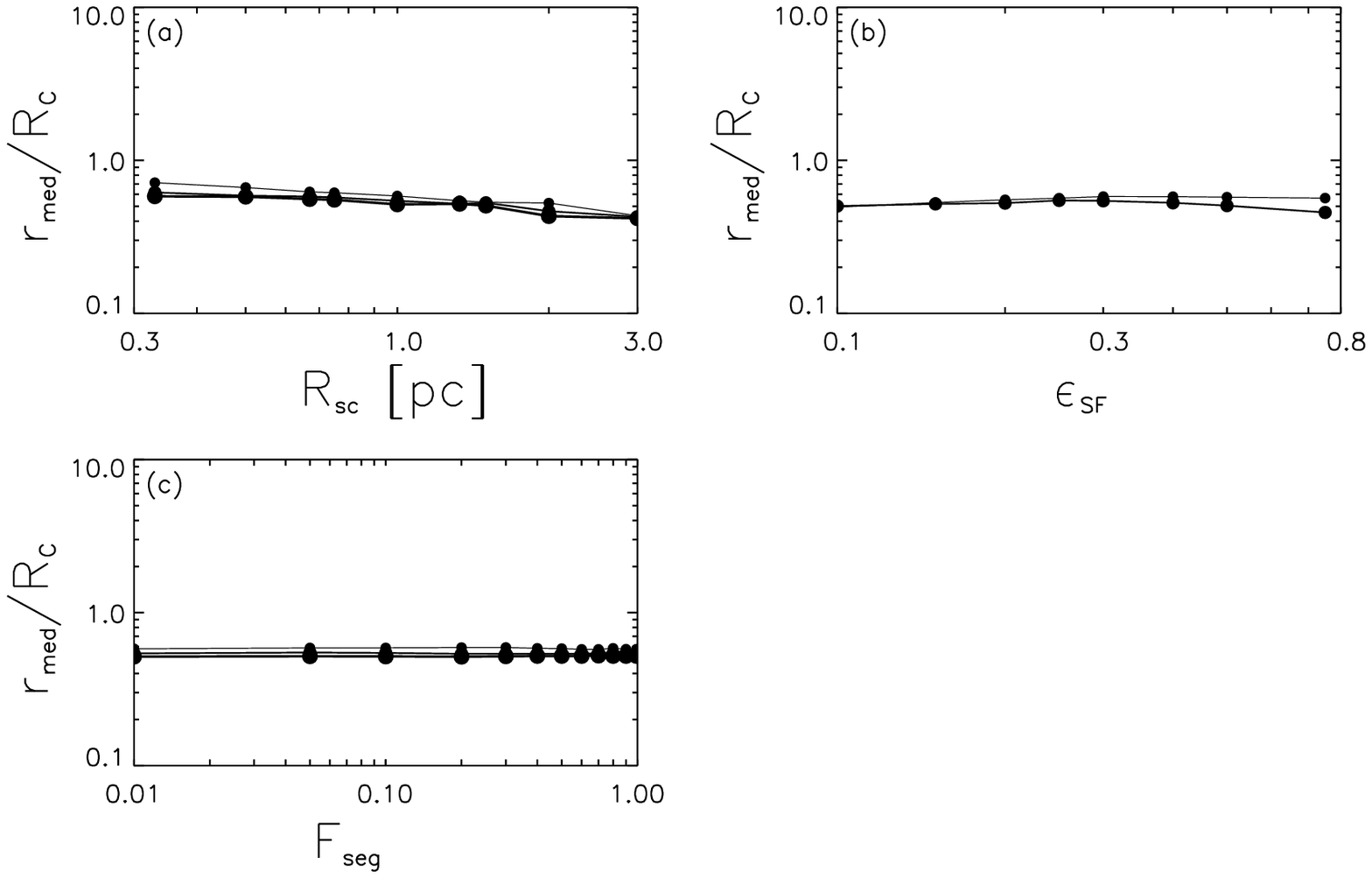} 
\caption{Continuation of Figure \ref{fig:MedianRadiusNorm}. The
  cluster parameter varied in each panel is as follows: (a) Cluster
  scaling radius $\rsc$, (b) Star formation efficiency $\sfe$, and (c)
  Degree of primordial mass segregation $\mseg$.  Clusters with $N$ =
  300, 1000, and 2000 are indicated by the small, medium, and large
  circles (\emph{thin, medium, and thick curves}), respectively,
  though there are no major differences between the clusters as a
  function of size.}
\label{fig:MedianRadiusNorm_02} 
\end{figure}

Although one can use the full distributions (see equations
[\ref{eq:MofR}, \ref{eq:NofR}]), it is sometimes useful to
characterize the distributions in terms of a single parameter.  
Toward this end, Figures \ref{fig:MedianRadius_01} and
\ref{fig:MedianRadius_02} present the median cluster radius $r_{med}$
calculated from the fits to the cumulative radial distributions for
the entire parameter space. The scale $r_{med}$ is defined as the
radius at which $N(r)/N = 0.5$ and thus represents the radius which,
on average, contains half of the cluster members.  As these plots
indicate, the median cluster radius $r_{med}$ scales with the initial
virial parameter $Q_i$, the gas removal timescale $\tgas$, the cluster
membership $N$, and the cluster scaling radius $\rsc$.  On the other
hand, the radius $r_{med}$ does not vary strongly with the either star
formation efficiency $\sfe$ or the degree of primordial mass
segregation $\mseg$.

Scaling the median radius $r_{med}$ by the initial cluster radius
$\rstar$ removes the dependency on this initial cluster parameter and
more readily identifies trends that are distinct from the initial
assumptions concerning cluster size.  Figure
\ref{fig:MedianRadiusNorm} displays the median radius $r_{med}$
normalized by the initial cluster radius $\rstar$.  Panel (a) of this
figure clearly shows that the median radius depends almost linearly on
the initial virial parameter $Q_i$ for $Q_i \leq 0.5$.  This result is
consistent with what we expect from the initial collapse associated
with the evolution of a cluster with subvirial velocities: The
quasi-equilibrium radius (that obtained after the initial dynamical
adjustment) scales linearly with the initial virial parameter.
Clusters with completely subvirial starting states have median radii
that are approximately $\sqrt{2}$ of the median radii of virial
clusters.

The median radius also decreases as a function of the gas removal time
$\tgas$. The data points connected by a solid line in Figure
\ref{fig:MedianRadiusNorm} panel (b) correspond to the time averaged
(over $0-10$ Myr) normalized median radius $r_{med}$ of clusters with
differing values of $\tgas$.  During the embedded phase, these
(initially subvirial) clusters remain bound and do not expand. As a
result, clusters that become unbound early in their history have
larger median radii simply due to time averaging (over the $10$ Myr
time interval of the simulations).  The data points connected by the
dashed curve correspond to the normalized median radius averaged over
the embedded stage of the cluster evolution ($0 - \tgas$ Myr).
Removing the apparent dependence on $\tgas$ that is actually due to
the time averaging, we find that in clusters with dispersal times
greater than $\sim2$ Myr, the cluster median radius does not depend 
sensitively on the gas removal time $\tgas$. 
 
Note that clusters with early gas dispersal times ($\tgas \lesssim 2$
Myr) have significantly larger median radii than clusters with later
gas dispersal times.  The average crossing time in a subvirial cluster
is $\sim1$ Myr, and thus gas removal within the first couple crossing
times prevents the cluster from approaching a state of virial
equilibrium.  This behavior also explains why the bound fractions in
clusters with $\tgas \lesssim 2$ Myr are very low (see Figure
\ref{fig:Bound_02} panel (a), and the discussion in Section
\ref{sec:boundfrac}). In other words, the process of gas dispersal in
a cluster that is not in virial equilibrium is more destructive to the
cluster than if gas removal occurs after the system approaches an
equilibrium state \citep{Goodwin2006MNRAS}.

Panel (a) in Figure \ref{fig:MedianRadiusNorm_02} displays the
normalized median cluster radius as a function of the initial scaling
radius $\rsc$ used in equation (\ref{eq:RofNparam}).  The
normalization of the median cluster radius includes the intrinsic
dependence on $\rsc$, and hence the trend observed in the normalized
median radius $r_{med}/\rstar$ must be accounted for by another
mechanism.  The larger normalized median radius observed in clusters
with smaller initial values of $\rsc$ can be understood in terms of
the higher interaction rates observed in these clusters; the
interactions keep the cluster cores slightly inflated. This trend
should be present, to some extent, in all clusters with high
interaction rates; however, it is easiest to observe in the $\rsc$
series of simulations because the interaction rates have the widest
dynamical range, varying by three orders of magnitude (see Figure
\ref{fig:Gamma0_01}, panel (c)).

Tables 14 -- 19 list the values of the parameters $(p, r_0, a)$ that
specify the mass and number profiles for the clusters considered
herein, where the profiles have the form given by equations
(\ref{eq:MofR}) and (\ref{eq:NofR}). Each table lists the fitted
parameters as a function of a given input variable, including the
stellar membership $N$ (Table 14), initial virial parameter $Q_i$
(Table 15), cluster scaling radius $\rsc$ (Table 16), star formation
efficiency $\sfe$ (Table 17), gas removal time $\tgas$ (Table 18), and
the mass segregation parameter $\mseg$ (Table 19). These tabulated
values, in conjunction with equations (\ref{eq:MofR}) and
(\ref{eq:NofR}), provide analytic descriptions of the mass and number
density distributions for a wide variety of clusters.  These analytic
forms, in turn, can be used to calculate related physical quantities.  
For example, the cluster median radius $r_{med}$ is given by the formula 
\bee
r_{med} = r_0 \left( 2^{1/p} - 1 \right)^{-1/a} \, , 
\label{eq:rmedian} 
\eee
where $(p, r_0, a)$ are the parameters that specify the number density
profile. As another example, the mass density of the cluster is given by
\bee
\rho(r) = \frac{1}{4 \pi r^2} \frac{d M}{dr} = 
\frac{M_{T\ast}}{4 \pi r^3} \frac{ap \, \xi^{ap}}{(1 + \xi^a)^{p+1}} 
\, = \frac{M (r)}{4 \pi r^3} \frac{ap}{1 + \xi^a} \, , 
\label{eq:density} 
\eee
where the parameters $(p, r_0, a)$ are those of the mass profile.
Similarly, the magnitude of the gravitational force is given by 
$|F| = G M(r)/r^2$, and the corresponding potential of the stellar
component can be obtained from the integral $\Psi_\ast \sim \int F dr$. 
Note that the force can be expressed in terms of elementary functions,
but the potential integral leads to hypergeometric functions for
general values of the indices. These analytic forms can be used in a
wide variety of applications to help determine the effects of cluster
environments on forming stars and young solar systems.

\section{Effects of Cluster Environment on Planetary Formation}
\label{sec:Effects}

In this section, we consider two mechanisms through which young
embedded clusters affect their constituent members.  Note that for
systems in the regime of parameter space considered here, the
background cluster environment has more effect on circumstellar disks
and planet formation than on the star formation process itself.  In
Section \ref{sec:Encounters}, we use the interaction rates calculated
from our ensemble of $N$-body simulations (Section \ref{sec:Gammas})
to determine the effectiveness of close encounters for disrupting
circumstellar disks and newly formed planetary systems. In Section
\ref{sec:Radiation}, we use the output number density profiles (from
Section \ref{sec:MassDists}) to determine how much UV radiation the
cluster provides to the circumstellar disks surrounding its members.

\subsection{Close Encounters}\label{sec:Encounters}

One way that planet formation may be compromised in stellar clusters
is through close encounters between planet-forming circumstellar disks
and other cluster members.  A close encounter will truncate a
circumstellar disk down to a radius that is roughly $1/3$ the distance
of closest approach \citep{Kobayashi2001Icar}.  We thus need a metric
to determine the importance of close encounters during the 10 Myr
embedded phase of cluster evolution (note that this timescale is
comparable to the observed lifetimes of circumstellar disks and the
expected time required for giant planet formation).  Using data from
our cluster simulations, we define the characteristic impact parameter
$b_C$ as the closest encounter that an average star will experience
over the course of a given time interval, taken here to be $10~\Myr$.
The value of $b_C$ is readily calculated from the distribution of
interaction rates. For the form given in equation (\ref{eq:IntRate}),
and we define the characteristic radius $b_C$ to be
\bee 
b_C \equiv 1000 \rm{AU} \left(t_0 \Gamma_0 \right)^{1/\gamma} \, , 
\eee
where $t_0$ = 10 Myr. 

\begin{figure}
\epsscale{1.0}
\plotone{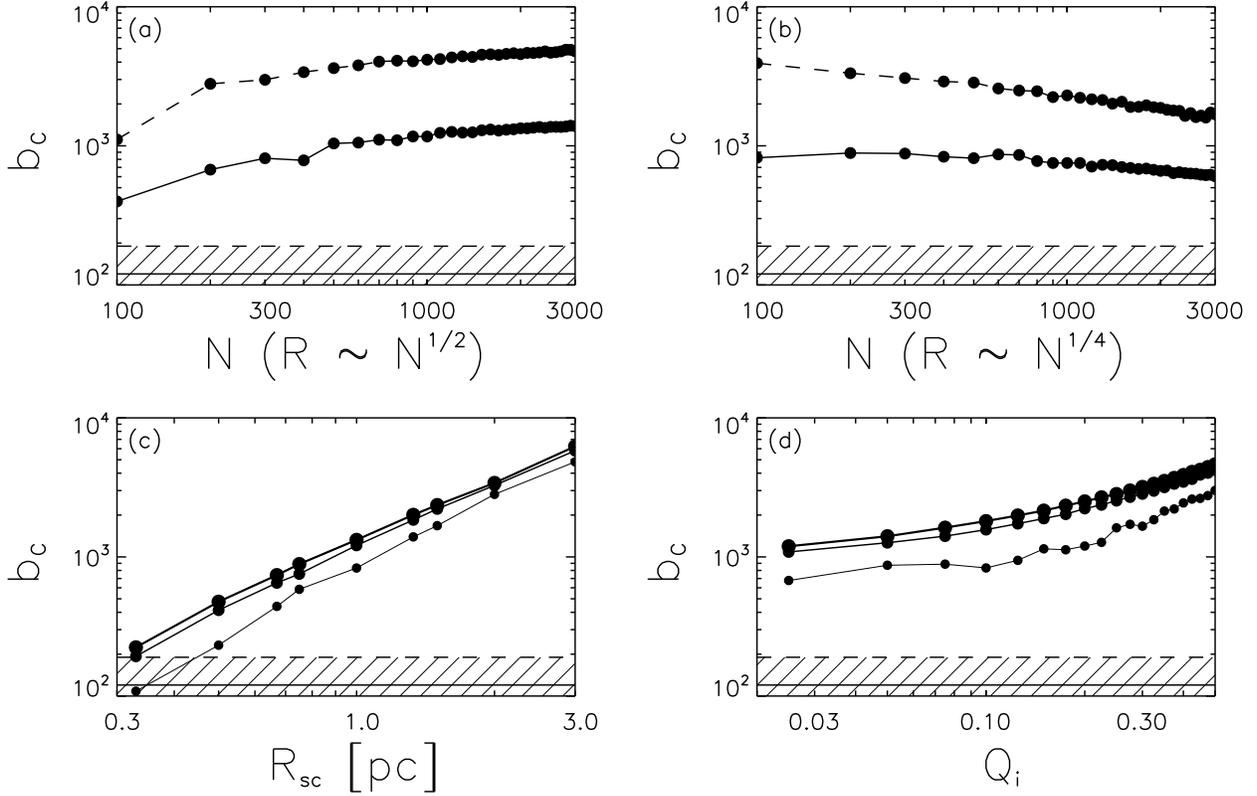} 
\caption{The characteristic impact parameter, $b_C$ (in AU) calculated
  from the close encounter profiles $\Gamma$ plotted as a function of
  cluster parameter, for all clusters included in the parameter space
  survey. The cluster parameter varied in each panel is as follows:
  (a) Stellar Membership $N$ for scaling relationship $\rstar \sim
  N^{1/2}$, (b) Stellar Membership $N$ for scaling relationship
  $\rstar \sim N^{1/4}$, (c) Cluster scaling radius $\rsc$, and (d)
  Initial virial ratio $Q_i$. In panels (a) and (b), subvirial
  clusters are indicated by the solid curve while virial clusters are
  indicated by the dashed curve. In panels (c) and (d) clusters with
  $N = 300, 1000$, and $2000$ are indicated by the small, medium, and
  large circles (\emph{thin, medium, and thick curves}), respectively.
  The dashed (solid) horizontal lines correspond to benchmark
  distances of 190 AU (120 AU). }
\label{fig:bc_01} 
\end{figure}

\begin{figure}
\epsscale{1.0}
\plotone{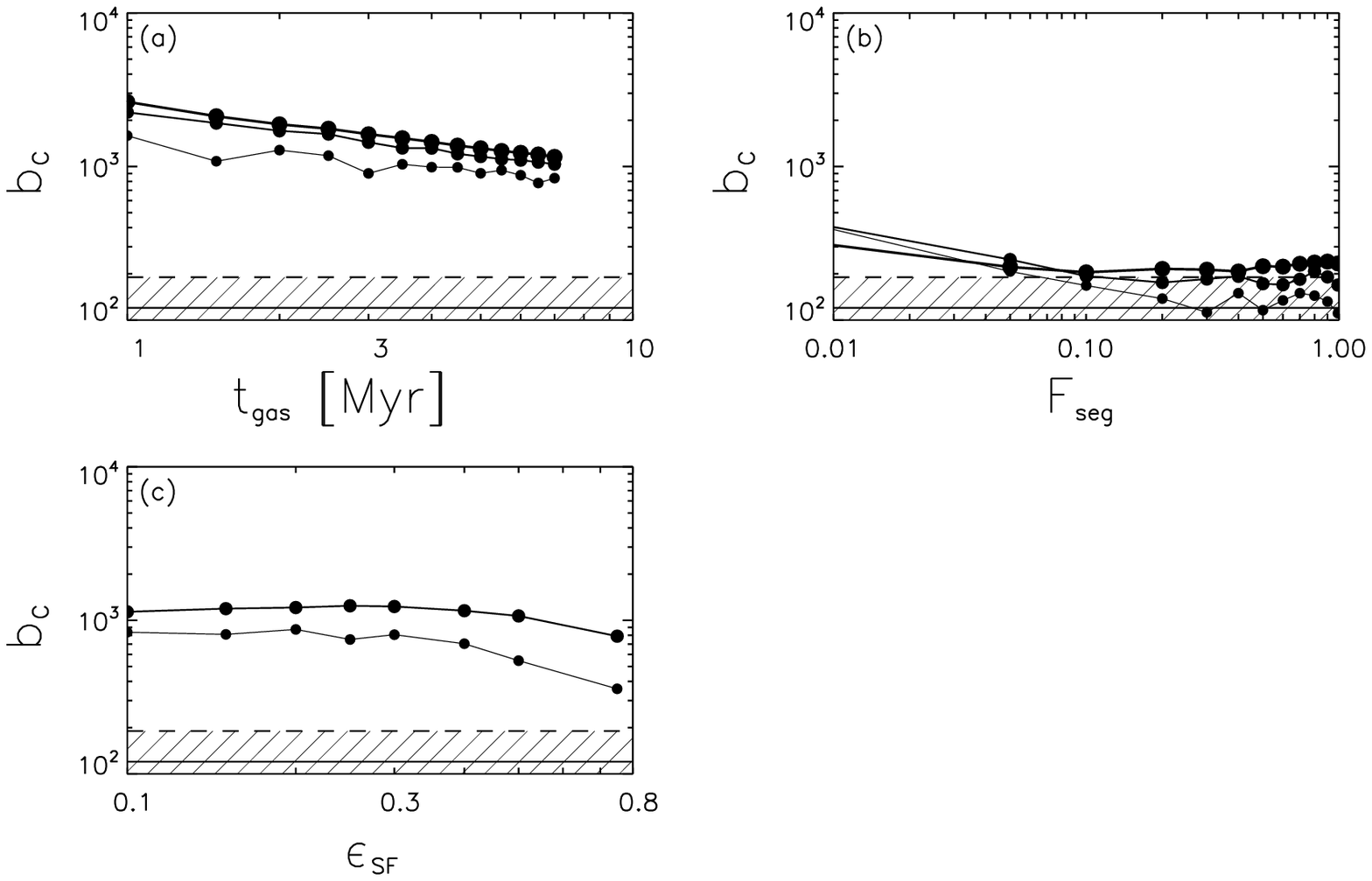} 
\caption{Continuation of Figure \ref{fig:bc_01}. The cluster parameter
  varied in each panel is as follows: (a) Gas removal time $\tgas$,
  (b) Degree of primordial mass segregation $\mseg$, and (c) Star
  formation efficiency $\sfe$. Clusters with $N = 300, 1000$, and
  $2000$ are indicated by the small, medium, and large circles
  (\emph{thin, medium, and thick curves}), respectively.  The dashed
  (solid) horizontal lines correspond to benchmark distances of 190 AU
  (120 AU). }
\label{fig:bc_02} 
\end{figure}

Figures \ref{fig:bc_01} and \ref{fig:bc_02} display the characteristic
impact parameter $b_C$ as a function of initial cluster parameter for
the clusters considered in this parameter space survey.  The
horizontal dashed and solid lines (at $190$~AU and $120$~AU) indicate
encounter distances that restrict planet formation in a circumstellar
disk to distances less than $55$~AU and $30$~AU, respectively
\citep{Kobayashi2001Icar}.  These latter distances correspond to the
outer edge of the Kuiper Belt and the orbit of the planet Neptune in
our solar system.

These figures indicate that most of the clusters in this parameter
space survey do not have interaction rates that are high enough to
seriously compromise planet formation within the $55$~AU Kuiper Belt
radius, or even within $100$~AU. Equivalently, disruptive encounters
at these close distances are predicted to be rare.  In the most
interactive clusters, those with the highest number densities due to
small values of $\rsc$ (Figure \ref{fig:bc_01}, panel (c)), and those
with significant amounts of mass segregation (Figure \ref{fig:bc_02},
panel (b)), the characteristic scale $b_C$ is small enough that
planet-forming disks may be truncated at radii $r_d \lesssim 100$~AU
and hence giant planet formation in this class of clusters can be
partially inhibited. Nonetheless, the cluster environment provides 
only moderate constraints on the planet forming process.

Interactions remain important even after planet formation in the
circumstellar disk has taken place. Close encounters with other
cluster members are capable of disrupting newly formed planetary
systems. In this context, the primary channel of ``disruption'' is to
increase the orbital eccentricity and/or the inclination angles of the
planetary orbits.  Sufficiently close encounters can eject planets
from their orbits entirely, and produce free floating planets in the
clusters.  The stellar interaction rates presented in this paper may
be combined with scattering calculations to investigate the rate of
solar system disruption in young clusters.

One way to characterize the possible effects of clusters on newly
formed planetary systems is to pose the following question: Under the
assumption that planet formation naturally produces systems similar to
our solar system (with planets in relatively circular orbits with
small inclination angles), how many planetary systems in a young
cluster will be noticeably different due to an encounter with another
cluster member?  Previous work provides cross sections for disrupting
solar systems in a variety of ways (e.g., Adams \& Laughlin 2001,
APFM); as expected, scattering interactions are most effective at
altering the orbits of the outermost planets.  For the sake of
definiteness, we consider a collection of solar systems that have the
same architecture as our own (same masses and semimajor axes for the
outer planets) except that the orbits are circular.  We then designate
a solar system to be ``noticeably different'' (due to an encounter)
when the analog of Neptune has its orbital eccentricity increased
(from zero) to values $\epsilon \ge 0.05$ (about twice the value
observed in our solar system). We note that nothing especially
dramatic happens when orbital eccentricities are increased by this
amount, only that the solar system is changed enough to be noticed.
Previous work shows that the cross section for increasing Neptune's
eccentricity to $\epsilon \ge 0.05$ during an encounter with a cluster
member is $\csec \approx$ 167,000 AU$^2$ \citep{Adams2001Icar}, which
corresponds to a closest approach distance of approximately 230 AU.
This cross section for increasing Neptune's eccentricity is about the
same as the cross section for increasing the spread in inclination
angles beyond that observed in our solar system ($\Delta i \ge
3.5^{\circ}$).  For completeness, we note that in these solar system
scattering calculations, the perturber is assumed to be a binary star
system, which is reasonable since a large fraction of stars are
members of multiple systems.

As shown in Figures \ref{fig:bc_01} and \ref{fig:bc_02}, the typical
interaction distance $b_C$ is somewhat larger than the $230$~AU
encounter distance required to ``disrupt'' the solar system. However,
some fraction of stars in a cluster will experience encounters that
are significantly closer.  For instance, consider a cluster with $N$ =
300 stars, subvirial initial conditions characterized by $Q_i$ = 0.04,
and an initial cluster radius $\rstar = 0.67$ pc.  In this cluster,
$b_c \approx 400$ (see Figure \ref{fig:bc_01}, panel (c)). The rate of
interactions with an encounter distance less than or equal to $b$ can
be determined from equation (\ref{eq:IntRate}) using the values
$\Gamma_0$ = 0.333 and $\gamma = 1.47$.  The rate of close encounters
with $b \leq 230$~AU is thus about $0.038$ interactions per star per 
Myr. As a result, in this cluster approximately 115 stars will
experience close encounters severe enough to change their planetary
systems.  In contrast, in a cluster similar in all respects except
with a slightly larger initial radius (i.e., $N=300$, $Q_i = 0.04$, and $\rstar
1$ pc), this number drops to $\sim50$ planetary systems ($\Gamma_0 =
0.1290, \gamma=1.39$).

This example illustrates the type of calculations that are possible
using the interaction rates and fitting parameters presented in this
numerical study.  The close encounter distributions may also be
combined with results presented by other scattering calculations to
determine the efficacy of circumstellar disk and planetary system
disruption in young stellar clusters \citep{Heller1993ApJ,
  Heller1995ApJ, deLaFuenteMarcos1997AA, deLaFuenteMarcos1999NewA,
  Kobayashi2001Icar, Adams2001Icar, Adams2006ApJ, pfalzner08}.

Before leaving this section, we can use the interaction rates
determined here to provide a consistency check.  Although this
ensemble of simulations does not include binary systems (only single
stars), the presence of binaries could affect the energy budget of the
system. As outlined above, relatively few stellar interactions take
place with distances of closest approach less than 100 AU, and the
characteristic radii $b_C$ are typically much larger ($\sim1000$ AU).
For comparison, for solar type stars, the peak of the binary period
distribution occurs at a period $P \approx 10^5$ days (Duquennoy \&
Mayor 1991), which corresponds to a distance of $\sim40$ AU. As a
result, the closest encounter experienced by the the vast majority of
stars (during the 10 Myr window of interest) is much wider than the
typical binary separation. These results thus vindicate the approach
of ignoring binarity for this class of simulations.  However, we note
that for more extreme regimes of parameters space (e.g., dense stellar
systems destined to become globular clusters), neglecting binaries is
not a good approximation.

\subsection{Radiation Fields}\label{sec:Radiation}

Another mechanism through which cluster environments may affect planet
formation is by photoevaporation of protoplanetary disks due to the
enhanced FUV radiation fields produced by massive young stars. It is
well known that radiation from the central host star can heat and
dissipate its surrounding planet-forming disk \citep{Shu1993Icar,
  Hollenbach1994ApJ}. Most stars are not massive enough to produce
large quantities of FUV radiation, the wavelength range that most
effectively photoevaporates the disk. In young clusters, however, the
most massive cluster stars can provide FUV radiation fields that are
strong enough to photoevaporate the disks associated with other
cluster members \citep{Johnstone1998ApJ, Adams2001ApJ} and dominate
over the radiation produced by the host star.

Recent studies have determined the FUV (and EUV) luminosities of
clusters as a function of cluster membership $N$, the mass function of
the cluster, and various amounts of extinction within the cluster
(e.g., Armitage 2000, APFM, Fatuzzo \& Adams 2008).  In this section,
we combine these previous determinations of the typical FUV background
luminosities in stellar clusters with the radial profiles presented in
Section \ref{sec:MassDists}. Taken together, these stellar positions
and the FUV luminosities determine the expected FUV flux that impinges
upon the circumstellar disks in young stellar clusters. In these
systems, this FUV flux places limits on both the timescale over which
planets may form and the region of the disk that has the potential to
form planets.

The total FUV luminosity of a cluster originates primarily from the
most massive stars in the cluster. As a result, the cluster FUV
luminosity is sensitive to the membership size $N$ of the cluster and
the mass function of the stars in the cluster.  APFM presented
detailed calculations of the expected FUV luminosity in a cluster of
size $N$ (see their Figure 6). For each cluster membership size $N$,
the distribution of possible FUV luminosities is quite wide due to
under-sampling of the stellar initial mass function; for clusters with
$N \lesssim 700$, the width of the FUV luminosity distribution is
larger than the mean (or median) value. As a result, clusters are
predicted to display substantial system to system variation in the
radiation fields they produce (especially for $N \lesssim$ 700).

Each individual cluster member will experience a FUV flux that is
time-dependent as it orbits through the cluster potential. For
purposes of this paper, we estimate the typical flux experienced by a
circumstellar disk in a simulated cluster by combining the FUV
luminosity $\lfuv(N)$ with the median radial position of the stars in
a simulated cluster.  The vast majority of the FUV radiation in a
cluster is produced by the most massive stars, which typically reside
near the cluster's center (see Section \ref{sec:MassSeg}).  The FUV
source is thus modeled as a point source at the center of the cluster.
For a given cluster with FUV luminosity $\lfuv(N)$ and median radius
$r_{med}$, the corresponding median FUV flux $\ffuv$ experienced by
the protoplanetary disks in the cluster is defined to be
\bee 
\ffuv \equiv \frac{\lfuv(N)}{4\pi r_{med}^2} \, , 
\label{eq:ffuv} 
\eee 
where $r_{med}$ is defined by the radial probability distribution of
the cluster (equation [\ref{eq:NofR}]) and $\lfuv(N)$ is taken from
the results presented in APFM.

In addition to the FUV flux provided by the background cluster, a
circumstellar disk is also subjected to FUV radiation from its host
star.  Observations of T Tauri stars provide estimates of the FUV flux
experienced by their disks at distances of $\sim100$~AU due to the
central star \citep{Bergin2004ApJ}.  For three T Tauri stars, these
authors determine FUV fluxes of $G_0$ = 240, 340, and 1500 where
$G_0=1$ corresponds to a benchmark value of $1.6 \times 10^{-3}$
~erg~s$^{-1}$~cm$^{-2}$ (close to the value of the interstellar
radiation field at FUV wavelengths). As a rough estimate, the FUV flux
associated with the host star can be taken to be $G_{\rm host} \approx
500$. When the background FUV radiation field of the cluster exceeds
this benchmark value, the environment can, in principle, affect the
evolution of circumstellar disks and planet formation. However, in the
outer regions of a cluster, where flux levels from the central massive
stars are relatively low, the FUV flux from the host star can be more
important than the FUV flux from the background cluster.

Figures \ref{fig:Radiation_01} and \ref{fig:Radiation_02} present the
median cluster FUV flux $\ffuv$ (defined by equation [\ref{eq:ffuv}])
as a function of the initial cluster parameters for the clusters
considered in this parameter space survey.  The dashed horizontal
lines correspond to FUV radiation levels of $G_0$ = 300 and 3000,
values for which the effects of photoevaporation on circumstellar
disks have been calculated in detail \citep{Adams2004ApJ}.  The
results of these photoevaporation models provide a rough scaling law:
Over the course of $10$~Myr, an FUV flux of $G_0 = 3000$ is capable of
truncating a circumstellar disk down to a radius $r_d$ given by
\bee
r_d \approx 36~{\rm AU}\left(\frac{M_\ast}{\rm M_{\odot}}\right) \, ,
\eee 
where the result depends on the stellar mass $M_\ast$. This nominal
radius is close to the size of our solar system. As a rough rule of
thumb, significant evaporation around solar type stars thus requires
flux levels $G_0 \gtrsim 3000$, with a weaker requirement for smaller
stars.

\begin{figure}
\epsscale{1.0}
\plotone{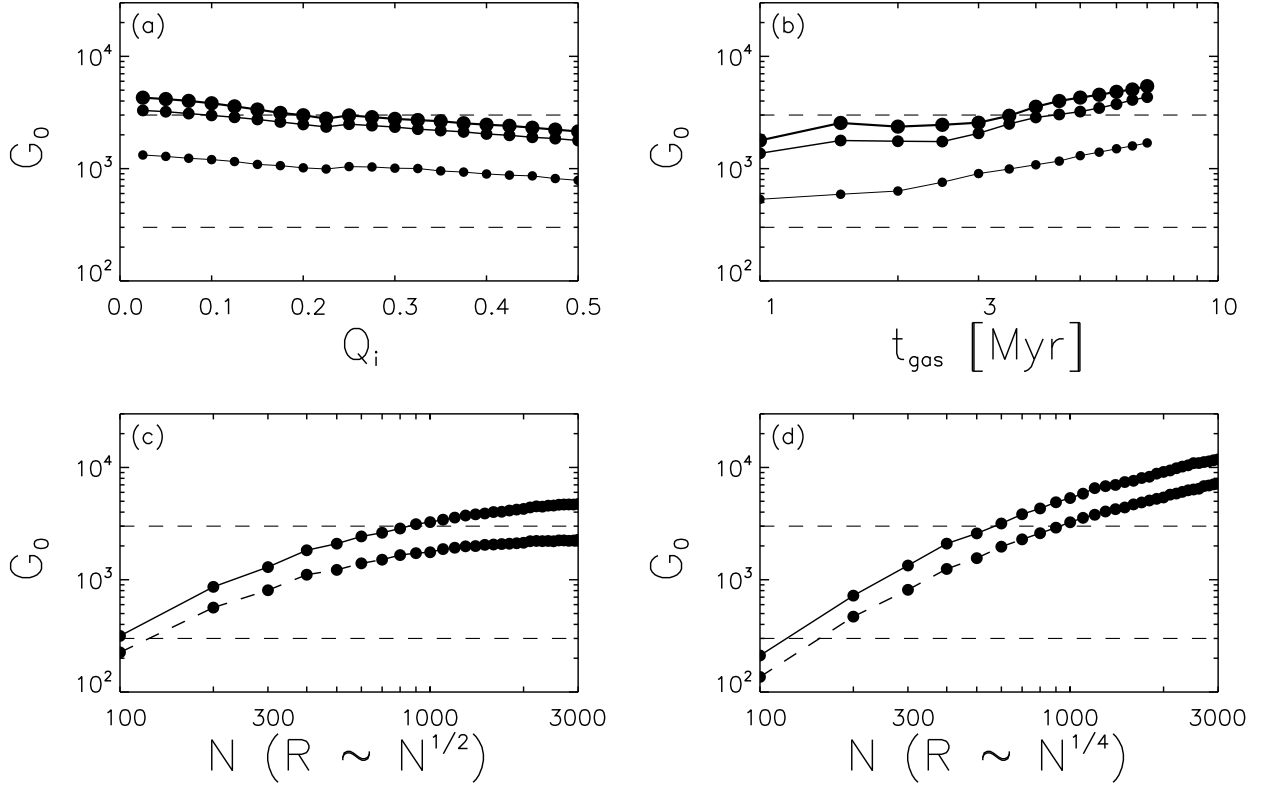} 
\caption{The FUV flux experienced due to the cluster environment for
  all clusters included in the parameter space survey. The cluster
  parameter varied in each panel is as follows: (a) Initial virial
  ratio $Q_i$, (b) Gas removal time $\tgas$, (c) Stellar Membership
  $N$ for scaling relationship $\rstar \sim N^{1/2}$, and (d) Stellar
  Membership $N$ for scaling relationship $\rstar \sim N^{1/4}$. In
  panels (a) and (b) clusters with $N = 300, 1000$, and $2000$ are
  indicated by the small, medium, and large circles 
  (\emph{thin, medium, and thick curves}), respectively. In panels (c)
  and (d), subvirial clusters are indicated by the solid curve while
  virial clusters are indicated by the dashed curve.  The horizontal
  lines at $G_0 = 300$ and $3000$ are benchmark values for which the
  effects of photoevaporation of circumstellar disks have been
  calculated \citep{Adams2004ApJ}. } 
\label{fig:Radiation_01} 
\end{figure}

\begin{figure}
\epsscale{1.0}
\plotone{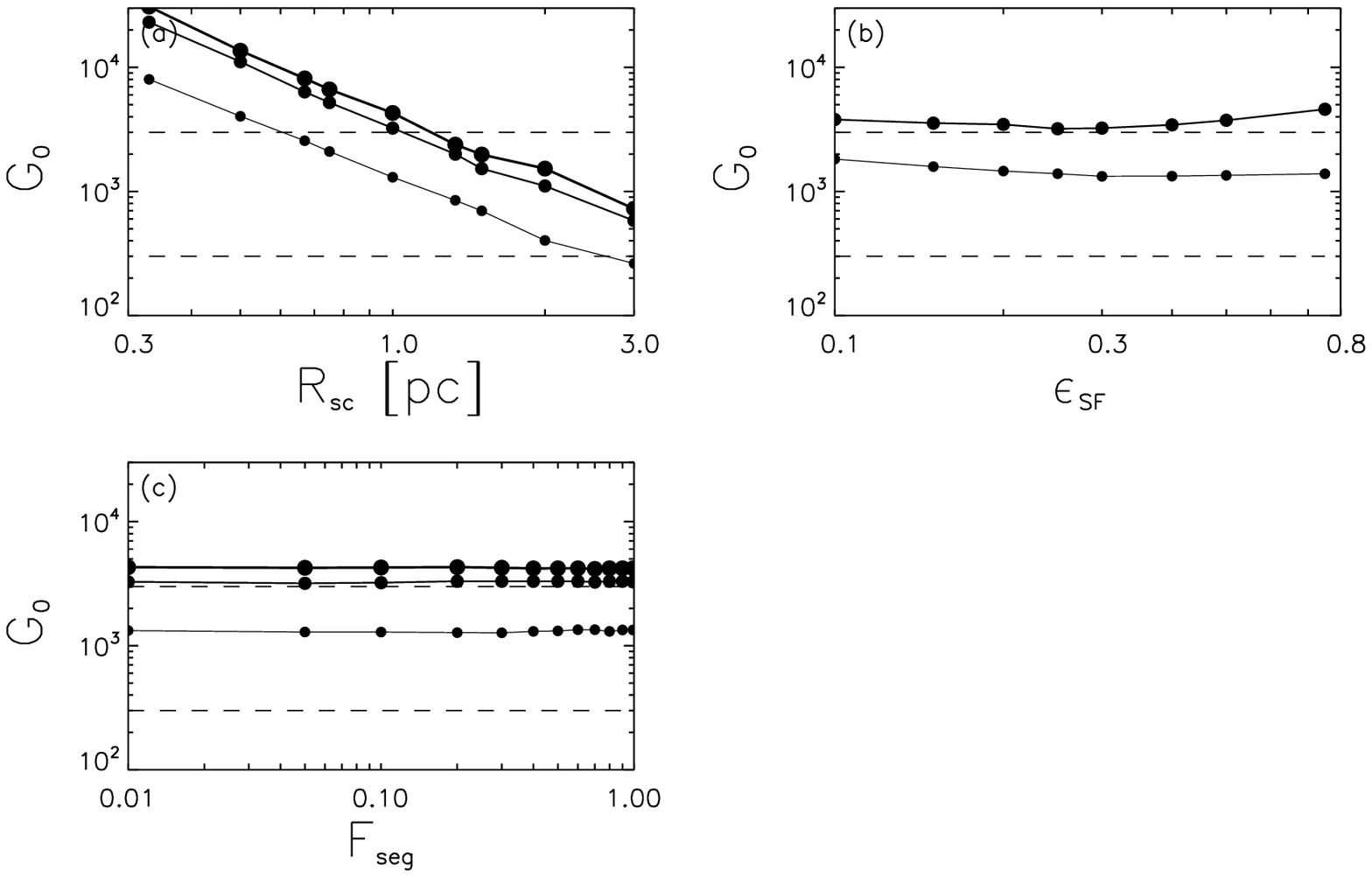} 
\caption{Continuation of Figure \ref{fig:Radiation_01}. The cluster
  parameter varied in each panel is as follows: (a) Cluster scaling
  radius $\rsc$, (b) Degree of primordial mass segregation $\mseg$,
  and (c) Star formation efficiency $\sfe$. Clusters with $N$ = 300,
  1000, and 2000 are indicated by the small, medium, and large circles
  (\emph{thin, medium, and thick curves}), respectively.}
\label{fig:Radiation_02} 
\end{figure}

The FUV luminosity of a cluster is an increasing function of the
cluster membership $N$, and this trend is visible in the FUV fluxes
shown in Figures \ref{fig:Radiation_01} and \ref{fig:Radiation_02}. In
subvirial clusters with more than $N \sim 1000$ members, circumstellar
disks are exposed to median FUV fluxes that are of order $G_0 \approx
3000$ (see Figure \ref{fig:Radiation_01}, panels (c) and (d)).  In
this class of clusters, significant photoevaporation of the outer disk
is predicted to take place. Many of the circumstellar disks in such
cluster systems will be truncated, so that planet formation is limited
to the inner $\sim35$~AU for solar type stars.  For the more numerous
red dwarf stars, planet formation is limited to the inner $\sim12$ AU.
In clusters with smaller memberships $N$, the radiation fields are
more modest and disks will not be significantly photoevaporated by
cluster radiation.  We note that in this latter class of clusters, the
FUV radiation from the host star is comparable to that of the
background cluster. Finally, we note that in clusters where gas
expulsion takes place on short time scales ($\tgas \leq$ 3~Myr), the
FUV fluxes are lower due to early expansion of the cluster and larger
values of $r_{med}$ (see panel (b) of Figure \ref{fig:Radiation_01}).

\section{Conclusion}\label{sec:Conclusion}

In this paper, we have presented the results from a large ensemble of
numerical simulations designed to study the dynamics young embedded
clusters over a wide range of conditions.  The choice of parameter
space was motivated by recent catalogs and surveys of star-forming
environments in the solar neighborhood.  We consider clusters with a
range of stellar memberships $N$, a variety of parameter values $\rsc$
and $\alpha$ that define the cluster membership-radius relation
(equation [\ref{eq:RofNparam}]), a range of gas removal timescales
$\tgas$, initial virial states $Q_i$, star formation efficiencies
$\sfe$, and amounts of primordial mass segregation $\mseg$. The range
of parameter space surveyed is summarized in Table \ref{tab:Initial}.

The results of this survey show how the evolutionary parameters that
describe the properties of evolving young clusters vary as a function
of the initial cluster conditions. Section \ref{sec:boundfrac}
considers how the cluster's bound fraction $\fbnd$ (generally 
evaluated at time $t = 10~\Myr$) varies with the input parameters.
The bound fraction depends most sensitively on the star formation
efficiency $\sfe$, but also depends quite strongly on the initial
virial state of the system. As a general rule, clusters in which the
stars are formed with subvirial velocities have higher bound fractions
than clusters in which stars are formed with initially virialized
velocities.

In Sections \ref{sec:Gammas} and \ref{sec:Velocity}, we considered the
distributions of close encounters between cluster members over the 
first 10 Myr of cluster evolution.  We find that the encounter rates
scale linearly with the average stellar density, so that $\Gamma \sim
\langle n \rangle$ as expected, and that the interaction rates
increase as the initial virial parameter $Q_i$ decreases.  Subvirial
clusters have interaction rates that are roughly $8$ times higher than
those found in clusters that start in a virialized state.  We also
show that the interaction rates are higher in clusters that have
larger amounts of primordial mass segregation. In addition, the
distributions of interaction velocities are nearly Gaussian (see
Figure \ref{fig:VelDist}). The interaction velocities themselves do
not vary strongly as a function the initial conditions in the cluster;
instead, the interaction velocities are always about twice the average
stellar velocity in the cluster.

In Section \ref{sec:MassDists}, we present the results of empirical
fits to the cluster radial profiles $M(r)$ and $N(r)$.  We find that,
in general, the median cluster radius $r_{med}$ scales as the initial
cluster radius $\rstar$, so that the ratio $r_{med}/\rstar$ is nearly
invariant.  In some clusters, however, significant interactions
between stellar members produce a break in the $r_{med} \sim \rstar$
relationship. More specifically, clusters that are initially subvirial
develop smaller median radii than those that are initially in virial
equilibrium.  In addition, if the embedding gas is removed from a
subvirial cluster early in the cluster's evolution ($\tgas \lesssim
2~\Myr$), the resulting median cluster radius is much larger than it
would be if gas removal occurred at a later time. This difference
occurs because the cluster is not in (or near) virial equilibrium at
the time of gas removal; as a result, gas dispersal is more
destructive in a cluster that has not yet approached virial
equilibrium.

We note that these general trends in the output parameters found in
this study (e.g., the interaction rates $\Gamma$ are proportional to
the mean density $\langle n \rangle$) are not unexpected. However,
this work puts these results on a firm, statistically significant
footing.  We find quantitative results, with accuracy and precision
well beyond that obtained from qualitative scaling arguments. In
addition, we obtain these results for varying values of the input
parameters that characterize the cluster. 

In summary, the principle contribution of this work is to provide a
collection of output parameters that describe cluster properties and
cluster evolution as a function of initial conditions (see Tables 2 --
19). The interaction rates for close encounters between cluster
members are given by the power-law form of equation (\ref{eq:IntRate}), 
where the parameters $(\Gamma_0, \gamma)$ are listed in Tables 8 -- 13. 
Taken together, these results provide us with an analytic description
of the interaction rates for a wide range of cluster input parameters.
Similarly, the radial profiles of mass and stellar number take the
forms given by equations (\ref{eq:MofR}) and (\ref{eq:NofR}), where
the parameters $(p, r_0, a)$ are listed in Tables 14 -- 19. We thus
also have an analytic description of cluster mass $M(r)$ and number 
$N(r)$ distributions. From these functions, additional physical
quantities can be derived, including mass density, number density,
stellar potential, median radius, and many others (see equations
[\ref{eq:rmedian}] and [\ref{eq:density}]).  Furthermore, all of these
analytic quantities are specified as a function of stellar membership
$N$, initial virial parameter $Q_i$, scale radius $\rsc$, star
formation efficiency $\sfe$, gas removal time $\tgas$, and degree of
mass segregation $\mseg$.  These results can be combined with
calculations of the radiation fields in young clusters and cross
sections for planetary system disruption to determine the effects of a
wide variety of cluster environments on planetary formation. In
Section \ref{sec:Effects}, we have provided a few examples of the
types of information that can be readily extracted from this data
set. These results --- along with analogous follow-up studies --- will
provide a more complete statistical description of cluster evolution
and can be used to help understand the impact of the cluster
environment on planet formation.

$\,$

\acknowledgements 

We thank Lori Allen, Rupali Chandar, Rob Gutermuth, Tom Megeath, and
Phil Myers for useful discussions.  This work was supported in part by
the Michigan Center for Theoretical Physics. EMP acknowledges support
of the University of Michigan Physics Department through a
dissertation fellowship. FCA is supported by the NASA Origins of Solar
Systems Program via grant NNX07AP17G, and by the NSF Division of
Applied Mathematics via grant DMS-0806756.

\beSStatTableQ{$\fbnd$}{$N$}{Nfbnd}{Bound Fraction}
     100&          0.502&          0.226&          0.562&          0.269\\
     200&          0.568&          0.224&          0.590&          0.238\\
     300&          0.590&          0.218&          0.589&          0.222\\
     400&          0.620&          0.207&          0.595&          0.200\\
     500&          0.630&          0.217&          0.595&          0.193\\
     600&          0.641&          0.207&          0.592&          0.188\\
     700&          0.632&          0.207&          0.591&          0.183\\
     800&          0.632&          0.212&          0.596&          0.177\\
     900&          0.624&          0.206&          0.600&          0.170\\
    1000&          0.609&          0.199&          0.598&          0.165\\
    1100&          0.604&          0.195&          0.600&          0.159\\
    1200&          0.599&          0.198&          0.603&          0.159\\
    1300&          0.593&          0.199&          0.605&          0.155\\
    1400&          0.595&          0.201&          0.605&          0.149\\
    1500&          0.593&          0.195&          0.607&          0.149\\
    1600&          0.594&          0.196&          0.606&          0.148\\
    1700&          0.597&          0.198&          0.609&          0.141\\
    1800&          0.596&          0.197&          0.609&          0.145\\
    1900&          0.595&          0.199&          0.608&          0.142\\
    2000&          0.598&          0.200&          0.614&          0.136\\
    2100&          0.593&          0.200&          0.613&          0.133\\
    2200&          0.600&          0.203&          0.608&          0.128\\
    2300&          0.599&          0.202&          0.610&          0.124\\
    2400&          0.603&          0.202&          0.611&          0.120\\
    2500&          0.602&          0.205&          0.605&          0.122\\
    2600&          0.602&          0.204&          0.609&          0.121\\
    2700&          0.603&          0.209&          0.605&          0.115\\
    2800&          0.608&          0.210&          0.604&          0.116\\
    2900&          0.614&          0.211&          0.603&          0.111\\
    3000&          0.614&          0.211&          0.603&          0.111\\
\label{tab:FirstStatTable}
\eTable

\beSStatTableNThree{$\fbnd$}{$Q_i$}{Qfbnd}{Bound Fraction}
0.025	&	0.599	&	0.627	&	0.616	\\
0.050	&	0.587	&	0.599	&	0.585	\\
0.075	&	0.575	&	0.570	&	0.553	\\
0.100	&	0.559	&	0.539	&	0.526	\\
0.125	&	0.529	&	0.508	&	0.503	\\
0.150	&	0.507	&	0.478	&	0.480	\\
0.175	&	0.487	&	0.447	&	0.457	\\
0.200	&	0.464	&	0.416	&	0.436	\\
0.225	&	0.433	&	0.393	&	0.416	\\
0.250	&	0.396	&	0.375	&	0.396	\\
0.275	&	0.367	&	0.355	&	0.368	\\
0.300	&	0.347	&	0.333	&	0.342	\\
0.325	&	0.317	&	0.311	&	0.315	\\
0.350	&	0.300	&	0.297	&	0.289	\\
0.375	&	0.274	&	0.279	&	0.267	\\
0.400	&	0.261	&	0.263	&	0.249	\\
0.425	&	0.247	&	0.245	&	0.233	\\
0.450	&	0.236	&	0.229	&	0.218	\\
0.475	&	0.231	&	0.217	&	0.208	\\
0.500	&	0.217	&	0.203	&	0.199	\\
\eTable

\beSStatTableNThree{$\fbnd$}{$\rsc$}{Rfbnd}{Bound Fraction}
0.33	&	0.414	&	0.562	&	0.586	\\
0.50	&	0.501	&	0.583	&	0.597	\\
0.67	&	0.546	&	0.597	&	0.592	\\
0.75	&	0.559	&	0.590	&	0.636	\\
1.00	&	0.591	&	0.609	&	0.592	\\
1.33	&	0.598	&	0.703	&	0.780	\\
1.50	&	0.628	&	0.780	&	0.822	\\
2.00	&	0.792	&	0.700	&	0.590	\\
3.00	&	0.507	&	0.410	&	0.443	\\
\eTable

\beSStatTableNTwo{$\fbnd$}{$\sfe$}{Sfbnd}{Bound Fraction}
0.10	&	0.102	&	0.130	\\
0.15	&	0.196	&	0.246	\\
0.20	&	0.326	&	0.354	\\
0.25	&	0.426	&	0.506	\\
0.30	&	0.520	&	0.580	\\
0.40	&	0.696	&	0.711	\\
0.50	&	0.782	&	0.864	\\
0.75	&	0.843	&	0.946	\\
\eTable

 \beSStatTableNThree{$\fbnd$}{$\tgas$}{Cfbnd}{Bound Fraction}
1.0	&	0.439	&	0.496	&	0.666	\\
1.5	&	0.519	&	0.397	&	0.331	\\
2.0	&	0.697	&	0.602	&	0.503	\\
2.5	&	0.636	&	0.761	&	0.685	\\
3.0	&	0.554	&	0.735	&	0.794	\\
3.5	&	0.564	&	0.633	&	0.747	\\
4.0	&	0.601	&	0.577	&	0.659	\\
4.5	&	0.596	&	0.588	&	0.600	\\
5.0	&	0.592	&	0.618	&	0.595	\\
5.5	&	0.591	&	0.647	&	0.603	\\
6.0	&	0.593	&	0.667	&	0.633	\\
6.5	&	0.589	&	0.657	&	0.670	\\
7.0	&	0.600	&	0.641	&	0.696	\\
\eTable

\beSStatTableNThree{$\fbnd$}{$\mseg$}{Mfbnd}{Bound Fraction}
1/$N$	&	0.598	&	0.634	&	0.610	\\
0.05	&	0.600	&	0.660	&	0.630	\\
0.10	&	0.601	&	0.675	&	0.651	\\
0.20	&	0.602	&	0.691	&	0.668	\\
0.30	&	0.604	&	0.700	&	0.671	\\
0.40	&	0.612	&	0.699	&	0.670	\\
0.50	&	0.609	&	0.703	&	0.675	\\
0.60	&	0.610	&	0.703	&	0.672	\\
0.70	&	0.606	&	0.703	&	0.675	\\
0.80	&	0.608	&	0.707	&	0.673	\\
0.90	&	0.604	&	0.705	&	0.673	\\
0.99	&	0.604	&	0.708	&	0.672	\\
\eTable
\clearpage

\beRminTableQ{$N$}{NRmin}
100	&	0.2580	&	1.03	&	0.0933	&	0.66	&	0.1240	&	1.09	&	0.0434	&	0.61	\\
200	&	0.1630	&	1.24	&	0.0263	&	1.30	&	0.1170	&	1.32	&	0.0297	&	1.01	\\
300	&	0.1320	&	1.35	&	0.0324	&	1.03	&	0.1200	&	1.43	&	0.0269	&	1.17	\\
400	&	0.1370	&	1.31	&	0.0188	&	1.37	&	0.1310	&	1.50	&	0.0230	&	1.38	\\
500	&	0.0939	&	1.58	&	0.0183	&	1.32	&	0.1370	&	1.54	&	0.0187	&	1.60	\\
600	&	0.0917	&	1.62	&	0.0127	&	1.55	&	0.1270	&	1.69	&	0.0262	&	1.41	\\
700	&	0.0846	&	1.64	&	0.0140	&	1.41	&	0.1300	&	1.74	&	0.0260	&	1.47	\\
800	&	0.0860	&	1.63	&	0.0151	&	1.34	&	0.1530	&	1.69	&	0.0231	&	1.62	\\
900	&	0.0766	&	1.72	&	0.0101	&	1.64	&	0.1620	&	1.70	&	0.0323	&	1.40	\\
1000	&	0.0770	&	1.69	&	0.0082	&	1.75	&	0.1640	&	1.74	&	0.0250	&	1.66	\\
1100	&	0.0679	&	1.80	&	0.0088	&	1.69	&	0.1660	&	1.78	&	0.0271	&	1.64	\\
1200	&	0.0656	&	1.82	&	0.0093	&	1.62	&	0.1830	&	1.76	&	0.0278	&	1.66	\\
1300	&	0.0674	&	1.80	&	0.0093	&	1.60	&	0.1770	&	1.82	&	0.0271	&	1.72	\\
1400	&	0.0667	&	1.81	&	0.0073	&	1.78	&	0.1800	&	1.84	&	0.0334	&	1.57	\\
1500	&	0.0619	&	1.86	&	0.0080	&	1.67	&	0.1900	&	1.83	&	0.0287	&	1.71	\\
1600	&	0.0602	&	1.88	&	0.0083	&	1.64	&	0.1960	&	1.84	&	0.0364	&	1.57	\\
1700	&	0.0627	&	1.85	&	0.0070	&	1.77	&	0.2030	&	1.85	&	0.0351	&	1.62	\\
1800	&	0.0605	&	1.88	&	0.0072	&	1.73	&	0.2020	&	1.87	&	0.0313	&	1.73	\\
1900	&	0.0599	&	1.87	&	0.0078	&	1.66	&	0.2100	&	1.87	&	0.0337	&	1.70	\\
2000	&	0.0575	&	1.90	&	0.0063	&	1.82	&	0.2180	&	1.88	&	0.0339	&	1.72	\\
2100	&	0.0578	&	1.89	&	0.0062	&	1.81	&	0.2170	&	1.90	&	0.0367	&	1.68	\\
2200	&	0.0563	&	1.91	&	0.0060	&	1.83	&	0.2330	&	1.87	&	0.0379	&	1.66	\\
2300	&	0.0547	&	1.93	&	0.0059	&	1.83	&	0.2300	&	1.91	&	0.0367	&	1.73	\\
2400	&	0.0569	&	1.90	&	0.0060	&	1.80	&	0.2370	&	1.91	&	0.0466	&	1.54	\\
2500	&	0.0547	&	1.92	&	0.0052	&	1.92	&	0.2410	&	1.92	&	0.0402	&	1.68	\\
2600	&	0.0547	&	1.92	&	0.0059	&	1.83	&	0.2430	&	1.92	&	0.0476	&	1.56	\\
2700	&	0.0549	&	1.92	&	0.0055	&	1.86	&	0.2510	&	1.93	&	0.0433	&	1.67	\\
2800	&	0.0540	&	1.93	&	0.0061	&	1.76	&	0.2550	&	1.93	&	0.0470	&	1.61	\\
2900	&	0.0522	&	1.95	&	0.0057	&	1.80	&	0.2540	&	1.95	&	0.0363	&	1.84	\\
3000	&	0.0533	&	1.93	&	0.0050	&	1.92	&	0.2640	&	1.93	&	0.0395	&	1.78	\\
\eTable

\beRminNThree{$Q_i$}{QRmin}
0.025	&	0.1660	&	1.30	&	0.0865	&	1.75	&	0.0714	&	1.90	\\
0.050	&	0.1210	&	1.39	&	0.0666	&	1.73	&	0.0527	&	1.87	\\
0.075	&	0.1165	&	1.28	&	0.0559	&	1.69	&	0.0400	&	1.89	\\
0.100	&	0.1231	&	1.14	&	0.0466	&	1.69	&	0.0328	&	1.88	\\
0.125	&	0.1066	&	1.14	&	0.0394	&	1.69	&	0.0279	&	1.86	\\
0.150	&	0.0850	&	1.21	&	0.0339	&	1.70	&	0.0242	&	1.85	\\
0.175	&	0.0875	&	1.12	&	0.0309	&	1.66	&	0.0206	&	1.86	\\
0.200	&	0.0820	&	1.10	&	0.0250	&	1.73	&	0.0181	&	1.86	\\
0.225	&	0.0774	&	1.06	&	0.0225	&	1.73	&	0.0158	&	1.87	\\
0.250	&	0.0547	&	1.25	&	0.0196	&	1.75	&	0.0155	&	1.79	\\
0.275	&	0.0523	&	1.20	&	0.0180	&	1.73	&	0.0135	&	1.82	\\
0.300	&	0.0580	&	1.07	&	0.0168	&	1.71	&	0.0127	&	1.78	\\
0.325	&	0.0514	&	1.08	&	0.0172	&	1.61	&	0.0109	&	1.83	\\
0.350	&	0.0399	&	1.21	&	0.0127	&	1.79	&	0.0106	&	1.78	\\
0.375	&	0.0401	&	1.15	&	0.0112	&	1.81	&	0.0104	&	1.73	\\
0.400	&	0.0336	&	1.22	&	0.0115	&	1.74	&	0.0089	&	1.78	\\
0.425	&	0.0323	&	1.18	&	0.0097	&	1.81	&	0.0088	&	1.73	\\
0.450	&	0.0358	&	1.06	&	0.0141	&	1.44	&	0.0076	&	1.79	\\
0.475	&	0.0334	&	1.08	&	0.0102	&	1.64	&	0.0071	&	1.78	\\
0.500	&	0.0311	&	1.06	&	0.0097	&	1.62	&	0.0068	&	1.75	\\
\eTable

\beRminNThree{$\rsc$}{RRmin}
0.33	&	2.0550	&	1.36	&	1.6120	&	1.69	&	1.4570	&	1.79	\\
0.50	&	0.7730	&	1.40	&	0.4800	&	1.78	&	0.4060	&	1.89	\\
0.67	&	0.3330	&	1.47	&	0.2120	&	1.75	&	0.1760	&	1.88	\\
0.75	&	0.2270	&	1.53	&	0.1620	&	1.71	&	0.1250	&	1.89	\\
1.00	&	0.1290	&	1.39	&	0.0716	&	1.75	&	0.0588	&	1.88	\\
1.33	&	0.0624	&	1.42	&	0.0342	&	1.75	&	0.0266	&	1.90	\\
1.50	&	0.0487	&	1.39	&	0.0245	&	1.77	&	0.0199	&	1.88	\\
2.00	&	0.0207	&	1.52	&	0.0126	&	1.75	&	0.0099	&	1.89	\\
3.00	&	0.0084	&	1.57	&	0.0047	&	1.74	&	0.0033	&	1.86	\\
\eTable

\beRminNTwo{$\sfe$}{SRmin}
0.10	&	0.136	&	1.73	&	0.0777	&	1.98	\\
0.15	&	0.138	&	1.53	&	0.0717	&	1.90	\\
0.20	&	0.123	&	1.52	&	0.0704	&	1.83	\\
0.25	&	0.146	&	1.32	&	0.0671	&	1.83	\\
0.30	&	0.134	&	1.36	&	0.0690	&	1.79	\\
0.40	&	0.155	&	1.25	&	0.0781	&	1.71	\\
0.50	&	0.198	&	1.13	&	0.0894	&	1.69	\\
0.75	&	0.288	&	1.03	&	0.1480	&	1.65	\\
\eTable

\beRminNThree{$\tgas$}{CRmin}
1.0	&	0.0618	&	1.04	&	0.0377	&	1.20	&	0.0196	&	1.68	\\
1.5	&	0.0922	&	1.02	&	0.0355	&	1.59	&	0.0259	&	1.79	\\
2.0	&	0.0721	&	1.33	&	0.0413	&	1.65	&	0.0306	&	1.87	\\
2.5	&	0.0799	&	1.38	&	0.0431	&	1.73	&	0.0344	&	1.88	\\
3.0	&	0.1131	&	1.20	&	0.0551	&	1.64	&	0.0410	&	1.84	\\
3.5	&	0.0953	&	1.42	&	0.0639	&	1.62	&	0.0455	&	1.85	\\
4.0	&	0.1014	&	1.45	&	0.0621	&	1.73	&	0.0503	&	1.86	\\
4.5	&	0.1019	&	1.51	&	0.0726	&	1.68	&	0.0557	&	1.86	\\
5.0	&	0.1160	&	1.47	&	0.0777	&	1.69	&	0.0599	&	1.87	\\
5.5	&	0.1090	&	1.56	&	0.0826	&	1.70	&	0.0644	&	1.88	\\
6.0	&	0.1220	&	1.53	&	0.0847	&	1.73	&	0.0678	&	1.89	\\
6.5	&	0.1430	&	1.45	&	0.0891	&	1.74	&	0.0709	&	1.90	\\
7.0	&	0.1310	&	1.55	&	0.0941	&	1.74	&	0.0759	&	1.89	\\
\eTable

\beRminNThree{$\mseg$}{MRmin}
1/$N$	&	0.187	&	1.23	&	0.134	&	1.42	&	0.193	&	1.16	\\
0.050	&	0.406	&	0.89	&	0.371	&	0.94	&	0.406	&	0.93	\\
0.100	&	0.464	&	0.86	&	0.446	&	0.91	&	0.444	&	0.94	\\
0.200	&	0.528	&	0.84	&	0.487	&	0.91	&	0.450	&	0.98	\\
0.300	&	0.588	&	0.81	&	0.480	&	0.93	&	0.455	&	0.98	\\
0.400	&	0.532	&	0.88	&	0.470	&	0.94	&	0.460	&	0.97	\\
0.500	&	0.585	&	0.82	&	0.495	&	0.91	&	0.446	&	1.00	\\
0.600	&	0.551	&	0.85	&	0.502	&	0.91	&	0.442	&	0.99	\\
0.700	&	0.532	&	0.88	&	0.482	&	0.93	&	0.430	&	1.00	\\
0.800	&	0.539	&	0.87	&	0.459	&	0.97	&	0.434	&	1.02	\\
0.900	&	0.559	&	0.85	&	0.474	&	0.94	&	0.428	&	1.02	\\
0.990	&	0.607	&	0.82	&	0.504	&	0.91	&	0.434	&	1.01	\\
\eTable
\clearpage

\beRTableQ{$N$}{NRProf}
100	&	0.80	&	0.44	&	2	&	0.65	&	0.55	&	3	&	0.86	&	0.51	&	2	&	0.70	&	0.68	&	3	\\
200	&	0.90	&	0.53	&	2	&	0.71	&	0.72	&	3	&	0.90	&	0.58	&	2	&	0.71	&	0.79	&	3	\\
300	&	0.94	&	0.62	&	2	&	0.71	&	0.89	&	3	&	0.94	&	0.61	&	2	&	0.72	&	0.88	&	3	\\
400	&	0.96	&	0.68	&	2	&	0.73	&	0.99	&	3	&	0.95	&	0.64	&	2	&	0.72	&	0.94	&	3	\\
500	&	0.97	&	0.75	&	2	&	0.73	&	1.12	&	3	&	0.96	&	0.68	&	2	&	0.72	&	1.00	&	3	\\
600	&	0.97	&	0.81	&	2	&	0.74	&	1.21	&	3	&	0.95	&	0.72	&	2	&	0.74	&	1.02	&	3	\\
700	&	0.98	&	0.87	&	2	&	0.75	&	1.30	&	3	&	0.96	&	0.73	&	2	&	0.73	&	1.07	&	3	\\
800	&	0.98	&	0.92	&	2	&	0.75	&	1.37	&	3	&	0.96	&	0.76	&	2	&	0.73	&	1.11	&	3	\\
900	&	0.99	&	0.95	&	2	&	0.75	&	1.46	&	3	&	0.97	&	0.77	&	2	&	0.73	&	1.14	&	3	\\
1000	&	1.00	&	0.99	&	2	&	0.75	&	1.55	&	3	&	0.97	&	0.79	&	2	&	0.75	&	1.14	&	3	\\
1100	&	1.00	&	1.03	&	2	&	0.76	&	1.59	&	3	&	0.98	&	0.80	&	2	&	0.74	&	1.17	&	3	\\
1200	&	0.97	&	1.09	&	2	&	0.76	&	1.66	&	3	&	0.98	&	0.80	&	2	&	0.74	&	1.20	&	3	\\
1300	&	0.96	&	1.13	&	2	&	0.76	&	1.72	&	3	&	0.97	&	0.83	&	2	&	0.74	&	1.22	&	3	\\
1400	&	0.95	&	1.18	&	2	&	0.76	&	1.80	&	3	&	0.97	&	0.86	&	2	&	0.74	&	1.25	&	3	\\
1500	&	0.96	&	1.21	&	2	&	0.77	&	1.84	&	3	&	0.97	&	0.87	&	2	&	0.74	&	1.28	&	3	\\
1600	&	0.96	&	1.24	&	2	&	0.76	&	1.92	&	3	&	0.96	&	0.90	&	2	&	0.74	&	1.29	&	3	\\
1700	&	0.92	&	1.32	&	2	&	0.76	&	1.98	&	3	&	0.97	&	0.90	&	2	&	0.74	&	1.31	&	3	\\
1800	&	0.93	&	1.34	&	2	&	0.77	&	2.03	&	3	&	0.97	&	0.92	&	2	&	0.74	&	1.33	&	3	\\
1900	&	0.93	&	1.37	&	2	&	0.76	&	2.10	&	3	&	0.97	&	0.92	&	2	&	0.74	&	1.35	&	3	\\
2000	&	0.94	&	1.39	&	2	&	0.76	&	2.15	&	3	&	0.97	&	0.93	&	2	&	0.74	&	1.37	&	3	\\
2100	&	0.93	&	1.43	&	2	&	0.77	&	2.18	&	3	&	0.97	&	0.95	&	2	&	0.74	&	1.38	&	3	\\
2200	&	0.94	&	1.44	&	2	&	0.77	&	2.23	&	3	&	0.96	&	0.96	&	2	&	0.74	&	1.40	&	3	\\
2300	&	0.94	&	1.48	&	2	&	0.77	&	2.29	&	3	&	0.97	&	0.96	&	2	&	0.74	&	1.41	&	3	\\
2400	&	0.94	&	1.50	&	2	&	0.77	&	2.34	&	3	&	0.97	&	0.97	&	2	&	0.74	&	1.42	&	3	\\
2500	&	0.95	&	1.52	&	2	&	0.77	&	2.40	&	3	&	0.97	&	0.97	&	2	&	0.74	&	1.44	&	3	\\
2600	&	0.95	&	1.54	&	2	&	0.77	&	2.43	&	3	&	0.97	&	0.99	&	2	&	0.74	&	1.46	&	3	\\
2700	&	0.94	&	1.58	&	2	&	0.77	&	2.48	&	3	&	0.97	&	1.00	&	2	&	0.75	&	1.44	&	3	\\
2800	&	0.94	&	1.61	&	2	&	0.77	&	2.54	&	3	&	0.96	&	1.02	&	2	&	0.75	&	1.46	&	3	\\
2900	&	0.95	&	1.63	&	2	&	0.77	&	2.58	&	3	&	0.97	&	1.02	&	2	&	0.75	&	1.47	&	3	\\
3000	&	0.95	&	1.65	&	2	&	0.78	&	2.59	&	3	&	0.97	&	1.03	&	2	&	0.75	&	1.49	&	3	\\
\eTable

\beRNThree{$Q_i$}{QRProf}
0.025	&	0.89	&	0.63	&	2	&	0.95	&	1.02	&	2	&	0.87	&	1.47	&	2 \\
0.050	&	0.95	&	0.61	&	2	&	1.01	&	0.99	&	2	&	0.96	&	1.39	&	2 \\
0.075	&	1.00	&	0.60	&	2	&	1.07	&	0.97	&	2	&	1.04	&	1.34	&	2 \\
0.100	&	1.02	&	0.60	&	2	&	1.10	&	0.97	&	2	&	1.09	&	1.33	&	2 \\
0.125	&	1.05	&	0.60	&	2	&	1.15	&	0.96	&	2	&	1.13	&	1.34	&	2 \\
0.150	&	1.07	&	0.61	&	2	&	1.17	&	0.97	&	2	&	1.16	&	1.36	&	2 \\
0.175	&	1.09	&	0.61	&	2	&	1.18	&	0.99	&	2	&	1.17	&	1.40	&	2 \\
0.200	&	1.10	&	0.62	&	2	&	1.19	&	1.01	&	2	&	1.20	&	1.41	&	2 \\
0.225	&	1.12	&	0.62	&	2	&	1.20	&	1.03	&	2	&	1.20	&	1.46	&	2 \\
0.250	&	0.65	&	0.81	&	3	&	0.71	&	1.34	&	3	&	0.71	&	1.89	&	3 \\
0.275	&	0.67	&	0.80	&	3	&	0.72	&	1.35	&	3	&	0.72	&	1.91	&	3 \\
0.300	&	0.67	&	0.81	&	3	&	0.73	&	1.36	&	3	&	0.73	&	1.93	&	3 \\
0.325	&	0.69	&	0.80	&	3	&	0.73	&	1.39	&	3	&	0.74	&	1.94	&	3 \\
0.350	&	0.69	&	0.82	&	3	&	0.74	&	1.40	&	3	&	0.75	&	1.95	&	3 \\
0.375	&	0.69	&	0.83	&	3	&	0.75	&	1.41	&	3	&	0.75	&	1.99	&	3 \\
0.400	&	0.70	&	0.84	&	3	&	0.75	&	1.44	&	3	&	0.75	&	2.02	&	3 \\
0.425	&	0.70	&	0.85	&	3	&	0.75	&	1.46	&	3	&	0.76	&	2.03	&	3 \\
0.450	&	0.71	&	0.85	&	3	&	0.75	&	1.49	&	3	&	0.76	&	2.07	&	3 \\
0.475	&	0.70	&	0.88	&	3	&	0.75	&	1.51	&	3	&	0.76	&	2.11	&	3 \\
0.500	&	0.71	&	0.89	&	3	&	0.75	&	1.54	&	3	&	0.76	&	2.15	&	3 \\
\eTable 

\beRNThree{$\rsc$}{RRProf}
0.33	&	0.72	&	0.30	&	2	&	0.87	&	0.41	&	2	&	0.91	&	0.53	&	2	\\
0.50	&	0.80	&	0.39	&	2	&	0.94	&	0.56	&	2	&	0.94	&	0.78	&	2	\\
0.67	&	0.87	&	0.46	&	2	&	0.96	&	0.73	&	2	&	0.98	&	0.98	&	2	\\
0.75	&	0.89	&	0.50	&	2	&	0.97	&	0.80	&	2	&	1.00	&	1.07	&	2	\\
1.00	&	0.92	&	0.62	&	2	&	0.99	&	1.00	&	2	&	0.94	&	1.39	&	2	\\
1.33	&	0.97	&	0.74	&	2	&	0.95	&	1.31	&	2	&	0.95	&	1.85	&	2	\\
1.50	&	0.98	&	0.81	&	2	&	0.94	&	1.51	&	2	&	0.97	&	2.00	&	2	\\
2.00	&	0.95	&	1.09	&	2	&	0.99	&	1.71	&	2	&	1.02	&	2.20	&	2	\\
3.00	&	0.97	&	1.33	&	2	&	0.99	&	2.36	&	2	&	0.99	&	3.26	&	2	\\
\eTable

\beRNTwo{$\sfe$}{SRProf}
0.10	&	0.86	&	0.55	&	2	&	0.91	&	0.98	&	2	\\
0.15	&	0.88	&	0.58	&	2	&	0.94	&	0.99	&	2	\\
0.20	&	0.91	&	0.59	&	2	&	0.97	&	0.98	&	2	\\
0.25	&	0.92	&	0.60	&	2	&	0.97	&	1.02	&	2	\\
0.30	&	0.93	&	0.61	&	2	&	0.99	&	1.00	&	2	\\
0.40	&	0.95	&	0.60	&	2	&	0.99	&	0.97	&	2	\\
0.50	&	0.94	&	0.60	&	2	&	0.96	&	0.95	&	2	\\
0.75	&	0.90	&	0.61	&	2	&	0.87	&	0.92	&	2	\\
\eTable

\beRNThree{$\tgas$}{CRProf}
1.0	&	0.96	&	0.94	&	2	&	0.92	&	1.62	&	2	&	0.97	&	2.11	&	2	\\
1.5	&	0.92	&	0.92	&	2	&	0.92	&	1.42	&	2	&	0.92	&	1.83	&	2	\\
2.0	&	0.92	&	0.89	&	2	&	0.93	&	1.42	&	2	&	0.90	&	1.93	&	2	\\
2.5	&	0.91	&	0.82	&	2	&	0.89	&	1.47	&	2	&	0.91	&	1.88	&	2	\\
3.0	&	0.91	&	0.75	&	2	&	0.90	&	1.34	&	2	&	0.91	&	1.84	&	2	\\
3.5	&	0.92	&	0.71	&	2	&	0.94	&	1.18	&	2	&	0.92	&	1.70	&	2	\\
4.0	&	0.92	&	0.68	&	2	&	0.97	&	1.08	&	2	&	0.92	&	1.55	&	2	\\
4.5	&	0.93	&	0.65	&	2	&	0.98	&	1.04	&	2	&	0.93	&	1.45	&	2	\\
5.0	&	0.94	&	0.61	&	2	&	0.98	&	1.01	&	2	&	0.93	&	1.40	&	2	\\
5.5	&	0.96	&	0.58	&	2	&	1.00	&	0.96	&	2	&	0.93	&	1.36	&	2	\\
6.0	&	0.96	&	0.56	&	2	&	1.02	&	0.91	&	2	&	0.95	&	1.30	&	2	\\
6.5	&	0.97	&	0.54	&	2	&	1.04	&	0.86	&	2	&	0.96	&	1.26	&	2	\\
7.0	&	0.98	&	0.52	&	2	&	1.05	&	0.83	&	2	&	0.98	&	1.20	&	2	\\
\eTable

\beRNThree{$\mseg$}{MRProf}
1/$N$	&	0.87	&	0.64	&	2	&	0.87	&	1.09	&	2	&	0.83	&	1.52	&	2	\\
0.05	&	0.77	&	0.71	&	2	&	0.77	&	1.21	&	2	&	0.75	&	1.65	&	2	\\
0.10	&	0.73	&	0.74	&	2	&	0.74	&	1.24	&	2	&	0.73	&	1.68	&	2	\\
0.20	&	0.71	&	0.76	&	2	&	0.72	&	1.25	&	2	&	0.70	&	1.73	&	2	\\
0.30	&	0.70	&	0.77	&	2	&	0.72	&	1.25	&	2	&	0.69	&	1.76	&	2	\\
0.40	&	0.70	&	0.76	&	2	&	0.72	&	1.25	&	2	&	0.69	&	1.77	&	2	\\
0.50	&	0.72	&	0.74	&	2	&	0.72	&	1.25	&	2	&	0.68	&	1.79	&	2	\\
0.60	&	0.71	&	0.74	&	2	&	0.72	&	1.25	&	2	&	0.68	&	1.79	&	2	\\
0.70	&	0.71	&	0.74	&	2	&	0.72	&	1.26	&	2	&	0.68	&	1.80	&	2	\\
0.80	&	0.70	&	0.76	&	2	&	0.72	&	1.25	&	2	&	0.68	&	1.79	&	2	\\
0.90	&	0.70	&	0.75	&	2	&	0.72	&	1.25	&	2	&	0.68	&	1.79	&	2	\\
0.99	&	0.70	&	0.75	&	2	&	0.71	&	1.27	&	2	&	0.68	&	1.79	&	2	\\
\eTable

\end{document}